\documentclass[12pt,english]{article}
\usepackage[T1]{fontenc}
\usepackage[utf8]{inputenc}
\usepackage{amsmath,amssymb}
\usepackage{amsfonts,mathrsfs}
\usepackage{hyperref}
\usepackage{here}
\usepackage{tba-math}

\usepackage[dvipdfmx]{graphicx}

\usepackage{subfigure,afterpage}
\usepackage{epsfig,color}
\usepackage{relsize,adjustbox}
\usepackage{simplewick}

\newif\ifusetikz
\usetikzfalse

\ifusetikz
\usepackage{xparse}
\usepackage{tikz}
\usetikzlibrary{arrows.meta,backgrounds,calc,external,graphs,matrix}
\pgfdeclarelayer{myback}
\pgfsetlayers{myback,background,main}
\usepackage{tikz-det}
\tikzset{wsquare/.style={rectangle, draw, fill=white},
darrow/.style={thick, arrows = {-Latex[width=6pt, length=6pt]}}
}
\fi

\renewcommand{\baselinestretch}{1.15}
\numberwithin{equation}{section}

\def\adg#1{{\color [rgb] {0.1,0.6,0.1} #1}}

\begin{document}
\null \hfill {\footnotesize 
OIQP-15-01}

\ \vskip 2cm

\begin{center}
\LARGE

\mbox{\bf Negative anomalous dimensions in $\cN=4$ SYM}


\vskip 2cm
\renewcommand*{\thefootnote}{\fnsymbol{footnote}}

\large
\centerline{Yusuke Kimura\footnote{{\tt londonmileend@gmail.com}} and Ryo Suzuki\footnote{{\tt Ryo.Suzuki@maths.ox.ac.uk}}}

\vskip 1cm

\normalsize
{\it
$^*$Okayama Institute for Quantum Physics (OIQP),\\
Kyoyama 1-9-1, Kita-ku, Okayama, 700-0015, JAPAN\\[2mm]
$^\dagger$Mathematical Institute, University of Oxford,\\
Andrew Wiles Building, Radcliffe Observatory Quarter,\\
Woodstock Road, Oxford, OX2 6GG, UK
}

\end{center}

\vskip 14mm

\centerline{\bf Abstract}

\vskip 6mm

We elucidate aspects of the one-loop anomalous dimension of $\alg{so}(6)$-singlet multi-trace operators in $\cN=4\ SU(N_c)$ SYM at finite $N_c$\,.
First, we study how $1/N_c$ corrections lift the large $N_c$ degeneracy of the spectrum, which we call the operator submixing problem. 
We observe that all large $N_c$ zero modes acquire anomalous dimensions starting at order $1/N_c^2$ with a non-positive coefficient and they mix only among the operators with the same number of traces at leading order.
Second, we study the lowest one-loop dimension of operators of length equal to $2N_c$\,.
The dimension of such operators becomes more negative as $N_c$ increases, which will 
eventually diverge in the double scaling limit.
Third, we examine the structure of level-crossing at finite $N_c$ in view of unitarity.
Finally we find out a correspondence between the large $N_c$ zero modes and completely symmetric polynomials of Mandelstam variables.

\vskip 2mm

\vfill
\thispagestyle{empty}
\setcounter{page}{0}
\setcounter{footnote}{0}
\renewcommand*{\thefootnote}{\arabic{footnote}}
\newpage

\noindent\rule{\textwidth}{0.2pt}

\vspace{-3mm}
\tableofcontents

\vspace{3mm}

\noindent\rule{\textwidth}{0.2pt}

\section{Introduction}\label{sec:intro}

Large $N_c$ gauge theories have been intensively studied for decades since the work of 't Hooft, who discovered that mesons are described by a string in the planar limit \cite{tHooft73a}.
Maldacena argued that a $d$-dimensional gauge theory can be described by a $(d+1)$-dimensional gravity theory, which is now called the AdS/CFT correspondence \cite{Maldacena97}.

The primary example of the AdS/CFT correspondence is the one between maximally supersymmetric Yang-Mills theory in four dimensions ($\cN=4$ SYM) and superstring on \AdSxS\ in the planar limit. Integrability is a powerful technique in this setup, which makes it possible to compute conformal dimensions of gauge-invariant operators in $\cN=4$ SYM and the energy of \AdSxS\ superstring states at any value of the 't Hooft coupling (see \cite{AdSCFTrev10} for a review).
In contrast, the non-planar problems are notoriously complicated and less well-understood.
The entire problem seems daunting, but we shall present a tractable corner of it.
The purpose of this paper is to look for simple structures in the non-planar problem and to provide a new point of view on the large $N_c$ problems.

Our motivation is explained in three ways.
The first one is to better understand the AdS/CFT correspondence at finite $N_c$\,.

The second one is to study operators with negative anomalous dimensions. Since all planar anomalous dimensions are non-negative, negative anomalous dimensions are inevitably a non-planar effect.
In addition, $1/N_c$ corrections to the dimension of double-trace operators are related to four-point functions.
The four-point functions on the gravity side are related to the phase shift of a high-energy scattering \cite{CCPS06a,CCPS06b,CCP07}, which must be positive in order to preserve the asymptotic causality of spacetime \cite{CEMZ14}.
Thus, the AdS/CFT and causality predict that certain double-trace operators should have negative anomalous dimensions.
We want to observe similar phenomena in $\cN=4$ SYM, though the operators of concern are different from \cite{CEMZ14}.

The third one is to inspect the large $N_c$ behavior of determinant-like operators with all non-planar corrections taken into consideration. 
It was found in \cite{BDHNPSS13,Hegedus15} that, if we take the na\"ive 't Hooft limit and apply integrability methods, then the dimensions of the double-determinant operators, which correspond to the energies of a pair of open string tachyons between a D-brane and anti-D-brane, becomes either divergent or complex at two-loops. 
Since any operators of $\cN=4$ SYM have perturbatively real and finite anomalous dimensions at finite $N_c$\,, such pathological behavior should be remedied with non-planar effects.

Given this situation, it is worthwhile to look for an alternative to integrability to study the non-planar problem. A promising approach is the group theoretical method 
initiated by \cite{CJR01}.
This method is based on the construction of a basis of local operators that diagonalizes 
the free two-point functions.
The work \cite{CJR01} dealt with the 1/2 BPS sector, and later it was extended to more general sectors. 
The basis of \cite{BBFH04,BCdM08} is suitable for describing open string excitations on giant gravitons, 
and the bases in \cite{KR07,Kimura09a,Kimura12} were designed by Brauer algebras. 
The flavor symmetry is manifested in the bases of \cite{BHR07,BHR08}. 
One common feature of the diagonal bases is that 
they are labeled by a set of Young diagrams. 
It was also observed that 
the one-loop mixing is highly constrained on these bases --- 
two operators can mix if their Young diagrams 
are related by moving a single box of the Young diagram 
\cite{dMSS07b,Brown08,dMMP10,dCdMJ10}. 
This property makes the problem tractable, and in certain situations it is possible to obtain the spectrum of the dilatation operator explicitly
\cite{CKL11,DGKdM11,dMDS11,dMR12,dMGM13,Lin14}. 
Although we have realized these bases are useful tools for the non-planar mixing problem, 
most results have been limited to the $\alg{su}(2)$ sector. 
One motivation of this work is 
to explore the non-planar mixing problem of the $\alg{so}(6)$ sector.\footnote{The $\alg{su}(2)$ sector consists of gauge-invariant local operators made out of two holomorphic scalars, which are closed under the operator mixing to all orders of perturbation theory in $\cN=4$ SYM. The $\alg{so}(6)$ sector is made out of six real scalars and closed at one-loop.}

The non-planar dilatation operator in the one-loop $\alg{so}(6)$ sector was written down in \cite{BKPSS02,BKPS02b},
\begin{equation}
\fD_\text{one-loop} = \frac{1}{N_c} :\! \left(
-\frac{1}{2} \, \tr [\Phi_{m},\Phi_{n}][\check{\Phi}^{m},\check{\Phi}^{n}]
-\frac{1}{4} \, \tr [\Phi_{m},\check{\Phi}^{n}] [\Phi_{m},\check{\Phi}^{n}]
\right) \! : \,,
\label{one-loop dilatation phi}
\end{equation}
and the general non-planar spectrum has been studied in many ways \cite{Kristjansen10rev,Velizhanin09a}.
In the planar limit, the Bethe Ansatz Equations give us the spectrum of all single-trace operators in the $\alg{so}(6)$ sector \cite{MZ02}. 
Naturally, one hopes to know the spectrum of all multi-trace operators at arbitrary $N_c$\,. 
This problem has not been studied thoroughly. 
One of the main difficulties is that the number of operators involved in the mixing grows factorially with respect to the operator length $L$. 
For example, there are 469 scalar $\alg{so}(6)$ singlet multi-trace operators at $L=10$, and 4477 operators at $L=12$.
By brute-force computation, we managed to compute explicitly the matrix elements of the non-planar mixing and obtained their eigenvalues up to $L=10$.
Various interesting properties of our results are explained below.

\bigskip
At $N_c=\infty$, the anomalous dimensions are highly degenerate, and this is why the planar mixing problem is integrable. 
Most of the large $N_c$ degeneracy are lifted by $1/N_c$ corrections. 
We focus on the degenerate eigenstate having the zero anomalous dimension at large $N_c$\,, which we call the {\it operator submixing problem}.
The submixing problem at $\cO(1/N_c^2)$ can be solved by diagonalizing the Hamiltonian
\begin{equation}
H_\text{sm}^\circ = \lim \limits_{\epsilon \to 0}
P_{\circ } \[ N_c \, \fD_1 - \fD_1 \( \fD_0 + \epsilon \)^{-1} \fD_1 \], \qquad
\fD_\text{one-loop} = \fD_{0} + \frac{1}{N_c} \, \fD_{1} \,,
\label{intro Hsm}
\end{equation}
where $P_{\circ}$ is the projector to the space of the large $N_c$ zero modes. 
We observed remarkable patterns among the solutions up to $L=10$, and conjecture that these patterns continue for any $L$.
In particular, each eigenstate is a sum of multi-trace operators with the same number of traces, and all eigenvalues are non-positive:
\begin{equation}
\[ \, H_\text{sm}^\circ \,, \# \(\text{traces}\) \, \] = 0, \qquad
\gamma_2 \le 0, \qquad H_\text{sm}^\circ \, \psi = \gamma_2 \, \psi \,,
\end{equation}
where $\gamma = \sum_{n=0} N_c^{-n} \, \gamma_n$ is the one-loop anomalous dimension.

Besides the dilatation operator, higher-point correlation functions offer an alternative route to compute the $1/N_c$ corrections to the operator dimension or string energy. For example, by studying the OPE expansion of the four-point functions of single-trace operators, one can compute the $1/N_c$ correction to the dimension of intermediate states, which are double-trace operators at large $N_c$\,.
This problem has been explored in the literature on the \AdSxS\ side \cite{DMMR99,AFP00a,Uruchurtu07,MP14,Goncalves14}, the SYM side \cite{BKRS99,ADOS02,APPSS02,APSS03}, and via the conformal bootstrap \cite{DO01,BDHO06,AB14,ABL14,KSS15}.
Following this line of development, we consider four-point functions of multi-trace operators in Section \ref{sec:rel correlators}, and argue that they constrain the eigensystem of submixing matrices.

In Section \ref{sec:spec FAN} we discuss the following aspects of the finite $N_c$ spectrum:

\smallskip \noindent 
$\bullet$ The one-loop dimension of determinant-like operators generally receives non-planar corrections even in the large $N_c$ limit, because their operator length is of order $N_c$\,. The proper way to take the large $N_c$ limit is to study their dimension at a finite $N_c \sim L$ and extrapolate the results to $N_c \gg 1$. This result can be different from the na\"ive limit, where we first take the large $N_c$ limit and 
identify a double-determinant-like operator as a state of an integrable open spin chain with boundaries.
We are interested in the operators with length $L=2N_c$\,, with the hope that non-planar corrections rescue the pathological behavior of the double-determinant operator found in \cite{BDHNPSS13}.
Indeed, we find an operator with $L=2N_c$ whose one-loop dimension is zero at any $N_c$\,, which may be regarded as a non-planar completion of the double-determinant operator.
In addition, we also find an operator with a lower dimension by studying dilatation eigenstates at $N_c=L/2$ up to $L=10$. 
The lowest-energy state has a negative one-loop dimension which decreases as $N_c$ increases. 
This anomalous dimension will eventually diverge in the large $N_c$ limit,\footnote{Of course, the tree-level dimension of the double-determinant is $2N_c$\,, which diverges in the large $N_c$ limit. Here we argue that the anomalous dimension in the $\alg{so}(6)$ sector diverges. This situation is different from the one in the $\alg{su}(2)$ sector, where the determinant operator $\det Z$ dual to giant graviton is BPS \cite{BBNS01}, and the determinant-like operators have finite anomalous dimensions in the large $N_c$ limit and are computed by the energy of an integrable open spin chain \cite{BV05,HM07}.}
\begin{equation}
\Delta - L \to - \infty, \qquad 
g_{\rm YM} \to 0, \ \ 
N_c = \frac{L}{2} \to \infty, \ \ 
\lambda = N_c \, g_{\rm YM}^2 : {\rm fixed} \,.
\end{equation}

\smallskip \noindent 
$\bullet$ We investigate level-crossing: whether the adjacent one-loop dimensions or energy levels cross at finite $N_c$\,. According to the non-crossing rule of von Neumann-Wigner \cite{vNW29}, there should be no level-crossing if the Hamiltonian of the system has no extra symmetry.
We find that most of the energy levels do not cross for $N_c \in [L, +\infty]$ but do collide for $N_c < L$, which can be explained by the non-Hermiticity of the operator mixing matrix and the finite $N_c$ constraints.

\smallskip \noindent 
$\bullet$ The one-loop dimensions exhibit eccentric behaviors for $N_c < L$, which makes it difficult to keep track of the dilatation eigenstates (or $\cN=4$ SYM operators) from large $N_c$ to small $N_c$\,.
In other words, when we regard the one-loop dimension as an analytic curve on the $(\gamma,N_c)$ plane, we are not able to tell which curve corresponds to which SYM operator unambiguously. To support this idea, we construct an automorphism which relates high energy states and low energy states.

\smallskip
Finally, Section \ref{sec:discussion} is devoted to Discussion and Outlook.

\bigskip
Appendices are organized as follows.
Our notation is explained in Appendix \ref{app:notation}. 
The one-loop spectrum at finite $N_c$ is summarized in Appendix \ref{app:foundations}.
The details of computations on correlation functions are given in Appendix \ref{app:large_Nf_correlator}.
The intriguing relation between the number of $\alg{so}(6)$ singlet large $N_c$ zero modes and the completely symmetric polynomial of Mandelstam variables is discussed in Appendix \ref{app:Mandelstam}.
A concise way of describing multi-trace operators is introduced in Appendix \ref{app:polynote}, where we briefly explain our idea behind {\tt Mathematica} implementation.
This paper is accompanied by a {\tt Mathematica} notebook and a data file which contain the operator mixing matrix at $L=10$ discussed in Appendix \ref{app:data}.

\section{Operator (sub)mixing problem}\label{sec:submixing}

\subsection{Synopsis}

We start with a brief review on the operator mixing problem.
In $\cN=4$ SYM at weak coupling, the two-point functions of scalar operators 
behave as\footnote{Note that $T_{AB}^{(0)} \neq 0$ in logarithmic CFT.}
\begin{equation}
\vev{\cO_A (x) \cO_B (0)} = \frac{S_{AB}^{(0)} 
+ \[ S_{AB}^{(1)} + T_{AB}^{(1)} \, \log (|x| \mu) \]}
{\abs{x}^{2 \Delta^{(0)}} } + \cO(g_{\rm YM}^4),
\label{def:SYM 2pt}
\end{equation}
where $S_{AB}^{(1)}$ and $T_{AB}^{(1)}$ are $\cO(g_{\rm YM}^2)$ and come from finite and singular counterterms, respectively \cite{APPSS02}. 
The quantity $\Gamma_{CB}^{(1)} = (S_{CD}^{(0)})^{-1} T_{DB}^{(1)}$ is related to the one-loop mixing matrix, or the one-loop dilatation operator.
By using the eigenstates of $\Gamma^{(1)}$ we can rewrite the two-point function \eqref{def:SYM 2pt} into the familiar form
\begin{equation}
\vev{\tilde \cO_A (x) \tilde \cO_B (0)} \simeq \frac{S_{AB}}{\abs{x}^{2 \( \Delta_A^{(0)} + \Delta_A^{(1)} \)} } \,, \qquad 
\Gamma_{AB}^{(1)} \, \tilde \cO_B = -2 \Delta_A^{(1)} \, \tilde \cO_A \,.
\end{equation}
Note that $\Gamma$ is generally not Hermitian even though $S, T$ are Hermitian \cite{GMR02b,BKPS02b}.

The dilatation operator is self-adjoint with respect to the two-point functions, which can be shown by the spacetime translational symmetry as
\begin{equation}
\( - x \frac{\partial}{\partial x} \) \pare{ \vev{\cO_A (x) \cO_B (0)} - \vev{\cO_A (0) \cO_B (-x)} } = 0 
\quad \Longrightarrow \quad
\vev{(\fD \cO_A) \cO_B} = \vev{\cO_A (\fD \cO_B)} .
\label{dil self-adjoint}
\end{equation}
If $\cO_A \,, \cO_B$ are the eigenstate of the dilatation operator $\fD$, we obtain
\begin{equation}
\( \Delta_A - \Delta_B \) S_{AB} = 0
\quad \Longrightarrow \quad
S_{AB} = 0 \ \  {\rm or} \ \ \Delta_A = \Delta_B\,.
\end{equation}
Below we investigate the case $\Delta_A = \Delta_B = \cO(1/N_c)$ and $S_{AB} \neq 0$, where the spectral degeneracy among the large $N_c$ zero modes is lifted by $1/N_c$ corrections.
This problem is what we called the operator submixing in Introduction.

We want to diagonalize the submixing matrix. This problem may look simple because the
dimension of the submixing matrix, namely the number of the large $N_c$ zero modes, is significantly smaller than the number of all operators.
Nevertheless, it contains rich information about non-planar interactions.

The submixing equation itself determines the eigenstates at the zeroth order of $1/N_c$ expansion.
We focus on the problem at $\cO(1/N_c^2)$, because in $\cN=4$ SYM, the degeneracy of the large $N_c$ zero modes is mostly lifted at this order.
Still, some eigenvalues may remain degenerate.

\subsection{Lifting the large $N_c$ degeneracy}\label{sec:lifting}

Let us derive submixing equations. 
We are mostly interested in the submixing among the large $N_c$ zero modes, namely the operators with the vanishing anomalous dimension at $N_c=\infty$. The large $N_c$ zero modes in the scalar $\alg{so}(6)$ singlets will be given explicitly in Section \ref{sec:LN zero modes}.

Let $\cH_\circ$ and $\cH_\bullet$ be the space of the large $N_c$ zero modes and non-zero modes, respectively. 
We split the dilatation operator into the planar and non-planar part as $\fD_\text{one-loop} = \fD_0 + N_c^{-1} \, \fD_1$\,, and regularize the planar part by $\fD_{\epsilon} = \epsilon \,  + \fD_0$ to make it invertible. 
Take a general linear combination of the large $N_c$ zero modes, and call it $\psi_0$\,. 
$\psi_0$ is regarded as the large $N_c$ limit of a finite $N_c$ eigenvector $\psi$. Then the non-planar operator mixing equation $\fD_\text{one-loop} \psi = \gamma \psi$ becomes
\begin{equation}
\( \fD_{\epsilon} + N_c^{-1} \, \fD_1 \) \psi = \(\epsilon + \sum_{i=0}^\infty N_c^{-i} \gamma_i \) \psi \,, \quad
\psi = \sum_{i=0}^\infty N_c^{-i} \psi_i\,, \quad 
(\psi_0 \in \cH_\circ\,,\ \gamma_0=0).
\label{eigenvalue eq with epsilon}
\end{equation}
Expand this equation in $N_c^{-1}$ and apply $\fD_\epsilon^{-1}$. The results are
\begin{equation}
\psi_{n} = \sum_{i=0}^{n} \gamma_{n-i} \, \fD_\epsilon^{-1} \psi_i + \epsilon \, \fD_\epsilon^{-1} \psi_{n} 
- \fD_\epsilon^{-1} \, \fD_1 \, \psi_{n-1}\,, \quad
(n \ge 0,\ \psi_{-1}=0).
\label{eigenvalue eqs epsilon}
\end{equation}

We take the limit $\epsilon \to 0$ and keep track of the singular terms. 
Denote the matrix elements of $\fD_{\epsilon}^{-1}$ and $\fD_1$ by
\begin{equation}
\( \fD_{\epsilon}^{-1} \)_{ij} = 
\begin{pmatrix}
\epsilon^{-1} & 0 \\
0 & \Lambda_{\bullet \epsilon}^{-1} \\
\end{pmatrix}, \quad
\( \fD_{1} \)_{ij} = 
\begin{pmatrix}
m_{\circ \circ} & m_{\circ \bullet} \\
m_{\bullet \circ} & m_{\bullet \bullet} \\
\end{pmatrix}, \quad
O_i = 
\begin{pmatrix}
O_\circ \\
O_\bullet
\end{pmatrix}, \quad
(O_{\circ/\bullet} \in \cH_{\circ/\bullet}),
\label{Dil-eps matrix elements} 
\end{equation}
where $\Lambda_{\bullet \epsilon} = \Lambda_{\bullet} + \epsilon$ are the non-zero eigenvalues of $\fD_\epsilon$\,.
The projector to $\cH_\circ$ is written as
\begin{equation}
P_\circ \equiv \lim \limits_{\epsilon \to 0} \ \epsilon \, \fD_{\epsilon}^{-1} \,, \qquad
P_\circ^2 = P_\circ \,.
\label{def:projector zero mode}
\end{equation}
The projector to $\cH_\bullet$ is $P_\bullet \equiv 1 - P_\circ$\,. We also use the notation $\psi^{\circ/\bullet} \equiv P_{\circ/\bullet} \, \psi$.

The equation \eqref{eigenvalue eqs epsilon} can be written as
\begin{equation}
\psi_{n}^\bullet = \lim \limits_{\epsilon \to 0} \( \sum_{i=0}^{n} \gamma_{n-i} \, \fD_\epsilon^{-1} \, \psi_i
- \fD_\epsilon^{-1} \, \fD_1 \, \psi_{n-1} \).
\label{eigenvalue eqs limit}
\end{equation}
The limiting behavior of the operator $\fD_\epsilon^{-1}$ depends on the operand; it is singular on $\cH_\circ$ and regular on $\cH_\bullet$\,. By evaluating \eqref{eigenvalue eqs limit} up to second order and using $\gamma_0=0$ and $\psi_0^\circ = \psi_0$\,, we obtain
\begin{align}
\psi_{0}^\bullet &= 0,
\label{eigenvalue epsilon O(1)} \\
\psi_{1}^\bullet &= \epsilon^{-1} \gamma_{1} \, \psi_{0}
- \fD_\epsilon^{-1} \, \fD_1 \, \psi_{0} \,,
\label{eigenvalue epsilon O(1/Nc)} \\
\psi_{2}^\bullet &= \epsilon^{-1} \gamma_{2} \, \psi_{0}
- \fD_\epsilon^{-1} \, \fD_1 \, \psi_{1} \,.
\label{eigenvalue epsilon O(1/Nc^2)}
\end{align}
By extracting the components of $\cH_{\circ/\bullet}$ in \eqref{eigenvalue epsilon O(1/Nc)} and \eqref{eigenvalue epsilon O(1/Nc^2)}, one finds
\begin{equation}
\epsilon^{-1} \gamma_{n} \, \psi_{0} = P_\circ \, \fD_\epsilon^{-1} \, \fD_1 \, \psi_{n-1} \,, \qquad
\psi_{n}^\bullet = - P_\bullet \, \fD_\epsilon^{-1} \, \fD_1 \, \psi_{n-1} \,, \qquad
(n=1,2).
\label{eigenvalue epsilon n all}
\end{equation}
The first equation determines the eigenvalue $\gamma_n$\,, which may lift the large $N_c$ degeneracy.
The second equation determines $\psi_{n}^\bullet$ but not $\psi_{n}^\circ$\,. 
No equations can constrain $\psi_{n}^\circ$ because $\psi_{n}^\circ$ are the zero modes of the planar dilatation. They cannot be fixed by perturbation at $n$-th order. 
These zero modes are not important at $\cO(N_c^{-2})$, because the eigenvalue $\gamma_2$ does not depend on $\psi_{1}^\circ$ as shown below.

\subsubsection{Matrix elements}

Let us rewrite the above equations in terms of matrix elements. 
We use small or capital letters in place of $\circ$ or $\bullet$ to label the components of $\cH_\circ$ or $\cH_\bullet$, respectively. An eigenstate is expanded as $\psi_{n} = u_{n,a} \, O_{\circ,a} + u_{n,A} \, O_{\bullet, A}$\,. 
Equations \eqref{eigenvalue epsilon n all} become,\footnote{Note that $\psi$ and $\gamma$ do not depend on $\epsilon$.}
\begin{equation}
\begin{pmatrix}
\epsilon^{-1} \gamma_{n} \, u_{0,a} 
\\[1mm] 
u_{n,A}
\end{pmatrix} =
\begin{pmatrix}
\epsilon^{-1} \( m_{ab} \, u_{n-1,b} + m_{aB} \, u_{n-1,B} \) + \cO(\epsilon^0)
\\[1mm]
\Lambda_A^{-1} \( m_{Ab} \, u_{n-1,b} + m_{AB} \, u_{n-1,B} \) + \cO(\epsilon)
\end{pmatrix} .
\end{equation}
Collecting the terms leading in the limit $\epsilon \to 0$ at $n=1$, we obtain, 
\begin{equation}
0 = \( \gamma_1 \delta_{ab} - m_{ab} \) u_{0,b} \,, \qquad
0= u_{1,A} + \frac{m_{A a}}{\Lambda_A } \, u_{0,a} \,.
\label{submixing 1st}
\end{equation}
These are the equations for degenerate perturbation at first order.
In Appendix \ref{app:data}, we observe that the large $N_c$ degeneracy of zero modes in $\cN=4$ SYM is not lifted at first order. Therefore, we assume
\begin{equation}
\gamma_1=m_{ab}=0.
\label{submixing 1st-2}
\end{equation}
At $n=2$, we find
\begin{align}
0 &= \( \frac{m_{aB} m_{Bc}}{\Lambda_B} + \gamma_2 \, \delta_{ac} \) u_{0,c} \,,
\label{submixing 2nd-1} \\[1mm]
0 &= \( \frac{m_{AB} m_{Bc}}{\Lambda_A \Lambda_B} \, u_{0,c} 
- \frac{m_{Ab}}{\Lambda_A} \, u_{1,b}
- u_{2,A} \).
\label{submixing 2nd-2}
\end{align}
The first line is the submixing equations because it lifts most of the large $N_c$ degeneracy at $\cO(N_c^{-2})$.
This equation is insensitive to the ambiguity of $\psi_{1}^\circ$. 
The second line determines the eigenstates $\psi_2^\bullet$.

\subsection{Large $N_c$ zero modes}\label{sec:LN zero modes}

We classify the large $N_c$ zero modes in the singlet representation of $\alg{so}(6)$.

The BPS states of $\cN=4$ SYM can be labeled by the irreducible representations of $\alg{psu}(2,2|4)$ \cite{AFSZ99} (see also \cite{DF02rev}). The conformal part of the algebra $\alg{psu}(2,2|4)$ can be neglected if we focus on primary operators.
Let us denote by $\cC^{(\ell)}$ the single-trace operator consisting of $\ell \ \alg{so}(6)$ scalars whose flavor indices are traceless and symmetric.
General half-BPS multiplets belong to the representation $[0,L,0]\ (L \ge 2)$. \footnote{Here $[p,q,r]$ denotes Dynkin labels of $\alg{su}(4)_R = \alg{so}(6)$\,.}
The primary operators can be written at large $N_c$ as
\begin{equation}
\cO_\text{half} = \[ \prod_{i=1}^m \cC^{(\ell_i)} \]_{[0, L, 0]} \,, \qquad 
\sum_{i=1}^m \ell_i = L, \qquad
L \ge 2,
\label{def:half BPS}
\end{equation}
Similarly, quarter- and eighth-BPS primary operators can be written at large $N_c$ as
\begin{alignat*}{9}
\cO_\text{quarter} &= \[ \prod_{i=1}^m \cC^{(\ell_i)} \]_{[h ,k, h]} \,, &\qquad 
\sum_i \ell_i &= k + 2 h, &\qquad h &\ge 1,
\\
\cO_\text{eighth} &= \[ \prod_{i=1}^m \cC^{(\ell_i)} \]_{[h ,k, h + 2 h']} \,, &\qquad
\sum_i \ell_i &= k + 2 h + 3 h', &\qquad h' &\ge 1.
\end{alignat*}
It is not easy to write down BPS operators explicitly at finite $N_c$ because one has to solve the mixing problem coming from the color structure \cite{BKRS99,Ryzhov01,DHHR03}. 
Recent attempts at solving the zero eigenvalue equation of the dilatation operator 
are available in \cite{Brown10a,Kimura10,PR10,DGKdM11,dMR12}.

One can prove that a large $N_c$ zero mode is always written as a product of $\cC^{(\ell)}$'s.
At large $N_c$\,, the anomalous dimension of a multi-trace operator is the sum of the anomalous dimension of the constituent single-trace operators. 
The dimensions of the single-traces are given by the planar dilatation, which is equal to an integrable spin-chain Hamiltonian.
The planar dilatation has the ground state given by the traceless symmetric single-trace operator, which is $\cC^{(\ell)}$, and its eigenvalues are non-negative. Note that the products of BPS operators are non-BPS in general.

We present examples of scalar multi-trace operators in $\alg{so}(6)$ singlet with the vanishing anomalous dimension at large $N_c$\,. For clarity, we introduce the flavor indices
\begin{equation}
\cC^{(\ell)} = \cC_{i_1 i_2 \dots i_\ell} = \tr ( \Phi_{(i_1} \Phi_{i_2} \dots \Phi_{i_\ell)} ),
\label{def:C-ell}
\end{equation}
where $(i_1 i_2 \dots i_\ell)$ means traceless and completely symmetric.
Then, the large $N_c$ zero modes with length $L=4,6,8$ are given by
\begin{equation}
\begin{gathered}
\{ \cC_{ij} \, \cC_{ij} \},
\\
\{ \cC_{ijk} \, \cC_{ijk} \,, \ 
\cC_{ij} \, \cC_{jk} \, \cC_{ki} \},
\\
\{ \cC_{ijkl} \, \cC_{ijkl} \,, \ 
\cC_{ijkl} \, \cC_{ij} \, \cC_{kl} \,, \ 
\cC_{ijk} \, \cC_{ijl} \, \cC_{kl} \,, \ 
\cC_{ij} \, \cC_{ji} \, \cC_{kl} \, \cC_{lk} \,, \ 
\cC_{ij} \, \cC_{jk} \, \cC_{kl} \, \cC_{li} \},
\end{gathered}
\label{products sym trsless}
\end{equation}
The number of the large $N_c$ zero modes with length $L$, denoted by $\cZ_L$\,, grows as in Table \ref{tab:cZL}.

\begin{table}[ht]
\begin{center}
\begin{tabular}{c|cccccc}\hline
$L$ & 2 & 4 & 6 & 8 & 10 & 12 \\\hline
$\cZ_{L}$ & 0 & 1 & 2 & 5 & 11 & 34 \\
$\dim \cH_{L}$ & 1 & 4 & 15 & 71 & 469 & 4477 \\\hline
\end{tabular}
\caption{$\dim \cH_{L}$ counts the numbers of all scalar multi-trace operators in $\alg{so}(6)$ singlet with length $L$. $\cZ_{L}$ counts the operators in $\cH_L$ whose anomalous dimension vanishes at large $N_c$\,.}
\label{tab:cZL}
\end{center}
\end{table}

Finding an explicit formula for $\cZ_L$ is a difficult problem of combinatorics.
Interestingly, this series coincides with the asymptotic number of completely symmetric polynomials of Mandelstam variables at degree $L/2$ subject to the massless momentum 
conservation \cite{OEIS,Boels13}, which will be further explained in Appendix \ref{app:Mandelstam}.
This coincidence may break down at $L \ge 14$ due to the finite\,-$N_f$ constraints, where $N_f=6$ in $\cN=4$ SYM.
An example of the finite\,-$N_f$ constraints is the following anti-symmetrization identity:
\begin{equation}
0 = \sum_{\sigma \in \cS_7} {\rm sign} (\sigma) \,
\cC_{i_1 i_{\sigma(1)}} \, \cC_{i_2 i_{\sigma(2)}} \, \cC_{i_3 i_{\sigma(3)}} \, 
\cC_{i_4 i_{\sigma(4)}} \, \cC_{i_5 i_{\sigma(5)}} \, \cC_{i_6 i_{\sigma(6)}} \, 
\cC_{i_7 i_{\sigma(7)}} \qquad
(i_k = 1,2, \dots, 6),
\label{finite Nf constraint}
\end{equation}
which reduces the number of independent large $N_c$ zero modes by one.

\subsection{Observations on submixing}\label{sec:observation}

\afterpage{
\begin{table}[t!]
\begin{center}
\begin{tabular}{c|cccc}\hline
$L/2$ & 2 & 3 & 4 & 5 \\\hline
$-\gamma_2$ & 20 & 60.2459 & 101.276 & 132.315 \\\hline
\end{tabular}
\caption{The coefficient $\gamma_2$ of double-trace operators for various $L$.}
\label{tab:gam2-dt}
\end{center}
\end{table}

\begin{figure}[h!]
\begin{center}
\includegraphics[scale=1]{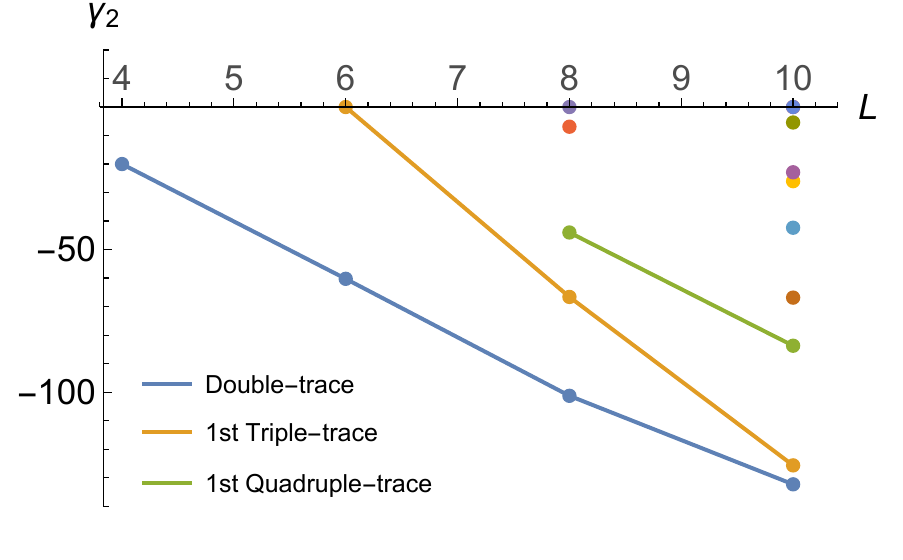}
\caption{Plot of $\gamma_2$ for the large $N_c$ zero modes against the operator length $L$ at $N_f=6$.
}
\label{fig:lam2}
\end{center}
\end{figure}
}

We present here interesting structure found in the submixing. The data in Appendix \ref{app:data} is summarized in Figure \ref{fig:lam2}, which shows the eigenvalues of the submixing Hamiltonian for $L=4,6,8,10$.
The one-loop anomalous dimension is given by $\gamma = \gamma_2/N_c^2 + \cO(N_c^{-3})$. 
The legend ``$m$-tuple trace" means that the operator consists of a mixture of $m$-trace operators at large $N_c$\,. At higher orders, they mix with everything else. Also, the numerical values of $\gamma_2$ for the operators starting from double-traces are shown in Table \ref{tab:gam2-dt}.

Let us rephrase our findings.
First, most of the large $N_c$ degeneracy of zero modes is lifted at second order. 
This statement is not trivial because the large $N_c$ degeneracy of non-zero modes is sometimes lifted at the first order in $1/N_c$\,.

Second, all the second-order eigenvalues are non-positive, $\gamma_2 \le 0$. 
This is again non-trivial despite the minus sing in \eqref{sm Ham elements},
because the matrix elements are not symmetric, $m_{aA} \neq m_{Aa}$\,. 
Note that the finite $N_c$ anomalous dimensions of the large $N_c$ zero modes can be positive. There exist operators with $\gamma_0=\gamma_1= \dots = \gamma_{n-1}=0$ and $\gamma_n > 0$ for some $n \ge 3$.

Third, when the large $N_c$ degeneracy of zero modes is lifted at $\cO(1/N_c^2)$, all eigenstates have a definite number of traces.
Thus, the submixing Hamiltonian should commute with the operator which counts the number of traces.
In contrast, the operators with different number of traces may submix for the large $N_c$ {\it non-zero} modes.

Fourth, if we define the density of submixing matrix elements by applying the Laplace transform to $\fD_0^{-1}$ as\footnote{There can be other integral representations of $\fD_0^{-1}$.}
\begin{equation}
h_\text{sm}^\circ (t)_{ac} \equiv 
- \sum_B m_{aB} \, e^{-t \Lambda_B} \, m_{Bc} \,, \quad
\int_0^\infty dt \, h_{ac} (t) = - \frac{m_{aB} m_{Bc}}{\Lambda_B} \,,
\end{equation}
the submixing densities at different $t$ satisfy
\begin{align}
0 &=
P_\text{min}^\circ \, \Bigl( h_\text{sm}^\circ (t)_{ab} h_\text{sm}^\circ (t')_{bc} - h_\text{sm}^\circ (t')_{ab} h_\text{sm}^\circ (t)_{bc} \Bigr)
\notag \\ 
&= 
\Bigl( h_\text{sm}^\circ (t)_{ab} h_\text{sm}^\circ (t')_{bc} - h_\text{sm}^\circ (t')_{ab} h_\text{sm}^\circ (t)_{bc} \Bigr) \, P_\text{min}^\circ = 0,
\label{hpm commutation}
\end{align}
where $P_\text{min}^\circ$ is the projector to $\cH_\text{min}^\circ$\,, which is the subspace of $\cH_\circ$ spanned by the products of length-two operators.
This ``projective commutation'' relation is equivalent to
\begin{align}
0 &= P_\text{min}^\circ \Big( m_{aB} \( \Lambda_B^n \) \, m_{Bc} \, m_{cD} \( \Lambda_D^m \) \, m_{De}
- m_{aB} \( \Lambda_B^m \) \, m_{Bc} \, m_{cD} \( \Lambda_D^n \) \, m_{De} \Big)
\notag \\
&= \Big( m_{aB} \( \Lambda_B^n \) \, m_{Bc} \, m_{cD} \( \Lambda_D^m \) \, m_{De}
- m_{aB} \( \Lambda_B^m \) \, m_{Bc} \, m_{cD} \( \Lambda_D^n \) \, m_{De} \Big) P_\text{min}^\circ,
\label{def:mut charges}
\end{align}
\adg{for $n,m \ge 0$}.
These relations imply that the submixing problem on $\cH_\text{min}^\circ$ is integrable.

It will be interesting to prove these findings and see how general these relations hold true, which will help to solve general submixing problem. The projection by $P_\text{min}^\circ$ may be removed by finding a better definition of the submixing density, or equivalently a better definition of mutually commuting charges \eqref{def:mut charges}.

\subsection{How to solve non-planar operator mixing}

Let us pause to explain how we obtained the aforementioned results on the operator submixing.
It is not a priori clear at which order the large $N_c$ spectral degeneracy is lifted, so we need to study the general non-planar operator mixing problem. Since the full non-planar problem is too complicated to work by hand, we used computer programs extensively to obtain concrete results. More specifically, we wrote {\tt Mathematica} codes and proceed as follows,
\begin{itemize}
\setlength{\itemsep}{0pt}
\setlength{\parskip}{0pt}
\item Generate a complete set of multi-trace operators with a given length, using the notation in Appendix \ref{app:polynote}.
\item Compute the action of one-loop dilatation operator on the complete set of multi-traces.
\item Extract the matrix elements, as analytic functions of $N_c$ and $N_f$\,.\footnote{$\cN=4$ SYM corresponds to $N_f=6$\,. We introduced a parameter $N_f$ as discussed in Appendix \ref{app:notation}.}
\end{itemize}
Our computation can be done symbolically with no numerical errors until this stage. See the attached {\tt Mathematica} code for details.

Having obtained explicit non-planar mixing matrices, we will study their eigensystems later in this paper. We will analyze
\begin{itemize}
\setlength{\itemsep}{0pt}
\setlength{\parskip}{0pt}
\item Operator submixing problem as $1/N_c$ perturbation
\item Eigenvalues as analytic functions of $N_c$ and $N_f$
\item Finite $N_c$ constraints on eigenvectors
\end{itemize}
There is no need to invent new techniques to solve these standard problems of linear algebra.
In order to manage large matrices (of size $\lesssim 5000$) efficiently, one should compute their eigenvalues numerically at ${\tt MachinePrecision}$. It will take an extremely long time to find all eigenvalues analytically if the matrix size is bigger than $\sim 30$.

It is not easy to inspect carefully the Mathematica codes, so we need to check that no mistakes were made during the computation, which can be done in two ways.
First, our mixing matrix eigenvalues at $L=4, 6$ agreed analytically with the literature \cite{BKS03,BRS03} as shown in Appendix \ref{app:data}. Second, all eigenvalues of the mixing matrix are real. This is a non-trivial check because we computed the mixing matrix elements using the basis in which the matrix is non-hermitian. Further discussion on the finite $N_c$ spectrum will be given in Section \ref{sec:spec FAN}.


\section{Correlation functions}\label{sec:correlator}

We will study the submixing problem by evaluating two-point functions.
First we evaluate $\gamma_2$ in terms of correlation functions of 
$\psi_i$, and derive the submixing Hamiltonian.
Then we discuss the sign and the $L$-dependence of $\gamma_2$\,.

\subsection{Preliminary}

In this subsection, we summarize preliminary facts about correlation functions. 

We will focus on 
color and flavor combinatorial factors 
of correlation functions, and we drop 
the space-time dependence, which is trivial to recover from conformal invariance. 
In our analysis, the difference between the $U(N_c)$ gauge group and 
the $SU(N_c)$ gauge group does not matter because we mainly discuss the leading large $N_c$ 
contribution. We use the $U(N_c)$ Wick-contraction
\begin{eqnarray}
\langle (\Phi_{a})^i_j (\Phi_{b})^k_m\rangle =\delta_{ab}\delta^i_m \delta^k_j.
\label{Wick-contraction_U(N)}
\end{eqnarray} 
It can be extended to the case of 
multi-fields as
\begin{eqnarray}
&&\langle :
(\Phi_{a_1})^{i_1}_{j_1} \cdots (\Phi_{a_L})^{i_L}_{j_L} ::
 (\Phi_{b_1})^{k_1}_{m_1}\cdots (\Phi_{b_L})^{k_L}_{m_L} :
\rangle 
\nonumber \\
&=&\sum_{\sigma \in S_L}
\delta_{a_1 b_{\sigma(1)}}\cdots \delta_{a_L b_{\sigma(L)}}
\delta^{i_1}_{m_{\sigma(1)}} \cdots  \delta^{i_L}_{m_{\sigma(L)}}
\delta^{k_1}_{j_{\sigma(1)}} \cdots  \delta^{k_L}_{j_{\sigma(L)}}
\nonumber \\
&\equiv&\sum_{\sigma \in S_L}
\delta_{a_1 b_{\sigma(1)}}\cdots \delta_{a_L b_{\sigma(L)}}
(\sigma)^I_M
(\sigma^{-1})^K_J \,.
\end{eqnarray}
where the normal ordering defines an operator without self-contractions. 
Because of the normal ordering,  
Wick-contractions apply only between $\Phi_{a_i}$ and $\Phi_{b_j}$\,.

Next gauge invariant operators are considered. 
Any multi-trace gauge invariant operator built from $L$ matrices 
can be characterized by 
an element of the symmetric group $\cS_L$ as 
\begin{eqnarray}
\trb{L}(\tau \Phi_{\vec{a}}^{\otimes L})
&\equiv&
(\tau)_I^J(\Phi_{\vec{a}}^{\otimes L})_J^I
\nonumber \\
&=&
\delta^{j_1}_{i_{\tau(1)}} \cdots  \delta^{j_n}_{i_{\tau(L)}}
(\Phi_{a_1})^{i_1}_{j_1} \cdots (\Phi_{a_L})^{i_L}_{j_L}.
\end{eqnarray}
Focusing only on the color structure by leaving out 
the flavor indices, 
this description has the symmetry 
\begin{eqnarray}
\trb{L}(\tau \Phi^{\otimes L}) =\trb{L}(g \tau g^{-1}\Phi^{\otimes L}),
\end{eqnarray}
where $g$ is any element in $\cS_L$. This implies that  
the multi-trace structure of gauge invariant operators is classified by the conjugacy classes of $\cS_L$, 
which are the partitions of $L$. 
It is then straightforward to obtain two-point functions for 
gauge invariant operators \cite{BHR08}, 
\begin{eqnarray}
\langle 
:\trb{L}(\tau_1 \Phi_{\vec{a}}^{\otimes L}):
:\trb{L}(\tau_2 \Phi_{\vec{c}}^{\otimes L}):
\rangle 
&=&
\sum_{\sigma \in \cS_L}
\delta_{a_1 c_{\sigma(1)}}\cdots \delta_{a_L c_{\sigma(L)}}
(\sigma)^I_L
(\sigma^{-1})^K_J
(\tau_1)_I^J (\tau_2)_K^L
\nonumber \\
&=&\sum_{\sigma \in \cS_{L}}
\delta_{a_1 c_{\sigma(1)}}\cdots \delta_{a_{L} c_{\sigma(L)}}
\trb{L}(\tau_1 \sigma^{-1}\tau_2 \sigma)
\nonumber \\
&=&
\sum_{\sigma \in \cS_{L}}
\langle \vec{a}|\sigma |\vec{c} \rangle
N_c^{C(\tau_1 \sigma^{-1}\tau_2 \sigma)} .
\label{twopt_gaugeinvariant}
\end{eqnarray} 
We introduced a shorthand notation
\begin{eqnarray}
\delta_{a_1 c_{\sigma(1)}}\cdots \delta_{a_{L} c_{\sigma(L)}}
=\langle \vec{a}|\sigma |\vec{c} \rangle, 
\end{eqnarray} 
and $C(\sigma)$ counts the number of cycles in the permutation $\sigma$, e.g.
$C(1)=L$, $C((12))=L-1$.

Finding the $L$-dependence of the correlator \eqref{twopt_gaugeinvariant}
is a complicated combinatorial problem 
with the color and flavor structures correlated. 
In Section \ref{define_correlator_group_theory}, 
we will concretely compute the 
two-point functions 
of the double-trace large $N_c$ zero mode, as an example to see 
how the correlators depend on $L$.

In the literature \cite{BHR07,BHR08,Kimura09a}, bases of gauge invariant operators in the $\alg{so}(6)$ scalar sector have been constructed, and they diagonalize the two-point functions 
including all $1/N_c$ corrections.
The readers can also refer to \cite{CJR01,KR07,BCdM08,KR08}, although we do not need those technologies in what follows.

When $\tau_1$ and $\tau_2$ are in the same conjugacy class
(i.e. they have the same trace structure), 
the two-point functions behave as $\sim N_c^{L}$. Its coefficient
is a symmetry factor determined by the number of permutations $\rho$ 
satisfying 
$\tau_1 \rho^{-1} \tau_2 \rho=1$,\footnote{
\label{symmetry_conjugacy_class}
The symmetry factor of the two-point functions can be associated with group theory.
Suppose $\tau$ is in the conjugacy class $c=[c_1,\cdots,c_L]$, 
where $c_i$ is the number of cycles of length $i$, i.e. $L=\sum_{i}ic_i$.
The number of permutations $\rho$ satisfying 
$\rho^{-1} \tau \rho=\tau$ is counted as follows. 
For a cycle with length $i$, there are exactly $i$ cyclic permutations that do not change the cycle. 
Hence, $i^{c_i}$ cyclic permutations can leave the cycles. 
Also, $\rho$ can permute $c_i$ cycles of the same length.
Then we have the formula
\begin{eqnarray}
Sym_L (c) = \dim \[ \bigoplus_{i=1}^L \( \cS_{c_i} \otimes \bb{Z}_i^{\otimes c_i} \) \]=\prod_i^L (i^{c_i}) (c_i!) .
\label{sym_formula}
\end{eqnarray} 
For $\tau$ in the conjugacy class $c_L=1$,  
we have $Sym_L (c)=L$. 
For $\tau$ in the conjugacy class $c_{L/2}=2$,  
we have $Sym_L (c)=L^2/2$. 
}
\begin{equation}
\langle :\trb{L}(\tau_1 \Phi_{\vec{a}}^{\otimes L}): :\trb{L}(\tau_2 \Phi_{\vec{c}}^{\otimes L}): \rangle 
= \sum_{\rho}\langle a |\rho |c  \rangle 
 \, N_c^L + \cO(N_c^{L-1}).
\label{expand-two-point}
\end{equation}
On the other hand, 
if $\tau_1$ and $\tau_2$ are not in the same conjugacy class, 
(i.e. they have different trace structure), 
the two-point functions behave as $N_c^{a}$ with $a<L$. 
In other words, 
for two normalized operators $\varphi_1$, $\varphi_2$
we have
\begin{eqnarray}
\langle \varphi_1\varphi_2\rangle \sim O(1) ,
\end{eqnarray} 
if they have the same trace structure, 
and 
\begin{eqnarray}
\langle \varphi_1\varphi_2\rangle \sim O(N_c^{-a})  \quad (a>1).  
\end{eqnarray}
if they have different trace structure.

We can also deduce the large $N_c$ behavior of matrix elements of the dilatation operator.
The planar dilatation operator does not change the number of traces while 
the non-planar dilatation operator does change the number of traces by one. 
In other words, the planar dilatation operator 
does not change trace structure of operators while 
the non-planar operator does. 
As a consequence we find
\begin{eqnarray}
\frac{\langle \phi \fD_0 \phi \rangle}
{\langle \phi \phi \rangle} \sim O(1),\quad 
\frac{\langle \phi \fD_1 \phi \rangle}
{\langle \phi \phi \rangle} \sim O(N_c^{-1}),
\label{expansion_assumption2}
\end{eqnarray}
and so on, where $\phi$ is a linear combination of operators with the same trace structure.

We expand a dilatation eigenstate as
\begin{eqnarray}
&&\psi =\psi_0+\frac{1}{N_c}\delta\psi:= \sum_{n=0}^\infty \frac{1}{N_c^{n}} \, \psi_n ,
\end{eqnarray}
where 
each field $\psi_i$ is allowed to be a linear combination of multi-traces. 
We will assume that the two-point function between 
$\psi_i$ and $\psi_{i+1}$ is subleading 
\begin{eqnarray}
\frac{\langle \psi_i \psi_{i+1} \rangle}{
\sqrt{\langle \psi_i \psi_{i} \rangle\langle \psi_{i+1} \psi_{i+1} \rangle}
} \sim O(N_c^{-1}) .
\label{expansion_assumption}
\end{eqnarray}
For example, if $\psi_0$ is a double-trace operator, $\psi_1$ can be
a linear combination of single-traces and triple-traces. 
This assumption is essential in the subsequent argument, and it is desirable to have a proof.

Note that correlation functions of $\psi$ like \eqref{expand-two-point} have a meaningful $1/N_c$ expansion only if $N_c \gg L$.
In general, the numerical factor appearing in each order of the expansion 
is a function of $L$ or $L_i$, where $L_i$ is a partition of $L$. 
Hence 
the expansion parameter in \eqref{expansion_assumption} and \eqref{expansion_assumption2} is $f(L_i)/N_c$. 
In order for this expansion to be a perturbative expansion, we need to assume $N_c \gg L$, which we do in what follows.
Then, two-point functions of $\psi$ reduce to those of $\psi_0$ in the planar limit.

\subsection{Correlator expressions of $\gamma$}

\subsubsection{$1/N_c$ perturbation revisited}

The dilatation operator can be divided into the planar and non-planar parts, 
\begin{equation}
\fD_\text{one-loop} = \fD_{0} + \frac{1}{N_c} \, \fD_{1} .
\label{def:Dnp psi}
\end{equation}
We will study the case  
$\psi_0$ is given by a linear combination of the large $N_c$ zero modes, which is characterized by 
\begin{equation}
\fD_{0}\psi_0=0. 
\label{def_zeromode_D0}
\end{equation}

Substituting the expansions into the eigenvalue equation
\begin{equation}
\fD_\text{one-loop}\psi=\gamma \psi, 
\end{equation}
we have 
\begin{equation}
\frac{1}{N_c} \( \fD_{0} \, \delta\psi + \fD_{1} \, \psi_{0} \) + \frac{1}{N_c^2} \, \fD_{1} \, \delta\psi
= \gamma \( \psi_{0} + \frac{1}{N_c} \, \delta\psi \).
\label{def:eveq one}
\end{equation}
Taking the two-point functions of this equation with $\psi_{0}$ leads to 
\begin{align}
\frac{1}{N_c} \Vev{\psi_{0} \( \fD_{0}\delta\psi +  \fD_{1}  \psi_{0} \) } 
+ \frac{1}{N_c^2} \Vev{ \psi_{0} \( \fD_{1}  \delta\psi \) }
&= \gamma \left( \Vev{ \psi_{0}   \psi_{0}} 
+ \frac{1}{N_c} \Vev{\psi_{0} \delta\psi} \right).
\end{align}
From the fact that the dilatation operator is self-adjoint 
with respect to the two-point function, 
we have 
\begin{align}
 \Vev{\psi_{0} \( \fD_{0}\delta\psi \) } 
 = \Vev{\(\fD_{0}\psi_{0}\) \delta\psi  } =0. 
\end{align}
Because $\delta \psi = \psi_1 + \psi_2/N_c +\cdots$, 
we obtain
\begin{align}
\gamma &= \frac{1}{N_c^2} \, \frac{N_c  \Vev{ \psi_{0}  \fD_{1}  \psi_{0} }
+ \Vev{\psi_{0}  \fD_{1}  \psi_1 }}{\Vev{\psi_{0}  \psi_{0} }} + \cO(N_c^{-3}).
\label{gamma_first_expression}
\end{align}
Assuming that $\psi_{\rm 0}$ and $\fD_{1} \psi_{\rm 0}$ have different trace structure, we see that the expansion of $\gamma$ starts from the order $1/N_c^2$.\footnote{This situation is parallel to our argument around \eqref{submixing 1st-2}.}  
Substituting \eqref{gamma_first_expression} into \eqref{def:eveq one}, we find
\begin{equation}
\fD_{0}\psi_1+\fD_{1} \psi_{0} = \frac{\eta}{N_c},
\label{constraint_for_psi_1_operator}
\end{equation}
where $\eta$ is a linear combination of multi-traces that are different from $\psi_1$. 
Considering the two-point functions of \eqref{def:eveq one} and $\psi_1$, we obtain 
\begin{equation}
\Vev{\psi_{1} \psi_{0} } =
\frac{1}{N_c^2 \gamma} \Bigl( \Vev{\psi_1 \eta } + \Vev{\psi_1 \fD_1 \psi_1} \Bigr)
- \frac{1}{N_c}\Vev{\psi_{1} \psi_{1} } = \cO (N_c^{-1}) \sqrt{\Vev{\psi_0 \psi_0} \Vev{\psi_1 \psi_1} } \,,
\end{equation}
which is consistent with our assumption \eqref{expansion_assumption}.

The equation (\ref{constraint_for_psi_1_operator})
can be used to get the following expression,
\begin{align}
\gamma &= \frac{1}{N_c^2} \, \frac{N_c \, \Vev{ \psi_{0} \, \fD_{1} \, \psi_{0} }
- \Vev{\psi_1 \, \fD_{\rm 0}  \psi_1 }}{\Vev{\psi_{0} \, \psi_{0} }} + \cO(N_c^{-3}),
\end{align}
or equivalently by expanding $\gamma=\sum_{n=0}N_c^{-n}\gamma_n$, 
\begin{eqnarray}
\gamma_0=0, \quad \gamma_1=0,\quad
\gamma_2 
= \frac{N_c \, \Vev{\psi_{\rm 0} \, \fD_{1} \, \psi_{\rm 0} } - \Vev{\psi_{\rm 1} \, \fD_{0} \, \psi_{\rm 1} }}{\Vev{ \psi_{\rm 0} \, \psi_{\rm 0}} } \,.
\label{gamma0_gamma1_gamma2}
\end{eqnarray}
This expression of $\gamma_2$ will be frequently used later.
Because $\gamma_2$ is the leading term of the $1/N_c$ expansion, we can use the planar limit to compute the correlation functions.

\bigskip
We will next give another form of $\gamma_2$. 
Following the methods of Section \ref{sec:lifting}, it is not difficult to rewrite the above expression of $\gamma_2$ as
\begin{equation}
\gamma_2 
= \lim_{\epsilon\rightarrow 0}
\frac{\Vev{\psi_{0} \( N_c \, \fD_1 - \fD_1  \fD_\epsilon^{-1} \fD_1 \) \psi_{0} }}{\Vev{\psi_{0} \, \psi_{0} }} \,.
\end{equation}
This simplifies further as
\begin{equation}
\gamma_2 
= - \lim_{\epsilon\rightarrow 0} 
\frac{\Vev{\Psi  \fD_\epsilon  \Psi}}{\Vev{\psi_{0} \, \psi_{0} }} \,, \qquad
\Psi =  \fD_\epsilon^{-1} \fD_1 \, \psi_{0} - \frac{N_c}{2} \, \psi_{0} \,,
\label{lambda_PsiD0Psi}
\end{equation}
which can be checked by using the self-adjoint property of $\fD$,  
and (\ref{def_zeromode_D0}), (\ref{constraint_for_psi_1_operator}).

\subsubsection{Submixing Hamiltonian}\label{sec:submixing Hamiltonian}

Let us derive the submixing Hamiltonian.
We use a Greek letter to label different large $N_c$ zero modes as 
\begin{equation}
\fD_\text{one-loop}\psi^{(\alpha)}=\gamma^{(\alpha)} \psi^{(\alpha)}, \qquad
\gamma_0^{(\alpha)} = 0.
\end{equation}
For simplicity, we assume that the eigenvalues are non-degenerate, 
$\gamma^{(\alpha)}\neq \gamma^{(\beta)}$ for $\alpha\neq \beta$.  
By generalizing the above argument, we obtain
\begin{eqnarray}
\gamma_2^{(\alpha)} \Vev{ \psi_{\rm 0}^{(\beta)} \, \psi_{\rm 0}^{(\alpha)}} 
&=& N_c \Vev{\psi_{\rm 0}^{(\beta)} \, \fD_{1} \, \psi_{\rm 0}^{(\alpha)} } 
- \Vev{\psi_{\rm 1}^{(\beta)} \, \fD_{0} \, \psi_{\rm 1}^{(\alpha)} } 
\nonumber \\
&=& \lim_{\epsilon\rightarrow 0} \Vev{\psi_{\rm 0}^{(\beta)} 
(
N_c \, \fD_1 - \fD_1  \fD_\epsilon^{-1} \fD_1
)
\psi_{\rm 0}^{(\alpha)} } . 
\label{correlator_submixing_eq}
\end{eqnarray}
The left-hand side is proportional to $\delta_{\alpha\beta}$\,, so the right-hand side is. Since $\{ \psi_0^{(\alpha)} \}$ spans the entire Hilbert space of the large $N_c$ zero modes, we can think of \eqref{correlator_submixing_eq} as an operator identity,
\begin{eqnarray}
H_\text{sm}^\circ \, \psi_0^{(\alpha)} = \gamma_2^{(\alpha)} \, \psi_0^{(\alpha)} \,, \qquad
H_\text{sm}^\circ=P_{\circ } \(
N_c \, \fD_1 - \fD_1  \fD_\epsilon^{-1} \fD_1 \), 
\label{def:op Hsm}
\end{eqnarray}
which is \eqref{def:submixing Hamiltonian}.

In Section \ref{sec:correlator} we will rewrite the submixing equation in the operator form,
\begin{gather}
\gamma_2 \, u_{0,a} \, O_{\circ,a} = - \( \frac{m_{aB} m_{Bc}}{\Lambda_B} \) u_{0,c} \, O_{\circ,c}  \quad \longleftrightarrow \quad
\gamma_2\, \psi_0 = H_\text{sm}^\circ \, \psi_0 \,,
\label{sm Ham elements} \\
H_\text{sm}^\circ = P_\circ \, H_\text{sm} \equiv P_\circ \Big( N_c \, \fD_1 - \fD_1 \, \fD_{\epsilon}^{-1} \, \fD_1 \Big).
\label{def:submixing Hamiltonian}
\end{gather}
We call $H_\text{sm}$ submixing Hamiltonian. The projector $P_\circ$ is needed because $H_\text{sm}$ maps $\cH_\circ$ to the whole space of operators, $\cH_\circ \oplus \cH_\bullet$\,.
Note that the first term in \eqref{def:submixing Hamiltonian} vanishes if we write down matrix elements as in Section \ref{sec:lifting} owing to \eqref{submixing 1st-2}. The (dis)appearance of the first term originates from the fact that the mixing matrix is not Hermitian whereas the dilatation operator is self-adjoint with respect to the two-point functions.

We can also rewrite the submixing density into an operator form as
\begin{equation}
h_\text{sm}^\circ (t) \equiv P_\circ \, h_\text{sm} (t), \qquad
h_\text{sm} (t) \equiv \( \frac{N_c}{2} \, \fD_\epsilon - \fD_1 \) e^{-t \, \fD_\epsilon }  \( \frac{N_c}{2} \, \fD_\epsilon - \fD_1 \) .
\end{equation}
The integral over $[0,\infty)$ converges if we consider $P_\circ h_\text{sm} (t) = h_\text{sm}^\circ (t)$ thanks to $\mathop{\lim}_{\epsilon \to 0} P_\circ \, \fD_\epsilon = 0$ and \eqref{submixing 1st-2}. Then we can safely take the limit $\epsilon \to 0$.

For the double-trace operator $\cC_{\vec{a}}\cC_{\vec{a}}$\,, the action of the submixing Hamiltonian can be schematically represented as
\begin{equation}
H_\text{sm}^\circ \quad \sim \quad
N_c \, \Biggl( \ \ 
\adjustbox{valign=c}{\includegraphics{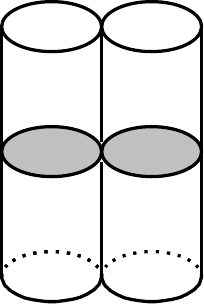}} 
\ \ \Biggr) \quad - \quad \Biggl( \ \ 
\adjustbox{valign=c}{\includegraphics{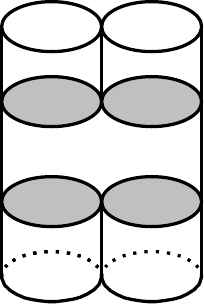}}
\ \ \Biggr),
\label{Hsm graphical}
\end{equation}
where each circle represents the single-trace operator $\cC_{\vec{a}}$\,. The $\fD_1$, represented by a gray circle, is an interaction vertex and $\fD_{\epsilon}^{-1}$ is a propagator.
The intermediate state in the second term must be a single-trace.

In the literature, the energy spectrum of string in pp-wave SFT was compared with the dimension of BMN-type operators in $\cN=4$ SYM. It was argued that we should include four-point contact terms in the pp-wave SFT for a better agreement and to cancel a divergent term from a cubic coupling \cite{CFHM02,GORSY05}. We can interpret the first term in \eqref{Hsm graphical} as the four-point contact term, and the second term as the cubic term squared, in consistent with the expectation from the pp-wave SFT.

Interestingly, we can remove the four-point contact term if we renormalize the wave-function as $\Psi' \equiv - \fD_\epsilon \Psi = \frac{N_c}{2} \, \psi_{0} - \fD_\epsilon^{-1} \fD_1 \, \psi_{0}$\,. From \eqref{lambda_PsiD0Psi} it follows that $H_\text{sm}^\circ \sim \Vev{\Psi' \, \fD_\epsilon^{-1} \, \Psi'}$, which contains only the propagator $\fD_\epsilon^{-1}$\,.

\subsubsection{On the sign of $\gamma_2$}

Here is another comment on the sign of $\gamma_2$. 
The eigenvalues of $\fD_0$ are non-negative, but it does not mean $\gamma_2$ is non-positive.
Let $V^I, V_0^I$ be an eigenbasis of $\fD, \fD_{0}$\,, respectively.
The state $\Psi$ can be expanded as
\begin{equation}
\Psi = F_I \, V^J = F_I \, U^I{}_J V_0^J . 
\end{equation}
Note that $U$ is not an orthogonal transformation, because $\{ V_0^I \}$ is not an orthonormal basis at finite $N_c$\,. It follows that
\begin{align}
\Vev{\Psi \, \fD_\epsilon \, \Psi} 
&= \Vev{(F_K U^K{}_L V_0^L) \, \fD_\epsilon \, F_I U^I{}_J V_0^J}
\notag \\
&= F_K U{}^K{}_L \Vev{V_0^L \, \gamma_0^J \, V_0^J}  F_I U^I{}_J 
\notag \\
&= F_K \Bigl( U^I{}_L\gamma_0^L (U^{T}){}^L{}_K \Bigr) F_I \,.
\end{align}
Since $\gamma_0$ and $(U \gamma_0 U^T)$ have different spectra at finite $N_c$\,,
$\Vev{\Psi \, \fD_\epsilon \, \Psi}$ can be positive or negative.

\subsection{Two characteristic classes of the large $N_c$ zero modes}
\label{define_correlator_group_theory}

As we mentioned in Section \ref{sec:observation}, 
the submixing pattern can be classified by the number of traces of the large $N_c$ zero modes.
We will focus on the two characteristic cases in the classification: the number of traces being minimal or maximal.
The first case is the double-trace operator 
$\cC_{a_1 \cdots a_{L/2}}\cC_{a_1 \cdots a_{L/2}}$\,, which is the unique operator of length $L$ with two traces.
The second case is a set of operators having the largest number of traces, namely $L/2$-tuple length-2 operators such as 
$\cC_{ij}\cC_{jk}\cC_{kl}\cdots \cC_{mi}$\,.
The number of such operators is equal to the number of partitions of $L/2$ excluding the partitions that contain 1. 
Let us introduce a shorthand notation 
$\cC^{[p]}=\cC_{a_{1}a_{2}}\cC_{a_{2}a_{3}}\cC_{a_{3}a_{4}}\cdots \cC_{a_{p}a_{1}}$.
There are two operators for $L=10$; 
$\cC^{[5]}$ and    
$\cC^{[3]}\cC^{[2]}$, and there are 
four operators for $L=12$; 
$\cC^{[6]}$,     
$\cC^{[4]}\cC^{[2]}$,  
$\cC^{[2]}\cC^{[2]}\cC^{[2]}$ and  
$\cC^{[3]}\cC^{[3]}$. 
For $p\ge7$, the constraint (\ref{finite Nf constraint}) has to be taken into account.

\bigskip
Below we want to see how the $L$-dependence arises from correlation functions, by taking the double-trace operator as the simplest example.
In general, the two-point functions of other large $N_c$ zero modes result in complicated combinatorial problems.

We write the singlet double-trace operator as
\begin{eqnarray}
\chi_L
&=&
: \tr(
\Phi_{(a_1} \cdots \Phi_{a_{L/2})}) \,
\tr( 
\Phi_{(a_1} \cdots \Phi_{a_{L/2})}):  
\nonumber \\
&=&
\langle \vec{a}|P_{[L/2]} |\vec{b} \rangle
:\trb{L}( \(\alpha_{L/2} \otimes \alpha_{L/2} \) 
\Phi_{\vec{a}}^{\otimes L/2}\otimes 
\Phi_{\vec{b}}^{\otimes L/2}):,
\label{define_double_trace_operator}
\end{eqnarray} 
where $\alpha_{L/2}$ represents the conjugacy class of cycle length $L/2$ in $\cS_{L/2}$\,.
We introduced the projection operator $P_{[L/2]}$ associated with 
the rank-$L/2$ symmetric traceless representation of $\alg{so}(6)$.

\begin{figure}[t]
\begin{center}
\includegraphics[scale=0.76]{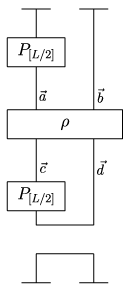}
\caption{The flavor structure of $\vev{\chi_L \chi_L}$. Two lines at the top and the bottom of the figure are identified. The contraction operator $\prod_i C_{ii}$ connects $\vec a$ with $\vec b$, and $\vec c$ with $\vec d$.}
\label{fig:flavor-tr}
\end{center}
\end{figure}

From the formula (\ref{twopt_gaugeinvariant}), 
the two-point functions of $\chi_L$ can be computed as
\begin{equation}
\vev{\chi_L \chi_L} = \sum_{\vec a, \vec b, \vec c, \vec d} \sum_{\rho \in \cS_L}
\vev{\vec a | P_{L/2} | \vec b} \vev{\vec c | P_{L/2} | \vec d}
\vev{\vec a \otimes \vec b | \rho | \vec c \otimes \vec d}
N_c^{C( (\alpha_{L/2} \otimes \alpha_{L/2}) \rho^{-1} (\alpha_{L/2} \otimes \alpha_{L/2}) \rho)} .
\end{equation}
To simplify it, let us use the diagrammatic computation presented in \cite{CR02,KR07} and Figure \ref{fig:flavor-tr}. We also introduce a trace over $V_f^{\otimes L}$, where 
$V_f$ is a 6-dimensional vector space for the flavor index.
It follows that\footnote{With the notation $V_f^{\otimes L} = V_f^{\otimes L/2} \otimes V_f^{\otimes L/2}$, $C_{ii}$ is the contraction between the $i$-th space in the former $V_f^{\otimes L/2}$ and the $i$-th space in the latter $V_f^{\otimes L/2}$.} 
\begin{align}
\langle \chi_L\chi_L \rangle 
&= \sum_{\rho \in S_{L}}
\trb{L}^{f} \((P_{[L/2]}\otimes 1)
\rho (P_{[L/2]}\otimes 1) C_{11}C_{22}\cdots C_{L/2L/2} \)
N_c^{C( (\alpha_{L/2} \otimes \alpha_{L/2}) \rho^{-1} (\alpha_{L/2} \otimes \alpha_{L/2}) \rho)}
\nonumber \\[1mm]
&=
2\sum_{\rho_1,\rho_2 \in S_{L/2}}
\trb{L}^{f} \((P_{[L/2]}\otimes 1)
(\rho_1\otimes \rho_2) (P_{[L/2]}\otimes 1) C_{11}C_{22}\cdots C_{L/2L/2} \)
\nonumber\\
& \quad \times 
N_c^{C(\alpha_{L/2} \rho_1^{-1}\alpha_{L/2}\rho_1 )+C(\alpha_{L/2} \rho_2^{-1}\alpha_{L/2}\rho_2 )}
+O(N_c^{L-1})
\nonumber \\[1mm]
&=
2 \, {\rm Dim}_{L/2}
\sum_{\rho_1^{(0)},\rho_2^{(0)} \in S_{L/2}}
N_c^{L}+O(N_c^{L-1})
\nonumber \\
&=
\frac{L^2}{2} \, N_c^{L}{\rm Dim}_{L/2}
+O(N_c^{L-1}),
\label{normalisation_double_trace}
\end{align}
where $\trb{L}^{f}$ is the trace over $V_f^{\otimes L}$.
The factor $2$ at the second line originated in the symmetry 
that exchanges two $\alpha_{L/2}$'s of the operator. 
At the third equality, $\rho_1^{(0)}$ and $\rho_2^{(0)}$ are permutations 
satisfying $\rho_i^{(0)} \alpha_{L/2} \rho_i^{(0)-1}=\alpha_{L/2}$ 
 (see footnote \ref{symmetry_conjugacy_class}), and 
we have used
\begin{eqnarray}
&&
\trb{L}^{f}((P_{[L/2]}\otimes 1)
(\rho_1\otimes \rho_2) (P_{[L/2]}\otimes 1) C_{11}C_{22}\cdots C_{L/2L/2})
\nonumber \\
&=&
\trb{L}^{f}((P_{[L/2]}\otimes 1) C_{11}C_{22}\cdots C_{L/2L/2})
\nonumber \\
&=&
\trb{L/2}^{f}(P_{[L/2]})
\nonumber \\
&=&
{\rm Dim}_{L/2},
\label{derive_Dim_L/2}
\end{eqnarray} 
where we used $\rho_i \, P_{[L/2]} = P_{[L/2]}$ and $P_{[L/2]}^2 = P_{[L/2]}$\,. The second equality can be explained by Figure \ref{fig:flavor-tr} without permutation $\rho$.
${\rm Dim}_{L/2} = \frac{1}{12}(L/2+1)(L/2+2)^2(L/2+3)$ 
is the dimension of the totally symmetric traceless representation of $\alg{so}(6)$.

\subsection{Evaluating $\gamma_2$}\label{lambda_2_evaluate_correlator}

We now evaluate $\gamma_2$ based on the expression (\ref{gamma0_gamma1_gamma2}).

First, we see that the dilatation operator $\fD_\text{one-loop}$ \eqref{one-loop dilatation phi} annihilates the single-trace operator $\cC_{a_1 ,\cdots ,a_{s}}=\tr(\Phi_{(a_1}\otimes \cdots \otimes \Phi_{a_s)})$.
The first term of the dilatation operator annihilates $\cC_{a_1 ,\cdots ,a_{s}}$
because the indices are symmetrized, and  
the second term annihilates because the indices are traceless. 
Then, the dilatation operator acts on the product $\cC_{\vec{a}}\cC_{\vec{b}} = \cC_{a_1 ,\cdots ,a_{s}}\cC_{b_1 ,\cdots ,b_{t}}$ as
\begin{eqnarray}
\fD_\text{one-loop}(\cC_{\vec{a}}\cC_{\vec{b}}) 
\sim (\fD_\text{one-loop}\cC_{\vec{a}})\cC_{\vec{b}}+ \cC_{\vec{a}} (\fD_\text{one-loop}\cC_{\vec{b}})
+(\partial \cC_{\vec{a}})(\partial \cC_{\vec{b}})
=(\partial \cC_{\vec{a}})(\partial \cC_{\vec{b}}). 
\end{eqnarray} 
The two derivatives in the dilatation operator do not act on the same single-trace.

With the above facts, we will turn to the correlator expression of $\gamma_2$. 
Suppose that the dilatation operator acts locally on the factor
\begin{eqnarray}
\varphi\equiv 
\cC_{a_1 ,\cdots ,a_{s}}\cC_{c_1 ,\cdots ,c_{t}}
\label{double-trace_example_dilatation_action_lambda_2}
\end{eqnarray} 
contained in $\psi_0$. 
The effect of the non-planar dilatation operator is to make a single-trace 
from the double-trace, which we find from the formula (\ref{D-action 1a})-(\ref{D-action 2d}), reducing 
the number of traces of $\psi_0$ by one. 
The total action of the dilatation is the sum of local actions on each pair of single-traces.

From the intuition of AdS/CFT, we may regard \eqref{double-trace_example_dilatation_action_lambda_2} as a two-string state. A single-string state is also created by the action of the dilatation. 
In this sense, the first term of $\gamma_2$ in (\ref{gamma0_gamma1_gamma2}) is considered to be 
the overlap between the two-string and single-string states, and
the second term of $\gamma_2$ can be interpreted to be the planar energy of the single-string state.

\bigskip
We next discuss the sign of $\gamma_2$ in (\ref{gamma0_gamma1_gamma2}). 
Recall that the numerical study for $L\le 10$ shows that $\gamma_2$
is always non-positive for $\psi_0$ given by a linear combination of the large $N_c$ zero modes. 
It is easy to see that the second term of $\gamma_2$ is non-positive because 
the eigenvalues of the planar dilatation operator are non-negative.  
The sign of the first term is non-trivial 
because the non-planar dilatation operator contains several terms with different signs.

In order to get more insight on $\gamma_2$\,,
we will analyze the correlation functions  
exploiting a large $N_f$ limit, which is defined by 
letting the flavor indices run over $1,\cdots,N_f$ and by taking $N_f \gg L$. 
(In Appendix \ref{app:large_Nf_correlator} we will realize that $1/N_f$ appears with a factor $L$.) 
This limit describes the case $N_f=6$ in a good numerical accuracy for small $L$ as will be shown in Figure \ref{fig:lam2 Nf}.
Thus, we expect that the anomalous dimension as given by the ratio of correlators is not very sensitive to the value of $N_f/L$.

Concrete computations are doable at large $N_f$, as will be given in Appendix \ref{app:large_Nf_correlator}. 
In Appendix \ref{proof_negative_firstterm_lambda2} 
we will show that the first term of $\gamma_2$ in (\ref{gamma0_gamma1_gamma2}),   
$\langle \psi_0 \fD_1 \psi_0 \rangle $, is non-positive. 
In Appendix \ref{lambda2_double_trace_order_estimate} we will compute 
$\gamma_2$ for the two classes of the large $N_c$ zero modes presented in the last subsection. 
We write $\gamma_2^{(2)}$ for the double trace operator and 
$\gamma_2^{(L/2)}$ for the $L/2$-trace operators. 

In general, the first term of $\gamma_2$ is of order $N_f$ while the second term is of order $N_f^2$. Only the second term, $-\langle \psi_1 \fD_0 \psi_1\rangle/\langle \psi_0 \psi_0\rangle$, is important at large $N_f$\,.
For the double-trace operator (\ref{define_double_trace_operator}), 
our result (\ref{appendix_doubletrace_gamma2}) reads
\begin{eqnarray}
\gamma_2^{(2)} \sim  -\frac{\alpha}{4}N_f^2 L^2 +O(N_f),
\label{gamma2-2 Ldep}
\end{eqnarray}
where $\alpha$ is an $N_f$-independent constant that may depend on $L$.
In view of the AdS/CFT correspondence, the double-trace operator would correspond to 
a non-supersymmetric two-string state, where each string has the angular momentum $\pm L/2$ with the opposite sign.
Then $\gamma_2^{(2)}$ measures the energy difference between the two-string state and that of the bound state. 
As $L$ becomes large, one needs the more energy to break the bound state into the oppositely-rotating two-string state. This interpretation is consistent with the numerical results suggesting that $\alpha L^2$ is an increasing function of $L$.

For $L/2$-trace operators, we have the following result (\ref{gamma_2_L/2_appendix}),  
\begin{eqnarray}
\gamma_2 ^{(T)}
\sim - \frac{1}{2}N_f^2 \sum_s L_s (1+\delta_{L_s,4}) \alpha_{(T;s)}+O(N_f) , 
\label{gamma2T-2 Ldep}
\end{eqnarray}
where $T$ specifies an eigenstate among various $L/2$-trace operators, and $\alpha_{(T;s)}$ is another $N_f$-independent constant.
More details are presented in Appendix \ref{lambda2_double_trace_order_estimate}. 
These operators should correspond to bound states of non-supersymmetric $L/2$-particles rather than strings.

Finding the exact $L$ dependence of the submixing eigenvalues is a future problem.
One of the main obstacles is to compute the inverse of the planar dilatation operator $\fD_\epsilon^{-1}$. 
The eigenstates of $\fD_0$ can be readily constructed by using the integrability methods, but one needs to solve combinatorial problems to expand the states $\fD_1 \, \psi_0$ and $\psi_1$ in terms of the planar eigenstates.
Group-theoretical methods will be useful for this purpose. The limit $N_c\gg L \gg 1$ with $N_f=6$ is also worth investigation. Our results imply that the expansion parameter, or the effective string coupling, in this region is $\sim \sqrt{\alpha} L/N_c$ for \eqref{gamma2-2 Ldep} and $\sim \sqrt{\alpha L}/N_c$ \eqref{gamma2T-2 Ldep}


\section{Submixing from four-point correlators}\label{sec:rel correlators}

We discuss the relation between operator submixing problem and four-point functions of multi-trace operators.

Consider the four-point function of half-BPS operators, $\Vev{ \prod_{i=1}^4 \cC^{(p)} (x_i) }$. 
The four-point function of $\cN=4$ SYM can be decomposed into a product of three-point couplings and superconformal blocks \cite{DO04}. By expanding the four-point function in Taylor series around $g_{\rm YM}=0$, we obtain a sum of anomalous dimensions averaged over all possible intermediate operators of given quantum numbers, where the weight of average is given by the product of three-point couplings.

\begin{figure}[t]
\begin{center}
\includegraphics[scale=0.7]{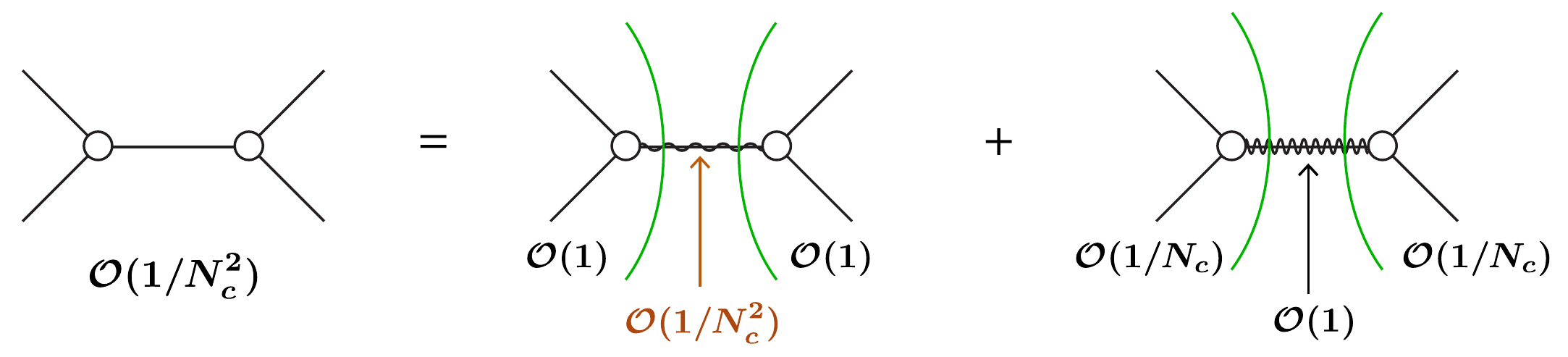}
\caption{OPE decomposition of a planar four-point function.}
\label{fig:4ptPlanar}
\end{center}
\end{figure}

If we normalize the two-point functions as $\cO(1)$, the perturbative part of the four-point function scales as $\cO(1/N_c^2)$ in the large $N_c$ limit \cite{GPS98, EHSSW98}. The three-point coupling scales as $\cO(1)$ if the intermediate operator has the same trace structure as $: \! \cC^{(p)} \cC^{(p)} \! :$, and $\cO(1/N_c)$ otherwise. For the former case, the anomalous dimension of the intermediate operator should scale as $\cO(1/N_c^2)$. If not, the anomalous dimension scales as $\cO(1)$. This structure is illustrated in Figure \ref{fig:4ptPlanar}.

The case of $p=2$ has been studied in detail in \cite{APPSS02}. From the four-point function $\vev{\prod_{i=1}^4 \cC^{(2)} (x_i)}$, we can extract the contribution of $\alg{so}(6)$ singlets which propagate along the internal line. There are four $\alg{so}(6)$ singlets at twist four, corresponding to the $L=4$ states \eqref{data mmop L=4}. If we denote the eigenvalues of the one-loop dilatation operator by $\{ \gamma^{(I)} \}$, we obtain
\begin{equation}
\sum_{I=1}^4 a_I^2 = 1, \qquad
\sum_{I=1}^4 a_I^2 \, \gamma^{(I)} = - \( \frac{N_c \, g_{\rm YM}^2}{8 \pi^2} \) \frac{4}{N_c^2}, \qquad
\sum_{I=1}^4 a_I^2 \, (\gamma^{(I)})^2 = \( \frac{N_c \, g_{\rm YM}^2}{8 \pi^2} \)^2 \frac{18}{N_c^2 \,},
\label{average anom dim}
\end{equation}
at $N_c \gg 1$. The negative sign of $\sum_{I} a_I^2 \, \gamma^{(I)}$ originates from the negative-mode of double-trace type, $: \! \cC_{ij} \cC_{ij} \!:$\,.

\bigskip
General submixing problem beyond the double-trace is related to the four-point functions of products of half-BPS operators. A simple example is
\begin{equation}
\Vev{ : \! \cC^{(2)} \cC^{(2)} \!:\! (x_1) \, \cC^{(2)} (x_2) \, \cC^{(2)} (x_3) \!:\! \cC^{(2)} \cC^{(2)} \!:\! (x_4) },
\end{equation}
whose OPE decomposition contains information about the triple-trace operator $:\! \cC^{(2)} \cC^{(2)} \cC^{(2)} \!:$. Although product operators like $: \! \cC^{(2)} \cC^{(2)} \!:$ are not dilatation eigenstates at finite $N_c$\,, we can neglect $1/N_c$ corrections by taking the large $N_c$ limit in what follows.

\begin{figure}[t]
\begin{center}
\includegraphics[scale=0.7]{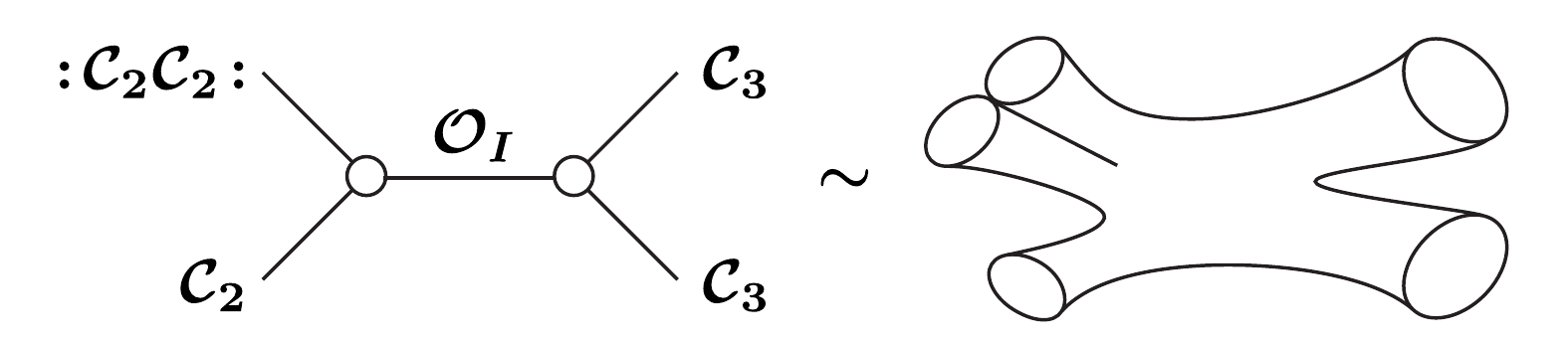}
\caption{A four-point function of $\cO(1/N_c^3)$. The right figure shows its color structure.}
\label{fig:mtr4pt}
\end{center}
\end{figure}

Some multi-trace four-point functions are $\cO(1/N_c^3)$ or less. In such situations, we can deduce constraints on the eigensystem of submixing matrices. Consider the multi-trace four-point function shown in Figure \ref{fig:mtr4pt},
\begin{equation}
F \equiv \Vev{ : \! \cC^{(2)} \cC^{(2)} \!:\! (x_1) \, \cC^{(2)} (x_2) \, \cC^{(3)} (x_3) \, \cC^{(3)} (x_4) }
\sim \sum_I \, C_{(2,2),2,I} \, C_{3,3,I} \, G_{\Delta_I, \ell_I} \Big( \frac{x_{12}^2 x_{34}^2}{x_{13}^2 x_{24}^2} \,, \frac{x_{14}^2 x_{23}^2}{x_{13}^2 x_{24}^2}  \Big).
\end{equation}
Since $F$ has the color structure of the five-point function of half-BPS operators, it scales as $\cO(1/N_c^3)$ at large $N_c$\,. Thus we find
\begin{equation}
F \big|_{\cO( g_{\rm YM}^{2n})} \sim 
\sum_I C_{(2,2),2,I} \, C_{3,3,I} \, (\gamma^{(I)} )^n \sim \cO(1/N_c^3) \qquad (\text{for any} \ n \ge 1),
\end{equation}
which implies
\begin{equation}
C_{(2,2),2,I} \, C_{3,3,I} \, \gamma^{(I)} \sim \cO(1/N_c^3) \qquad (\text{for any} \ I).
\label{suppression eq}
\end{equation}
Is there a situation where both $C_{(2,2),2,I}$ and $C_{3,3,I}$ are $\cO(1)$? It can happen if an intermediate operator is given by a linear combination of the double-trace and triple-trace large $N_c$ zero modes,
\begin{equation}
\cO_I^{\rm (hyp)} = c_1 \!:\! \cC^{(2)} \cC^{(2)} \cC^{(2)} \!:\! + \, c_2 \!:\! \cC^{(3)} \cC^{(3)} \!: \,.
\label{hypothetical op}
\end{equation}
If $\cO_I^{\rm (hyp)}$ is an eigenstate of the submixing Hamiltonian, then its dimension must be $\cO(1/N_c^3)$ from \eqref{suppression eq}. Indeed, our computation with $\alg{so}(6)$ singlets at $L=6$ shows that either $c_1$ or $c_2$ vanishes in \eqref{hypothetical op}, and one of the three-point couplings is $\cO(1/N_c)$. Moreover, the triple-trace large $N_c$ zero-mode has zero anomalous dimensions at any $N_c$\,.

Certainly, this argument can be generalized beyond $\alg{so}(6)$ singlets, and to the states with $L > 6$. Typically, a multi-trace four-point function is $\cO(1/N_c^3)$ if the total number of traces is different between two sides of the OPE decomposition. Then, from \eqref{suppression eq} the large $N_c$ zero-modes which propagate the internal line should have a definite number of traces to avoid $\cO(1)$ three-point couplings, or their anomalous dimensions should be $\cO(1/N_c^3)$. This result is consistent with our observation in Section \ref{sec:observation}.

It will be interesting to check this argument by perturbative results. It is conjectured in \cite{DP08} that the one-loop $n$-point correlators of the chiral primaries satisfy a generalized factorization formula at large $N_c$\,. A similar structure may be found for the multi-trace four-point functions considered here.

\section{One-loop spectrum at finite and analytic $N_c$}\label{sec:spec FAN}

In this section, we want to study the finite $N_c$ physics intrinsically, not by summing $1/N_c$ corrections using the large $N_c$ theories.
We will regard an anomalous dimension as a real analytic curve on the $(\gamma, N_c)$ plane and study the properties of the curves.

\subsection{Double determinant operators}\label{sec:ddet}

Let us start with the finite $N_c$ problem. In particular, we consider the spectrum of determinant-like operators based on our spectral data at arbitrary $N_c$\,.
Foundations of the following discussion are summarized in Appendix \ref{app:foundations}.

When the length of an operator is comparable to $N_c$\,, the mixing between single-trace and multi-trace operators is no longer negligible even at large $N_c$\,. One cannot a priori guarantee that the dimensions of such operators can be computed by integrability methods. In \cite{BDHNPSS13}, one of the authors studied the following double-determinant operators,\footnote{Here $X_k = \Phi_{2k-1} + i \Phi_{2k} = (W,Y,Z)$ are complex scalars of $\cN=4$ SYM.}
\begin{equation}
\cO_{Y,\olY} [Z^L, Z^{L'}] \sim \epsilon_{i_1 i_2 \dots i_N} \epsilon^{j_1 j_2 \dots j_N}
\epsilon_{k_1 k_2 \dots k_N} \epsilon^{l_1 l_2 \dots l_N}
Y^{i_1}_{j_1} \dots Y^{i_{N-1}}_{j_{N-1}} (Z^L)^{i_N}_{l_N}
\olY^{k_1}_{l_1} \dots \olY^{k_{N-1}}_{l_{N-1}} (Z^{L'})^{k_N}_{j_N} \,,
\label{def:YYbar}
\end{equation}
which should correspond to a pair of open strings ending on giant- and anti-giant-graviton branes.
The dimension of the $Y\olY$ operator was computed by integrability methods and perturbative $\cN=4$ SYM techniques, which precisely agreed for $L, L' \ge 2$.
However, at finite values of the 't Hooft coupling, the dimensions of these operators exhibit pathological behavior, which can be interpreted as the existence of tachyons \cite{BDHNPSS13,Hegedus15}.
Also, when $L$ or $L'=1$, an unexpected divergence is found at two loops. Thus, we should take the large $N_c$ limit more carefully because the operator \eqref{def:YYbar} is not the exact eigenstate of the one-loop dilatation.

In the heuristic argument of \cite{BDHNPSS13}, it was assumed that the anomalous dimension of the $Y\olY$ operator \eqref{def:YYbar} becomes non-trivial starting from the wrapping order. In particular, its anomalous dimension must be zero at one-loop at large $N_c$\,. 
To obtain the desired result by solving the non-planar operator mixing, we can look for an operator of length $\sim 2N_c$ whose one-loop dimension is zero at any $N_c$\,.
Since the length of double determinants exceeds $N_c$\,, it is important to guarantee that such a state survives under the finite $N_c$ reduction discussed in Appendix \ref{app:finite Nc constr}.

\paragraph{Zero mode at finite $N_c$\,.}

For simplicity, we study double determinant operators without insertion of $Z^L$'s instead of the $Y\olY$ operators \eqref{def:YYbar},
\begin{equation}
\cO_{Y,\olY} = \epsilon_{i_1 i_2 \dots i_N} \epsilon^{j_1 j_2 \dots j_N}
\epsilon_{k_1 k_2 \dots k_N} \epsilon^{l_1 l_2 \dots l_N}
Y^{i_1}_{j_1} \dots Y^{i_{N-1}}_{j_{N-1}} Y^{i_N}_{l_N}
\olY^{k_1}_{l_1} \dots \olY^{k_{N-1}}_{l_{N-1}} \olY^{k_N}_{j_N} + \dots,
\label{def:YYbar pure}
\end{equation}
which corresponds to a pair of giant and anti-giant graviton branes without open strings.
The symbol $+ \dots$ represents the terms induced by the operator mixing, such as the one given by the interchange $Y \leftrightarrow \olY$ or $Y \olY \leftrightarrow X_i X_i^\dagger$.

Our goal is to find completion of \eqref{def:YYbar pure} with zero anomalous dimension at any $N_c$\,.
Holomorphic-antiholomorphic operators are good candidates.
They belong to the so-called $k=0$ sector of \cite{KR07}, and their explicit forms are
\begin{equation}
\cO_{\rm hol} = \Bigg( \prod_{a=1}^M \tr Y^{\ell_a} \Biggr) 
\Biggl( \prod_{b=1}^N \tr \olY^{\bar \ell_b} \Biggr), \qquad
\sum_a \ell_a = \sum_b \bar \ell_b = \frac{L}{2} \,.
\label{def:hol op}
\end{equation}
The one-loop dilatation \eqref{one-loop dilatation phi} may break the holomorphic-antiholomorphic structure as 
\begin{equation}
\tr Y^{\ell_a} \, \tr \olY^{\bar \ell_b} \ \mapsto \ \Bigl\{ \tr (Y^{\ell_a} \olY^{\bar \ell_b}),\ 
\tr (Y^{\ell_a-1} X_i \, \olY^{\bar \ell_b-1} X_i^\dagger), \ 
\tr (Y^{\ell_a-1} X_i X_i^\dagger \, \olY^{\bar \ell_b-1}), \ \dots \Bigr\}.
\label{reducing hol-antihol}
\end{equation} 
Once broken, the holomorphic-antiholomorphic structure cannot be restored by further dilatation actions. Thus, the operators \eqref{def:hol op} do not appear on the right-hand side of $\fD \cdot \cO_I = M_{IJ} \cO_J$\,.
As a result, if an eigenstate of the dilatation operator contains a holomorphic-antiholomorphic element, its eigenvalue must be zero:
\begin{equation}
\psi \sim \cO_{\rm hol} + \dots \ \ \implies\ \ 
\gamma [ \psi ] = 0.
\label{hol mix cond}
\end{equation}
Similarly, the $k=0$ operators in the $\alg{su}(2)$ sector have zero anomalous dimension \cite{Kimura10}.

Consider the finite $N_c$ reduction on the holomorphic-antiholomorphic operators. 
If $L$ is a multiple of four, then the following holomorphic-antiholomorphic operator does not become null for $N_c \ge 2$,\footnote{If $L \equiv 2 \ ({\rm mod} \ 4)$, consider $\cO_{\rm hol}' = \cC_Y^{(2)} \, \cO_{\rm hol}$ with $\cC_Y^{(2)} = \tr (Y\olY) - \frac13 \, \tr (X_i X_i^\dagger)$.}
\begin{align}
\cO_{\rm hol} = \Bigl( \tr (Y^2) \Bigr)^{L/4}  \Bigl( \tr (\olY^2) \Bigr)^{L/4} 
= 2^{L/2} \Bigl( Y_{21} Y_{12} + Y_{2,2}^2 \Bigr)^{L/4} \( \olY_{21} \olY_{12} + \olY_{2,2}^2 \)^{L/4} + \dots ,
\label{YYbar elem}
\end{align}
One can show that the mixing $\tr (Y^2) \tr (\olY^2) \leftrightarrow (\tr (Y\olY))^2$ cannot cancel \eqref{YYbar elem} completely. Thus the term \eqref{YYbar elem} survives at $N_c=2$. And if an eigenstate is non-trivial at $N_c=2$, then it must not be null for $N_c \ge 2$.

\paragraph{Closed subsector of one-loop dilatation.}

Let us briefly explain how the dimensions of $Y\olY$ operators are related to those of $\alg{so}(6)$ singlets which we studied earlier.

Under the dilatation actions, the $Y\olY$ operators like \eqref{def:YYbar pure} mix with $\alg{so}(6)$ singlets as in \eqref{reducing hol-antihol}.
Since the dilatation itself is $\alg{so}(6)$ singlet, the $\alg{so}(6)$ singlet operators mix among themselves.
Let us define the characteristic polynomial of the mixing matrices for two subsectors,
\begin{equation}
\fP_{Y\olY} (\gamma) = {\rm det} \, (M_{IJ} - \gamma \, \delta_{IJ}) \Big|_\text{$Y\olY+$ singlet} \,, \qquad
\fP_\text{singlet} (\gamma) = {\rm det} \, (M_{IJ} - \gamma \, \delta_{IJ}) \Big|_\text{singlet} \,.
\end{equation}
Since the mixing matrix for $Y\olY$ contains the mixing matrix for singlets as a subset, the polynomial $\fP_{Y\olY} (\gamma)$ is divisible by $\fP_\text{singlet} (\gamma)$.

Our calculation at $L=4,6$ reveals that the most negative eigenvalue belongs to $\fP_\text{singlet} (\gamma)$, although $\fP_{Y\olY} (\gamma)/\fP_\text{singlet} (\gamma)$ contains some negative modes. Assuming that this is a generic pattern, we will study the lowest eigenvalue among $\alg{so}(6)$ singlets in detail.

\paragraph{The lowest eigenvalue at finite $N_c$\,.}

\begin{figure}[ht]
\begin{center}
\includegraphics[scale=0.6]{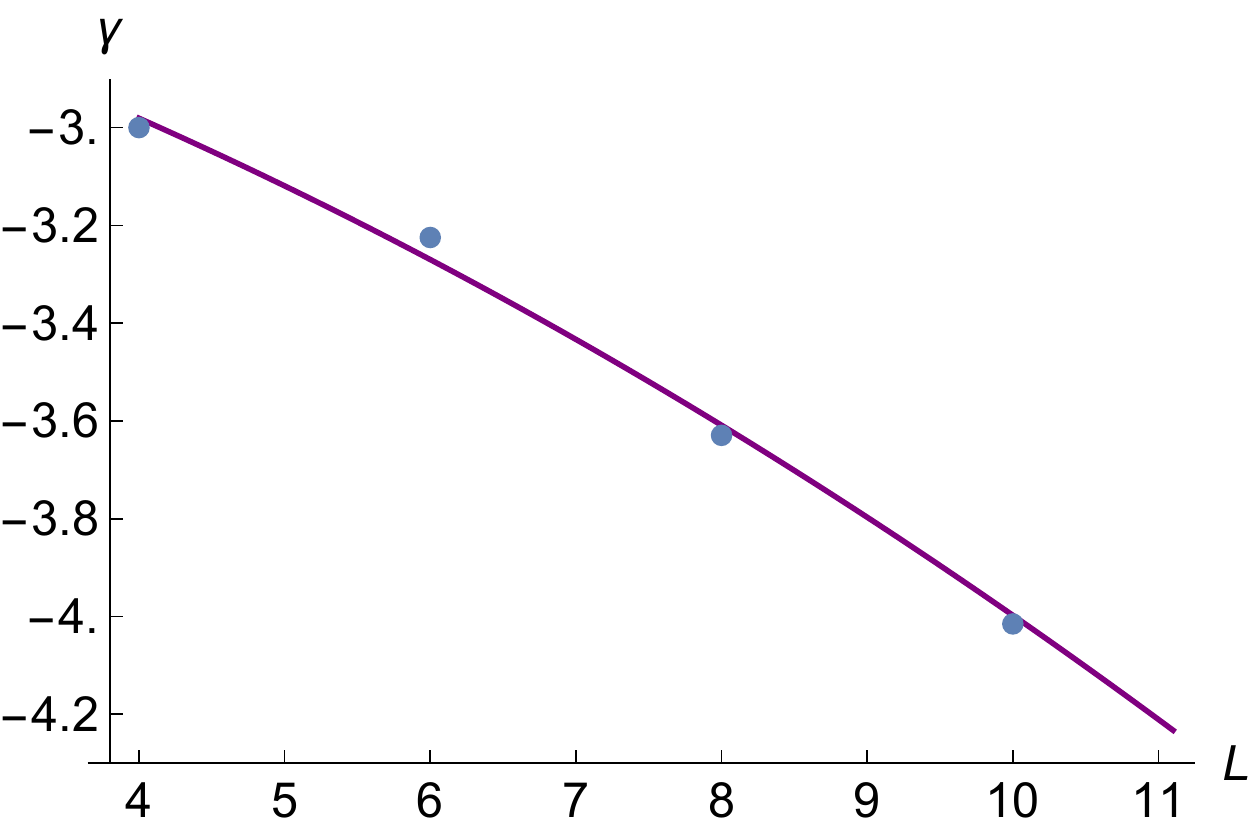}
\caption{Plot of the lowest anomalous dimension at $N_c=L/2$ in Table \ref{tab:gam2-lowest}. The quadratic fit is given by $\gamma = -2.70242 -0.031668 L -0.01006 L^2$.}
\label{fig:DDbar}
\end{center}
\end{figure}

At large $N_c$\,, the state with the smallest $\gamma_2$ shown in Figure \ref{fig:lam2}, which is the double-trace operator for $L \le 10$, has the lowest eigenvalue among all $\alg{so}(6)$ singlets at a fixed $L$.
Moreover, we observed that this lowest eigenstate does not become null for $N_c \ge L/2$.
The numerical values of the lowest eigenvalue in the $\alg{so}(6)$ singlets at $N_c=L/2$ are shown in Table \ref{tab:gam2-lowest} and Figure \ref{fig:DDbar}.

\bigskip
\begin{table}[ht]
\begin{center}
\begin{tabular}{c|cccc}\hline
$L$ & 4 & 6 & 8 & 10 \\\hline
$\gamma$ & $-3$ & $-3.22476$ & $-3.62944$ & $-4.01516$ \\\hline
\end{tabular}
\caption{The smallest one-loop anomalous dimension $\Delta_1 = \( \frac{N_c \, g_{\rm YM}^2}{8 \pi^2} \) \gamma$ at $N_c=L/2$.}
\label{tab:gam2-lowest}
\end{center}
\end{table}

The results show that $|\gamma_2|$, the coefficient of the one-loop anomalous dimension, increases faster than linearly as $N_c$ increases. If we extrapolate this behavior at large $N_c$\,, the lowest-energy state will have the dimension
\begin{equation}
\Delta = 2N_c - \frac{N_c \, g_{\rm YM}^2}{8 \pi^2} \( \alpha_0 + \alpha_1 N_c + \alpha_2 N_c^2 \) + \dots,
\label{ground state dim}
\end{equation}
neglecting $\alpha_i$ for $i \ge 3$.
At large $N_c$\,, this expression hits the unitarity bound $\Delta=0$ around $g_{\rm YM} = 0$ very quickly. 
In other words, the operators with the anomalous dimension \eqref{ground state dim} cannot be studied in the 't Hooft limit.

The relation between the divergence of the lowest eigenvalue in \eqref{ground state dim} and the pathological behavior of $Y\olY$ operators found in \cite{BDHNPSS13,Hegedus15} is unclear.
The problem in the na\"ive planar limit has disappeared by including all non-planar corrections (i.e. by identifying the ground state of integrable spin chain with boundaries as a holomorphic-antiholomorphic operator). However, there is another operator whose dimension diverges in the large $N_c$ limit with all non-planar corrections included.

\bigskip
Let us recall that there are other situations where the integrability methods such as the generalized L\"uscher formula or the mirror TBA disagree with perturbative field-theory calculation.
Not all the disagreements are related to non-planar effects.
One of the most puzzling examples is the single-particle state $\tr (XZ)$ in the $\beta$-deformed theory \cite{AdvT10,FSW13b}.
The field theory calculation shows that this operator is protected owing to the so-called prewrapping effect. 
However, the TBA computation based on a na\"ive asymptotic Bethe Ansatz neglecting the prewrapping effect predicts a divergent answer.
It is not known how to incorporate the prewrapping effect in the integrability methods.

\subsection{Level-crossing and level-pairing}\label{sec:level cross}

Below we regard one-loop dimensions as locally real analytic functions of $N_c$\,, instead of collection of real numbers evaluated at integer values of $N_c$\,.
It allows us to keep track of each eigenstate from large $N_c$ to small $N_c$\,.

\subsubsection{Level-crossing}\label{sec:level crossing}

We ask the question when two operators have exactly the same one-loop dimensions at finite $N_c$\,.

Consider the characteristic polynomial of the non-planar mixing matrix $M_{IJ}$\,. This polynomial factorizes into
\begin{equation}
\fP(\gamma) = \det (M_{IJ} - \gamma \, \delta_{IJ}) = \prod_a \fP_a (\gamma),
\label{def:ch poly}
\end{equation}
where $\fP_a(\gamma)$ is a prime polynomial over $\bb{C}$ at generic values of $N_c$\,.

The roots of different prime polynomials are unrelated, and nothing prevents the level-crossing.
For example, at $L=6$ there exists an apparently non-BPS operator which is protected at any values of $N_c$\,.\footnote{It is known that this operator remains protected at two-loop order \cite{BRS03}.}
At $L=8$ we find two non-BPS operators whose anomalous dimensions are simple functions of $N_c$\,.
These energy levels do cross with the other energy levels.\footnote{We use the words ``energy level'' and ``one-loop anomalous dimension'' interchangeably.}

In contrast, the roots of the same prime polynomial rarely collide unless $N_c$ is small, which can be roughly explained as follows.
The roots of the same prime polynomial come from a submatrix whose off-diagonal components are non-zero, and the non-zero off-diagonal components keep the eigenvalues separated.

Let us argue more accurately by writing down the condition that the one-loop dimensions of two operators coincide at finite $N_c$\,.
We denote the dilatation operator at $N_c=N_\bullet$ and $N_\bullet - \delta N$ by $\fD_\bullet$ and $\fD_\bullet - \delta \fD$, respectively. We want to solve the eigenvalue equation at $N_c=N_\bullet - \delta N$ perturbatively around $\delta D=0$.
The operator mixing equation can be written as
\begin{equation}
\( \fD_\bullet - \delta \fD \) \psi = \gamma \, \psi \quad \Leftrightarrow \quad
\psi = \( \fD_\bullet - \gamma \)^{-1} \, \delta \fD \, \psi .
\end{equation}
If we take the limit $\delta \fD \to 0$, only the states with $\( \fD_\bullet - \gamma \) \psi = \cO(\delta \fD)$ contribute to the right-hand side.
In particular, if the energy levels of two states are sufficiently close, the other energy levels are neglected.
We denote the one-loop dimensions of the two states at $N_c=N_\bullet \,, N_\bullet - \delta N$ by $\gamma_{\bullet (\pm)} \,, \gamma_{(\pm)}$\,, respectively.
The condition that these two states become degenerate at $N_c = N_\bullet - \delta N$ is given by 
\begin{equation}
\gamma_{(+)} - \gamma_{(-)} = \sqrt{( \epsilon + V_{11} - V_{22} )^2 + 4 \, V_{12} V_{21}} = 0, \qquad 
\gamma_{\bullet (+)} - \gamma_{\bullet (-)} = \epsilon ,
\end{equation}
where $\delta \fD \cdot \cO_I = V_{IJ} \, \cO_J$\,.
If $V_{IJ}$ is Hermitian, then the level-crossing is possible only if the off-diagonal components $V_{12} = V_{21}^*$ vanish. 
The off-diagonal components vanish, for example, when two states belong to different irreducible representations of the symmetry group of the model. This non-crossing rule is a famous statement by von Neumann and Wigner \cite{vNW29}.

The matrix $V_{IJ}$ is not Hermitian in the non-planar mixing problem of $\cN=4$ SYM.
Two roots of the same prime polynomial can collide and become a pair of complex conjugate roots.
Of course, any gauge-invariant operators of $\cN=4$ SYM must have real dimensions, at least to all orders of perturbation in $g_{\rm YM}$\,. 
It suggests that whenever two eigenvalues collide and become complex, the corresponding eigenvectors should be nullified by the finite $N_c$ reduction.
As a corollary, the roots of a prime polynomial will never collide for $N_c \ge L$.\footnote{This fact can also be used as a consistency check on the computation of the non-planar mixing matrix.}

A few remarks are in order.
Since the levels repel each other, the energy levels are quite dense for $N_c > L \gg 1$, and the one-loop dimensions can change little as we vary $N_c$ as long as $N_c > L$.
Also, the $1/N_c$ expansion of the one-loop dimensions fails to converge at the points when complex roots show up.

\subsubsection{Specifying branches} \label{sec:specify branch}

Let us consider the non-planar eigenvalue problem from a mathematical point of view.

The dimension of a physical operator is a root of the characteristic polynomial \eqref{def:ch poly} evaluated at integer points of $N_c$\,. Since the polynomial is an analytic function of $N_c$\,, the dimension can be analytically continued at any $N_c$\,, which we call an eigenvalue curve.
By keeping track of eigenvalue curves, we can see which finite $N_c$ eigenstate is connected to which of the large $N_c$ eigenstate.

\begin{figure}[t]
\begin{center}
\includegraphics[scale=0.8]{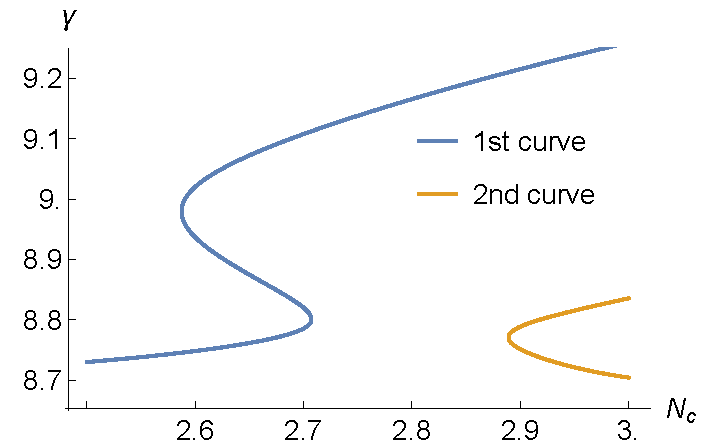}
\hspace{3mm}
\includegraphics[scale=0.8]{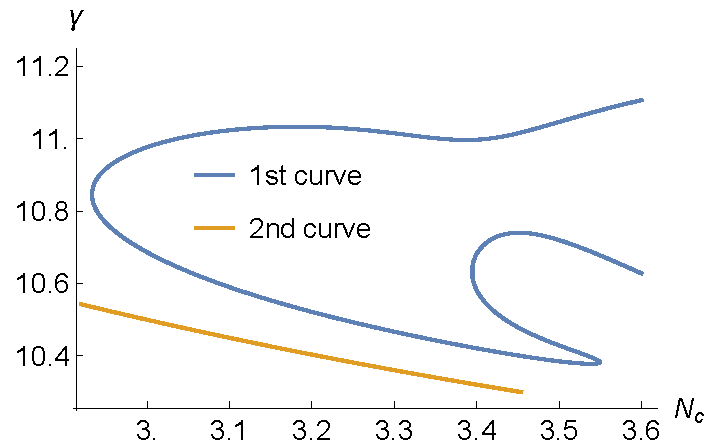}
\caption{The S-shape trajectory of eigenvalue curves based on $L=8$ data.}
\label{fig:S-shape}
\end{center}
\end{figure}

Not all large $N_c$ eigenstates can be extended to small $N_c$\,. As $N_c$ decreases, a pair of adjacent energy levels collide and create a pair of complex conjugate energy levels. In other words, if we start from a small $N_c$ theory and increase $N_c$\,, a new pair of states are created at finite $N_c$\,. 
There also exists a pair-annihilation, where two adjacent energy levels collide as $N_c$ increases.
The combination of pair-creations and pair-annihilations makes it quite non-trivial to keep track of eigenvalue curves as a function of $N_c$\,. 
Some eigenvalue curves draw an S-shape trajectory on the $(\gamma,N_c)$ plane via a pair-creation and pair-annihilation, as shown in Figure \ref{fig:S-shape}.
Inside the S-shaped region, an eigenvalue curve is no longer a single-valued function of $N_c$\,.

Mathematically, specifying an operator diagonalizing two-point functions of $\cN=4$ SYM, is equivalent to specifying one branch of the algebraic curve defined by the characteristic polynomial \eqref{def:ch poly}. 
We can define a branch of the algebraic curve algebraically or geometrically. Algebraically, a branch is a root of the characteristic polynomial $\gamma^{(\alpha)} (N_c)$, where $\gamma^{(\alpha)} (N_c)$ is a single-valued function of $N_c$\,.
Geometrically, a branch is a connected component of the eigenvalue curves on the $(\gamma,N_c)$ plane among the collection of eigenvalue curves.
Although the two definitions are closely related, it turns out that an eigenvalue curve is not always a single-valued function of $N_c$\,.\footnote{Moreover, this definition depends on the choice of coordinates on the $(\gamma,N_c)$ plane.}
In short, the geometric definition is more useful than the algebraic one at finite $N_c$.
Then, how many connected components are there?
How precisely can we specify an operator of $SU(N_c) \ \cN=4$ SYM?

\bigskip
To answer this question, we must define the meaning of ``connected''.
It is convenient to exclude a point or points where almost all branches meet. In our definition of the one-loop dilatation operator \eqref{one-loop dilatation phi}, we should exclude the points $(\gamma, N_c)=(\pm \infty,0)$ and study the curve in the region $N_c > 0$. 
Then, each branch can be specified by prescribing their large $N_c$ behavior,
\begin{equation}
\fP (\gamma, N_c) = 0, \qquad 
\lim \limits_{N_c \to + \infty} \gamma (N_c) 
= \gamma^{(\alpha)}_0 + N_c^{-1} \gamma^{(\alpha)}_1 + N_c^{-2} \gamma^{(\alpha)}_2 + \dots 
\qquad (N_c > 0).
\label{specify branch gamma}
\end{equation}
We define a connected curve by identifying the adjacent branches which collide at finite $N_c$\,.

Alternatively, we can also use the rescaled eigenvalue $\gamma = \tilde \gamma/N_c$\,.
Then, the large $N_c$ dimensions can be read off from the asymptotic slope of the curves in $(\tilde \gamma, N_c)$ plane, and we exclude the points $(\tilde \gamma, N_c)=(\infty,\infty)$.\footnote{The large $N_c$ zero modes asymptotes to $(\tilde \gamma, N_c)=(0,\infty)$. We should also exclude these points.}
More importantly, the $\tilde \gamma$'s stay finite around $N_c=0$, and each eigenvalue curve can be smoothly extended to $N_c < 0$.

The reader may be upset because one cannot define gauge-invariant operators for $N_c \le 1$\,. 
This difficulty can be circumvented by replacing $U(N_c)$ gauge group with $U(N_c+k|k)$. 
The one-loop mixing matrix does not change if we modify ${\rm tr} (1) = N_c$ to ${\rm str} (1) = N_c$.\footnote{We neglect the decoupling of $U(1)$ for simplicity. The $\cN=4$ SYM scalar $\Phi$ with $SU(N_c)$ gauge group should be extended to the scalar $\hat \Phi$ with $PSU(N_c+k|k)$ gauge group by imposing $\tr(\hat \Phi) = {\rm str} \, (\hat \Phi) = 0$.}
For a sufficiently large $k$, the supergroup theory is not subject to any finite $N_c$ constraints and one gets non-unitary AdS/CFT correspondence realized by ghost D-branes \cite{Vafa14,OT06}.
In this setup, the dilatation eigenvalues at negative $N_c$ are well-defined, and they can be complex due to the loss of unitarity.

\bigskip
Now let us have a closer look at the rescaled one-loop dimensions at negative $N_c$\,.

If we take the limit $N_c \to - \infty$, we should recover the spectrum of planar dilatation operator, neglecting the flipped sign. The invariance under $N_c \to - N_c$ can be understood as the invariance under the interchange of $M \leftrightarrow N$ in the $U(M|N)$ gauge group.
Another explanation is that the 't Hooft limit of $SU(N_c)$ theory is universal in the sense that one cannot discern if $N_c$ is positive or negative.\footnote{The sign is important to 
distinguish the $SO(N_c)$ and $Sp(N_c)\ \cN=4$ SYM theory \cite{CKZ10, CdMK13,Kemp14b}.}

Let us give yet another proof that the dilatation spectra at $N_c=\pm \infty$ are identical. 
From the block structure of the dilatation operator, one can show that the characteristic polynomial is compatible with the following $\bb{Z}_2$ symmetry:
\begin{equation}
\fP (\gamma, N_c) = \fP (\gamma, -N_c) \qquad \Leftrightarrow \qquad
\fP \( \tilde \gamma \, N_c , N_c \) = \fP \( (-\tilde \gamma) (-N_c), -N_c \).
\label{chP identity}
\end{equation}
To explain the block structure, let us take monomial multi-trace operators as a basis of the Hilbert space and collect the operators together according to their trace structure.
The matrix elements of the dilatation operator within the same trace structure (i.e. block-diagonal parts) are of order $N_c^0$, while those between different trace structures (i.e. block-off-diagonal parts) are of order $N_c^{-1}$. 
Now recalling the definition of the characteristic polynomial,
\begin{equation}
\fP (\gamma, N_c) = \sum_{\sigma \in \cS_d} {\rm sign} (\sigma) \, 
(M_{1 \, \sigma(1)} - \gamma \, \delta_{1 \, \sigma(1)})
(M_{2 \, \sigma(2)} - \gamma \, \delta_{2 \, \sigma(2)})
\dots 
(M_{d \, \sigma(d)} - \gamma \, \delta_{d \, \sigma(d)}),
\end{equation}
one finds that, for each $\sigma$, all terms come with an even power of $N_c$\,. 
Hence the identity \eqref{chP identity} follows. Note that this identity does not imply that all eigenvalues are expanded in powers of $N_c^{-2}$, because there may be a pair of eigenvalues obeying $\gamma_\pm = a \pm b/N_c + c/N_c^2 + \dots$.

In terms of the rescaled eigenvalues, the above identity induces an automorphism
\begin{equation}
\iota \,:\, (\tilde \gamma, N_c) \ \mapsto \ (-\tilde \gamma, -N_c),
\label{def:iota}
\end{equation}
which shows that the eigenvalue curves and their mirror-images are identical on the $(\tilde \gamma, N_c)$ plane.
In general, the automorphism $\iota$ relate different eigenvalue curves. To see this, consider the relation between the highest and the lowest eigenvalues.
\begin{equation}
\tilde \gamma^\text{(lowest)} (-N_c) = - \tilde \gamma^\text{(highest)} (N_c)
= \iota \cdot \[ \tilde \gamma^\text{(lowest)} (-N_c) \] .
\label{iota high-low}
\end{equation}
Our data in Appendix \ref{app:data} shows that the outermost eigenvalues never collide with the adjacent eigenvalues, and remain real for real $N_c$\,. Thus, the two energy levels are disjoint on the $(\tilde \gamma, N_c)$ plane.
Incidentally, the highest energy state in the $\alg{so}(6)$ sector is the product of Konishi operators, whose one-loop dimension at large $N_c$ behaves as $\gamma = 3L + N_c^{-2} \, L(L-2)/8 + \dots$.

Generally, by using $\iota$ we can group together the low-energy and high-energy eigenvalue curves as
\begin{equation}
\tilde \gamma^\text{($n$-th lowest)} (-N_c) = - \tilde \gamma^\text{($n$-th highest)} (N_c)
= \iota \cdot \[ \tilde \gamma^\text{($n$-th lowest)} (-N_c) \].
\label{iota high-low 2}
\end{equation}
Thus, the upper (or lower) half of the spectrum is redundant. We can throw away a half of the eigenvalue curves, or a half of the operators in $U(N_c+k|k)$ SYM theory by using $\iota$. We call it level-pairing.
The level-pairing structure is evident from Figure \ref{fig:top L=4,6}, where we plotted the rescaled eigenvalues for $N_c \in \bb{R}$ at $L=4,6$.
The plots for $L=8$ is shown in Appendix \ref{app:data L8}.

\begin{figure}[t]
\centering
\subfigure{\includegraphics[scale=0.5]{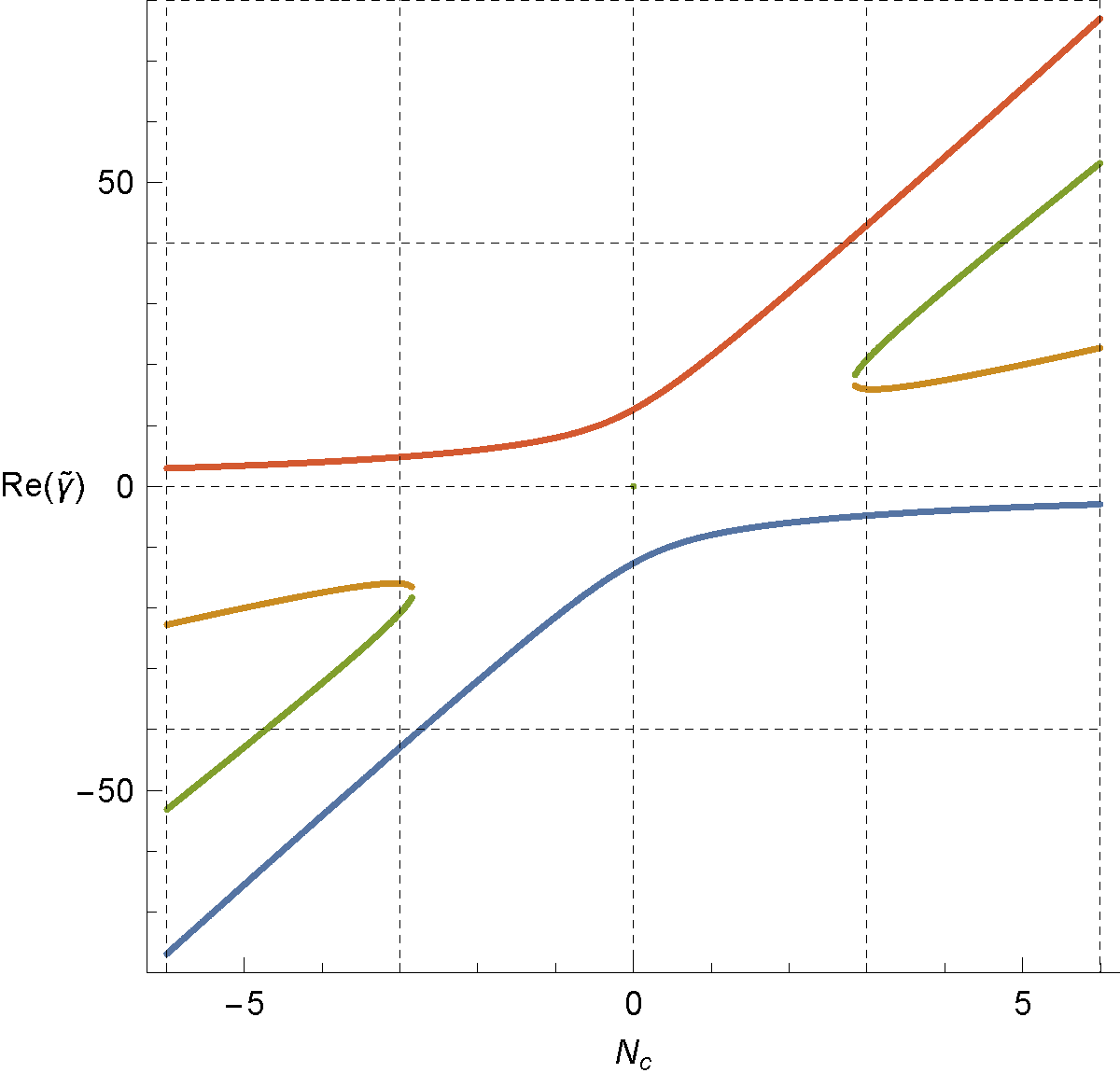}}
\hspace{4mm}
\subfigure{\includegraphics[scale=0.5]{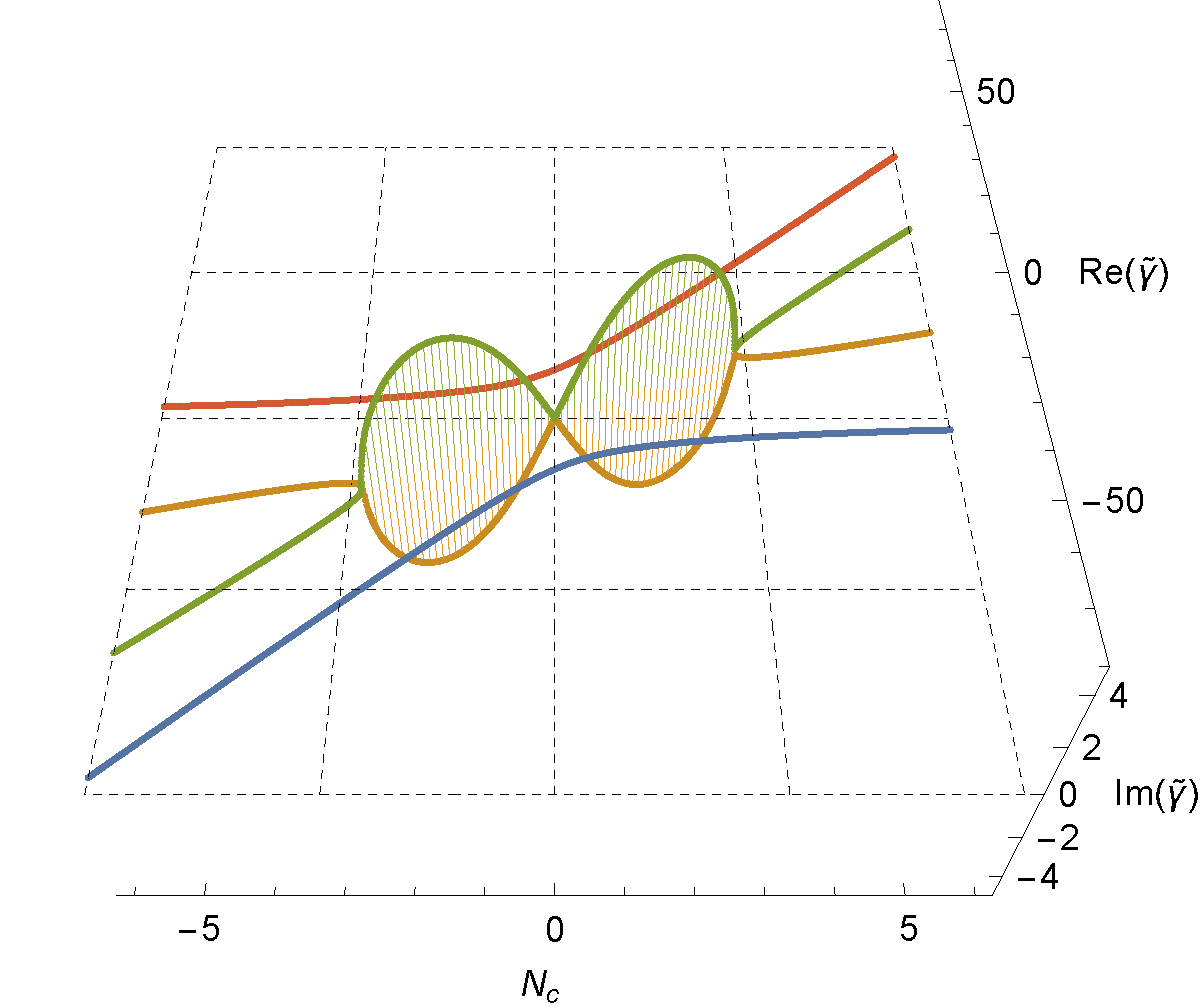}}
\\[4mm]
\subfigure{\includegraphics[scale=0.5]{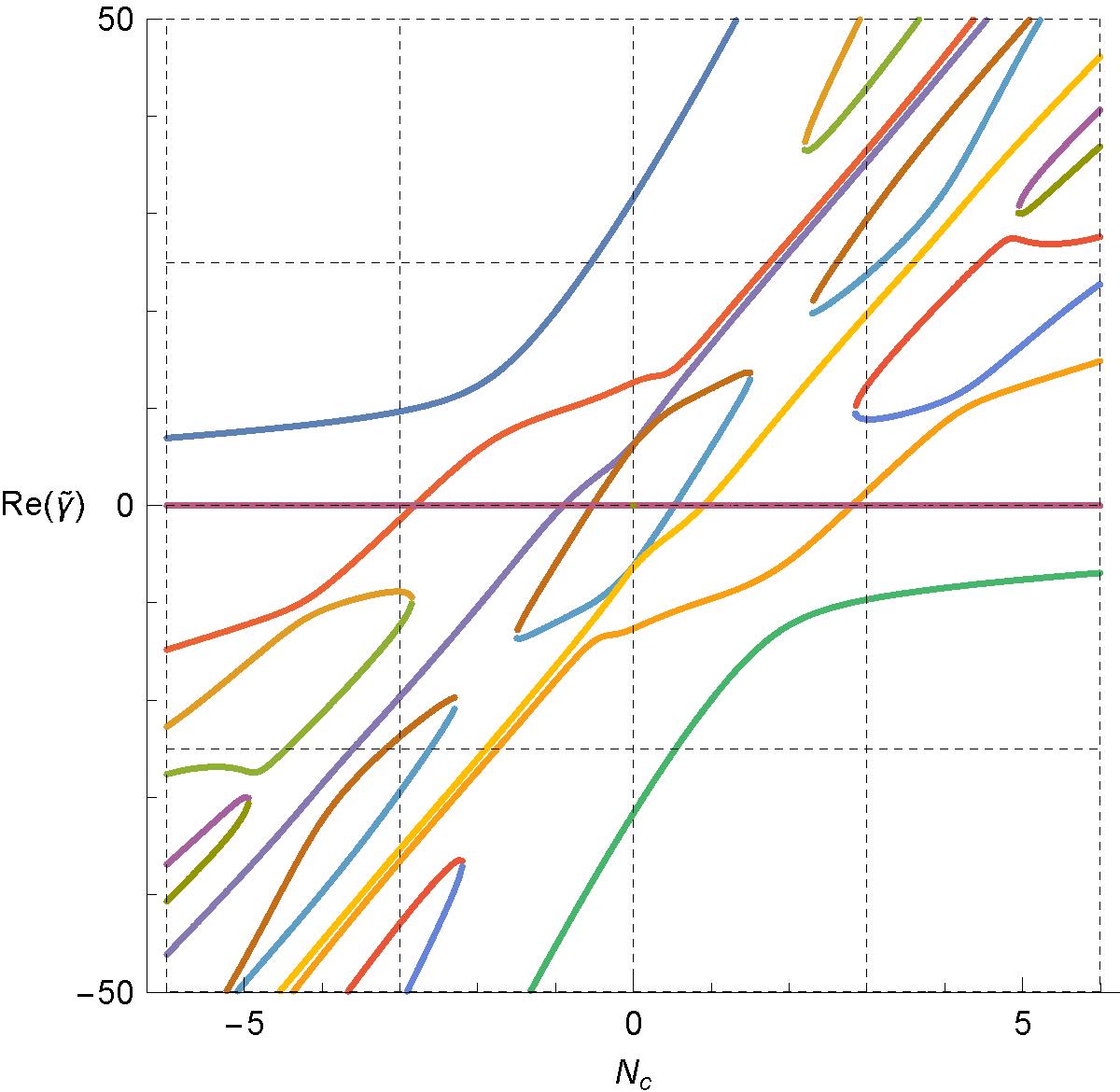}}
\hspace{4mm}
\subfigure{\includegraphics[scale=0.5]{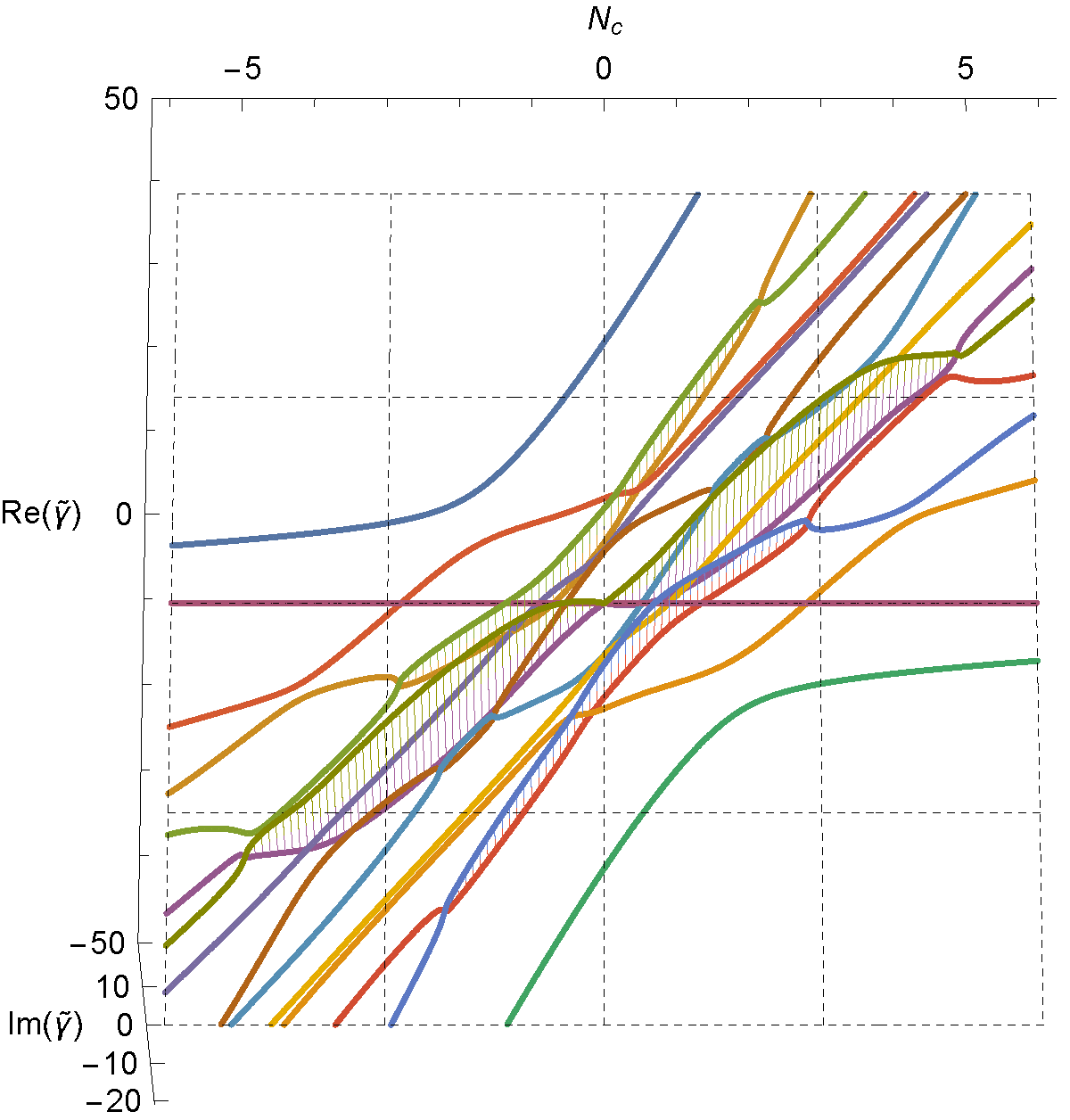}}
\caption{The rescaled one-loop anomalous dimensions of $\alg{so}(6)$ singlets of length $L=4$ (upper) and $L=6$ (lower) at finite $N_c$\,. Left figures show the eigenvalues on the real plane ${\rm Im} \, \tilde \gamma = 0$. Right figures are an aerial view showing both ${\rm Re} \, \tilde \gamma$ and ${\rm Im} \, \tilde \gamma$.
We included the real eigenvalues which are not continuously connected to the large $N_c$ data on the real plane ${\rm Im} (\tilde \gamma) = 0$ in both figures.
The eigenvalues do not collide off the real plane because of the dimensionality; two straight lines in more than two-dimensional space usually passes over or under the other.} 
\label{fig:top L=4,6}
\end{figure}

\clearpage

Here we list some properties of the level-pairing. 
First, $\iota$ maps a $1/N_c$\,-protected states to itself, because $\tilde \gamma = \gamma N_c$ is a straight line on the $(\tilde \gamma, N_c)$ plane.
Second, to find the level-pairing one must consider all multi-trace operators, even in the planar limit.
Third, the level-pairing preserves $\alg{so}(6)$ charges.

The automorphism $\iota$ induces a non-trivial map between the conformal dimensions. It maps $\Delta = L + \frac{g_{\rm YM}^2}{8 \pi^2} N_c \tilde \gamma + \cO(g_{\rm YM}^4)$ to $\iota \circ \Delta = L - \frac{g_{\rm YM}^2}{8 \pi^2} N_c \tilde \gamma + \cO(g_{\rm YM}^4)$. Then one needs to analytically continue $N_c$ to $-N_c$ along the same eigenvalue curve to obtain $\Delta' \equiv \text{a.c.} \, (\iota \circ \Delta) = L + \frac{g_{\rm YM}^2}{8 \pi^2} N_c \tilde \gamma' + \cO(g_{\rm YM}^4)$. It is not easy to quantify the relation between $\Delta$ and $\Delta'$ explicitly.

Let us make a few speculative comment about the automorphism $\iota$.
The $\bb{Z}_2$ symmetry \eqref{chP identity} should persist at higher orders in $g_{\rm YM}$ because this is a corollary of the large $N_c$ factorization (see \cite{Makkenko00a} for a review). Then we can ask whether the level-pairing can be interpreted as certain duality on \AdSxS\ superstring in the planar limit. 
Such a transformation should act on multi-string states on \AdSxS, and map a low-energy state to a high energy state. It should also satisfy other expected properties, namely to map a low-energy state to a high energy state, and a $1/N_c$\,-protected state to itself with the same $\alg{so}(6)$ charges.

Since $\iota$ is a relation between multi-string states, a worldsheet duality (such as the worldsheet T-duality \cite{IK07,HOSV07,BM08,KT12}) cannot be the precise counterpart of $\iota$.
Nevertheless, it may happen that a collection of multi-string states can be described by a single-string coherent state if the number of strings is sufficiently large. 
For example, if the automorphism $\iota$ corresponds to the worldsheet T-duality, then the product of Konishi operator in the limit $L \to \infty$ would be dual to a hoop-like string.\footnote{An infinitely winding hoop string potentially suffers from two types of instability: $\alpha'$ corrections and $g_s$ corrections. The leading $\alpha'$ correction to the energy of a hoop string is complex \cite{AA08}, which represents the instability of a hoop shrinking into the north or south pole. It is likely that the $g_s$ correction is also complex. Since the hoop intersects with itself infinitely many times, each piece of the hoop can recombine itself to string bits without consuming energy. The resulting state may be dual to the Konishi product.}

\section{Discussion and Outlook}\label{sec:discussion}

In this paper, we studied the spectrum of the non-planar one-loop dilatation operator among $\alg{so}(6)$ singlets by computing the matrix elements of operator mixing explicitly up to $L=10$.

At large $N_c$ we considered the submixing problem, which concerns how the large $N_c$ degeneracy is lifted by $1/N_c$ corrections, and observed interesting patterns.
First, the $1/N_c$ corrections to the dimensions of the large $N_c$ zero modes start appearing at the order $1/N_c^2$.
Second, the coefficient $\gamma_2$ is always non-positive.
Third, the operators with a different number of traces do not mix at large $N_c$\,, and those with the same number of traces do mix.
Fourth, the submixing density satisfies the projective commutation relations.

We have given some expressions of $\gamma_2$ in terms of correlation functions and derived the submixing Hamiltonian.
Using the correlator expression of $\gamma_2$, we have shown 
that $\gamma_2$ is non-positive in the large $N_f$ approximation.  
We studied in detail $L/2$-trace operators and the double-trace operator, which is the simplest operator
to work because it does not submix the other large $N_c$ zero modes. 
We also estimated how $\gamma_2$ depends on the operator length $L$.

From the AdS/CFT point of view, the negative sign of $\gamma_2$ can be interpreted as follows.
Multi-trace operators can be regarded as multi-string states on \AdSxS, and the product of BPS operators in $\cN=4$ SYM correspond to the product of BPS string states.
In general a product of BPS states is not protected by supersymmetry.
When multi-string states start interacting, they attract each other and form a bound state due to gravitational interaction.
The negative sign should also be related to the causality constraint on the \AdSxS\ side \cite{CEMZ14}. To identify the precise relation, we need to clarify two issues.
First, we studied $\alg{so}(6)$ singlets which are different from higher-spin operators of \cite{CEMZ14}. Second, we found operators which carry small positive anomalous dimensions at $\cO(1/N_c^3)$ or higher.

When the rank of the gauge group is comparable to the operator length, we are studying determinant-like operators with all non-planar corrections taken into account.
In AdS/CFT, the $Y\olY$ double-determinant operators should correspond to a pair of giant- and anti-giant-graviton D-branes. Such $D\ol{D}$-brane configuration is unstable and should decay into the vacuum of \AdSxS. 
At weak coupling, we find two interesting operators of $L \sim 2 N_c$ whose behaviors are similar to those of the $D\ol{D}$-brane system. The holomorphic-antiholomorphic operator can be regarded as non-planar completion of the $Y\olY$ double-determinant operator whose dimension is protected. The lowest-energy eigenstate in the $\alg{so}(6)$ sector has a negative anomalous dimension, which can be regarded as the non-planar ground state.
We conjecture that the anomalous dimension of the lowest-energy state diverges to $-\infty$ in the double scaling limit.\footnote{It is not clear what the gravity dual of the lowest-energy state with $L \sim 2 N_c$ is. The dual object might be a composite of another D-brane system, e.g. \cite{Witten98c}. RS thanks Nadav Drukker for discussion on this point.}

Furthermore, we have clarified the structure of level-crossing. 
When adjacent energy levels collide, the corresponding eigenstates should become null to prevent complex eigenvalues.
We found it quite non-trivial to keep track of the operator dimensions as we vary $N_c$\,. As a by-product, we discussed there is a natural pairing between different energy levels, particularly in the non-unitary $U(M|N)$ theories.

\bigskip
Our methods are mostly based on brute-force computation, whose power is limited only to relatively small $L$.
The dimension of the mixing matrix grows factorially with respect to $L$, and we encounter a challenging problem even numerically. It takes a long computational time to study the property of a huge matrix, and the results become less reliable due to the accumulation of numerical errors.

It is therefore important to look for another approach to the finite $N_c$ problem.
In Section \ref{sec:rel correlators} we related the submixing to the four-point functions of (products of) BPS operators, which will be an interesting future direction.
Examining the integrability in higher-point functions can be insightful, and rewriting the correlators in the Mellin space may also be useful \cite{Mack09a,Mack09b,Penedones10}.

Another promising approach to work on the finite $N_c$ physics is 
to exploit group representation theory. 
Operators can be labeled by a set of Young diagrams, and the operator mixing problem is
neatly described by group theoretic quantities. 
Yet, the study of the operator mixing problem in the $\alg{so}(6)$ sector 
has not been very active so far 
because of the complexity caused by the flavor structure. 
The $\alg{so}(6)$ singlet sector is much simpler than the full $\alg{so}(6)$ sector.
As we have seen in this paper, 
the singlet sector contains interesting physics 
that have not been observed in the $\alg{su}(2)$ sector. 
It would be one of the next directions 
to look closely into the $\alg{so}(6)$ sector at finite $N_c$.

\subsubsection*{Acknowledgements}

The authors acknowledge Agnese Bissi, Christoph Sieg, Gleb Arutyunov, Keisuke Okamura, Lionel Mason, Matthias Wilhelm, Luis Fernando Alday, Sanjaye Ramgoolam, Robert de Mello Koch, Tomek Lukowski for stimulating discussions.
We also thank Nadav Drukker for detailed comments on the manuscript.

The work of RS is supported by Marie Curie Intra-European Fellowship of the European Community's Seventh Framework Programme under the grant agreement number 327996. RS is RMCM of Kellogg College, University of Oxford.

We acknowledge useful conversations with workshop participants at the ESF and STFC supported workshop ``Permutations and Gauge String duality''.

\appendix

\section{Notation}\label{app:notation}

Let us define real scalars of $\cN=4$ SYM by
\begin{equation}
(\Phi_a)^i_j \equiv \sum_{A=1}^{\dim G} (T^A)^i_j \Phi_a^A \,, \qquad
(\vv{\Phi}_a)^i_j \equiv \sum_{A=1}^{\dim G} (T^A)^i_j \, \vv{\Phi}_a^A \,, \qquad
\end{equation}
where $G$ is the gauge group, $a=1,2, \dots, 6$ and $i,j=1,2, \dots, N_c$\,. As for $G=U(N_c)$, the symbol $\vv{\Phi}$ differentiates $\Phi$ as
\begin{equation}
(\vv{\Phi}_a)_i^j \, (\Phi_b)_k^l \equiv \frac{\partial}{\partial (\Phi_a)^i_j} (\Phi_b)_k^l
= \delta_{ab} \, \delta^l_i \, \delta^j_k \,.
\label{U(N) contraction}
\end{equation}
This rule can be used to compute the one-loop mixing matrix of $SU(N_c)$ theory as well by imposing the traceless condition $\tr(\Phi_a)=0$ when constructing a basis of operators.
The splitting-and-joining rules follow straightforwardly:
\begin{equation}
\contraction{\tr (A}{\vv{\Phi}_a}{) \, \tr (}{\Phi_b}
\contraction{\tr (A \vv{\Phi}_a ) \, \tr (\Phi_bB) = \delta_{ab} \, \tr (AB), \qquad \tr (A }{\vv{\Phi}_a}{B}{\Phi_b}
\tr (A \vv{\Phi}_a ) \, \tr (\Phi_bB) = \delta_{ab} \, \tr (AB), \qquad
\tr (A \vv{\Phi}_a B \Phi_b) = \delta_{ab} \, \tr(A) \, \tr (B).
\label{sj rules}
\end{equation}

The dilatation operator and conformal dimensions can be expanded in series of $g_{\rm YM}$ as\footnote{The dimension of Konishi multiplet is $\Delta = \Delta_0 + 6 \( \frac{N_c \, g_{\rm YM}^2}{8 \pi^2} \) + \dots$.}
\begin{equation}
\fD_\text{total} = \fD_\text{tree} + \( \frac{N_c \, g_{\rm YM}^2}{8 \pi^2} \) \fD_\text{one-loop} + \dots ,\quad 
\Delta = \Delta_0 + \Delta_1 + \dots, \quad
\Delta_1 = \( \frac{N_c \, g_{\rm YM}^2}{8 \pi^2} \) \gamma .
\label{def:Delta's}
\end{equation}
The non-planar dilatation operators in the scalar sector at one-loop is 
\begin{align}
\fD_\text{one-loop} = \frac{1}{N_c} :\! \left(
-\frac{1}{2} \tr [\Phi_{m},\Phi_{n}][\check{\Phi}^{m},\check{\Phi}^{n}]
-\frac{1}{4} \tr [\Phi_{m},\check{\Phi}^{n}] [\Phi_{m},\check{\Phi}^{n}]
\right) \! : \,,
\label{one-loop dilatation app}
\end{align}
which commute with all $\alg{so}(6)$ generators. It also commutes with the projection operator to a union of eigenspaces $\cH$,
\begin{equation}
[P_X, \fD_\text{total}] = 0, \qquad
P_X = {\bf 1} - \sum_{I \in \cH} \ket{\psi_I} \, \bra{\psi_I} \,, \qquad
\fD_\text{total} \ket{\psi_I} = \Delta \ket{\psi_I} .
\end{equation}

The first term of \eqref{one-loop dilatation app} acts on multi-trace operators as
\begin{alignat}{9}
&: \! \tr [\Phi_{m} \Phi_{n} \check{\Phi}^{m} \check{\Phi}^{n}] \! : \tr (\Phi_a A) \, \tr (\Phi_b B)
& &= \tr ( B A \Phi_b \Phi_a + A B \Phi_a \Phi_b ),
\label{D-action 1a} \\
&: \! \tr [\Phi_{m} \Phi_{n} \check{\Phi}^{m} \check{\Phi}^{n}] \! : \tr (\Phi_a A \Phi_b B)
& &= \tr (A) \, \tr (B \Phi_b \Phi_a) + \tr (B) \, \tr (A \Phi_a \Phi_b ),
\label{D-action 1b} \\%
&: \! \tr [\Phi_{m} \Phi_{n} \check{\Phi}^{n} \check{\Phi}^{m}] \! : \tr (\Phi_a A) \, \tr (\Phi_b B)
& &= \tr ( B A \Phi_a \Phi_b + A B \Phi_b \Phi_a ),
\label{D-action 1c} \\
&: \! \tr [\Phi_{m} \Phi_{n} \check{\Phi}^{n} \check{\Phi}^{m}] \! : \tr (\Phi_a A \Phi_b B)
& &= \tr (A) \, \tr (B \Phi_a \Phi_b) + \tr (B) \, \tr (A \Phi_b \Phi_a ),
\label{D-action 1d}
\end{alignat}
and the second term as
\begin{alignat}{9}
&: \! \tr [\Phi_{m} \check{\Phi}^{n} \Phi_{m} \check{\Phi}^{n}] \! : \tr (\Phi_a A) \, \tr (\Phi_b B)
& &= 2 \, \delta_{ab} \, \tr (\Phi_m A \Phi_m B),
\label{D-action 2a} \\
&: \! \tr [\Phi_{m} \check{\Phi}^{n} \Phi_{m} \check{\Phi}^{n}] \! : \tr (\Phi_a A \Phi_b B)
& &= 2 \, \delta_{ab} \, \tr (\Phi_m A) \, \tr (\Phi_m B),
\label{D-action 2b} \\
&: \! \tr [\Phi_{m} \Phi_{m} \check{\Phi}^{n} \check{\Phi}^{n}] \! : \tr (\Phi_a A) \, \tr (\Phi_b B)
& &= \delta_{ab} \, \tr [\Phi_m \Phi_m (AB+BA) ],
\label{D-action 2c} \\
&: \! \tr [\Phi_{m} \Phi_{m} \check{\Phi}^{n} \check{\Phi}^{n}] \! : \tr (\Phi_a A \Phi_b B)
& &= \delta_{ab} \pare{ \tr (A) \, \tr(\Phi_m \Phi_m B) + \tr (B) \, \tr(\Phi_m \Phi_m A) }.
\label{D-action 2d} 
\end{alignat}

\subsection{Notation for operator mixing}

To highlight the structure of the non-planar operator mixing, we introduce a formal parameter $N_f$ so that the global symmetry becomes $\alg{so}(N_f)$. The $\cN=4$ SYM corresponds to $N_f=6$.
There can be several ways to define the dilatation operator at general $N_f$\,, but the difference is not much important at the fully non-planar level. We use the simplest generalization; namely the operator identical to \eqref{one-loop dilatation phi} except that the flavor indices run $m,n=1,2, \dots, N_f$\,.\footnote{Another possible generalization is
\begin{equation*}
\fD'_\text{one-loop} = \frac{1}{N_c} :\! \left(
-\frac{1}{2} \, \tr [\Phi_{m},\Phi_{n}][\check{\Phi}^{m},\check{\Phi}^{n}]
-\frac{1}{2(N_f-1)} \, \tr [\Phi_{m},\check{\Phi}^{n}] [\Phi_{m},\check{\Phi}^{n}]
\right) \! : \,,
\label{one-loop dilatation Nf gen2}
\end{equation*}
which is integrable at large $N_c$ \cite{Reshetikhin83a,Reshetikhin85c,MZ02}.}

The matrix of operator mixing is defined as follows. We use monomial multi-trace operators as the basis, and denote them by $\{ \cO_I \}$. By applying the one-loop dilatation \eqref{one-loop dilatation phi} to them, we obtain the mixing matrix 
\begin{equation}
\fD \cdot \cO_I = M_{IJ} \, \cO_J \,.
\label{def:mixing matrix}
\end{equation}
Then we look for the eigenvector of the form $\psi_\alpha = c_{\alpha I} \, \cO_I$\,, which implies $M_{JI} \, c_{\alpha J} = \gamma \, c_{\alpha I}$\,. Note that this mixing matrix is the transpose of \eqref{def:mixing matrix}, and in {\tt Mathematica} the eigenvectors are given by {\tt Eigenvectors[Transpose@M].Table[O[i],\{i,d\}]}.
The matrix elements of operator submixing will be defined in the same way, namely
\begin{equation}
H_\text{sm}^\circ \cdot \cO_I = (M_\text{sm})_{IJ} \, \cO_J \,.
\end{equation}
Explicit matrix elements are computed via \eqref{sm Ham elements}.

\section{Foundations of finite $N_c$ calculation}\label{app:foundations}

\subsection{Finite $N_c$ reduction}\label{app:finite Nc constr}

When the rank of the gauge group is smaller than the operator length, gauge-invariant operators become linearly dependent.
This is a well-known phenomenon, and the linear relations are called finite $N_c$ constraints.\footnote{This is a non-perturbative effect on the string theory side, and called the stringy exclusion principle \cite{MS98}.}
Any finite $N_c$ constraints are written as the identity for an antisymmetric tensor 
\begin{equation}
0 = T_{[i_1 i_2 \dots i_{N_c+1}]} \,, \qquad
(i_k = 1,2, \dots, N_c).
\label{finite Nc constr}
\end{equation}
As $N_c$ decreases, the dimension of the Hilbert space of states shrinks by finite $N_c$ reduction.

The two-point functions at tree level can be diagonalized at any $N_c$ by using group-theoretical bases \cite{CJR01,KR07,BCdM08,BHR08}. 
In these bases, finite $N_c$ constraints are typically written as
$c_1(R)\le N_c$, where $c_1(R)$ is the number of rows of a Young diagram $R$. 
A set of finite $N_c$ constraints in the $\alg{so}(N_f)$ scalar sector was given in \cite{BHR08}.
In general there are several ways to express the finite $N_c$ constraints in a given sector. 
(This is the same as that there are several orthogonal bases in a sector.) 
The reason 
can be clarified from the existence of some sets of conserved charges at tree level \cite{KR08}.

At the loop level, there is no freedom in choosing the basis in which the two-point functions are orthonormal, except for degenerate cases.
Still, the finite $N_c$ constraints reduce the dimension of the Hilbert space of gauge-invariant operators.
When the finite $N_c$ constraints are imposed, we should find either (i) the eigenvalues remain unchanged, or (ii) the eigenvector becomes null.
Written explicitly, this means
\begin{equation}
\fD_\text{one-loop} \, \psi_I = \gamma_I \, \psi_I  \ \ \stackrel{\text{finite $N_c$}}{\implies} \ \ 
\pare{\fD_\text{one-loop} \, \psi'_I = \gamma_I \, \psi'_I \quad {\rm or} \quad \psi'_I = \vec 0}.
\label{Red ebasis}
\end{equation}
where $\psi'_I$ is the eigenstate subject to the finite $N_c$ constraints. 

Let us make a few remarks.
First, if an eigenvector $\psi_I$ becomes null, all correlators involving $\psi_I$ also vanish.\footnote{See \cite{BERS02} for an example.}
Second, if an eigenvector becomes null at $N_c=h$, then it remains null for $N_c \le h-1$, as follows from \eqref{finite Nc constr}.

\bigskip
A simple way to determine which eigenvector survives at finite $N_c$ is to substitute $N_c \times N_c$ real traceless matrices to the fields $\Phi_a$\,.
In other words, we regard multi-trace operators as $GL(N_c)$\,-invariant polynomials of the matrix elements $(\Phi_a)_{ij}$\,. When $N_c$ decreases by one, we remove the last row and column of the matrix elements. Schematically, this can be depicted as removing gray region of the following matrix:
\ifusetikz
\begin{equation}
\Phi_a = \begin{tikzpicture}[baseline=-\the\dimexpr\fontdimen22\textfont2\relax ]
\matrix (m)[matrix of math nodes,left delimiter=(,right delimiter=)]{
\phantom{0} & & & & & \\
& \phantom{0} & & & & \\
& & \phantom{0} & & & \\
& & & \phantom{0} & & \\
& & & & \phantom{0} & \\
& & & & & \phantom{0} \\
};
\begin{pgfonlayer}{myback}
\fhighlightDDD{m-1-1}{m-6-6}
\fhighlightDD{m-1-1}{m-5-5}
\fhighlightD{m-1-1}{m-4-4}
\fhighlight{m-1-1}{m-3-3}
\fhighlightW{m-1-1}{m-2-2}
\end{pgfonlayer}
\end{tikzpicture} , \qquad
(\Phi_a)_{11} = - \sum_{i=2}^{N_c} (\Phi_a)_{ii} \,.
\label{finite Nc Phi}
\end{equation}
\else
\begin{equation}
\Phi_a = \adjustbox{valign=c}{\includegraphics{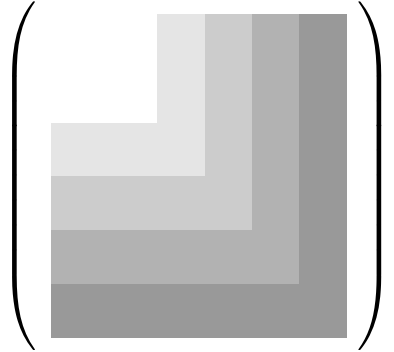}} , \qquad
(\Phi_a)_{11} = - \sum_{i=2}^{N_c} (\Phi_a)_{ii} \,.
\label{finite Nc Phi}
\end{equation}
\fi

By exploiting this idea, one can associate a Young diagram to each eigenstate as follows.
Instead of removing the last row and column, we can rescale the entries in the last column and the last row by $\epsilon_{N_c}$\,. The diagonal element $(\Phi_a)_{N_c N_c}$ are rescaled by $\epsilon_{N_c}^2$\,.
Then we measure how an eigenstate scales in the limit; $\psi \sim \cO (\epsilon_{N_c}^{\kappa_{N_c}} )$ as $\epsilon_{N_c} \to 0$. 
We can continue this procedure as
\begin{equation}
\lim \limits_{\epsilon_h \to 0} \ \dots \ 
\lim \limits_{\epsilon_{N_c-1} \to 0} \ 
\lim \limits_{\epsilon_{N_c} \to 0} \ 
\epsilon_{h+1}^{-\kappa_{h+1}} \dots \epsilon_{N_c-1}^{-\kappa_{N_c-1}}  \epsilon_{N_c}^{-\kappa_{N_c}} \, 
\psi = \cO (\epsilon_{h}^{\kappa_{h}} ),
\end{equation}
until $h=2$. This assigns to each eigenstate $\psi$ a sequence of integers $(\kappa_2, \kappa_3, \dots, \kappa_L)$ which is a partition of $L$.

The finite\,-$N_c$ constraints put restrictions on the matrix elements of non-planar operator mixing.
Let $\cV_h$ be the space of the $GL(N_c)$\,-invariant polynomials of $(\Phi_a)_{ij}$ which vanish for $N_c \le h$. 
In other words, $h$ is the smallest integer satisfying $\kappa_i=0$ for all $i>h$.
Each $\cV_h$ is closed under the dilatation operator. However, the mixing matrices are generally not block diagonal on the basis of $\{ \cV_h \}$. To see it, expand the eigenvector which becomes null at $h$ as
\begin{equation}
\psi (N_c) = \sum_i \tilde c_{i} (N_c) \, \tilde v_i + \sum_j c_{j} (N_c) \, v_j \,, \qquad
( \tilde v_i \in \cV_h \,, \ v_j \in \cV_2 \setminus \cV_h ).
\end{equation}
The null condition $\psi (h) = 0$ gives $c_{j} (h) = 0$. Since the coefficients are non-trivial functions of $N_c$\,, they do not identically vanish.

That said, the same reasoning allows us to remove some elements of the mixing matrix. If we use monomial multi-traces as the basis of operators, then the elements of mixing matrix are at most linear in $N_c^{-1}$; see \eqref{D-action 1a}-\eqref{D-action 2d}. This property remains unchanged unless we rotate the basis by a $N_c$\,-dependent matrix. Then, the condition $c_{j} (N_c=h) = c_{j} (N_c=h-1) = 0$ implies $c_j=0$, which excludes off-diagonal elements from $\cV_{h'}$ to $\cV_{h}$ for $h' \le h-2$.

Here is a side remark. 
In Liouville theory, the one-point function on a torus is equal to the four-point function on a sphere with degenerate fields \cite{FLNO09,Poghossian09,HJS09}. It would be interesting to seek for similar relations for the correlation functions of the null states in $\cN=4$ SYM.

\subsection{Spectral data}\label{app:data}

We summarize the basic properties of the finite $N_c$ spectrum, and the spectral data of one-loop anomalous dimensions for the scalar $\alg{so}(6)$ singlet operators at finite $N_c$ with length $L=4,6,8,10$. These results are obtained by explicit computation using {\tt Mathematica}.
The data include plots of the finite $N_c$ spectrum, the submixing matrix and the basis of operators. The eigenvalues of submixing matrices are given at general values of $N_f$ for $L \le 8$.

\subsubsection{$L=4$}

\newsavebox\fnbfour
\begin{lrbox}\fnbfour
{\scriptsize \verb|{{4,-2,-2*Nf/Nc,2/Nc},{-2,3+Nf,-2/Nc,Nc^(-1)+Nf/Nc},{2/Nc-2*Nf/Nc,-2/Nc+2*Nf/Nc,0,2},{-12/Nc,12/Nc,0,2*Nf}}|}
\end{lrbox}

\begin{figure}[t]
\begin{center}
\subfigure{\includegraphics[scale=1]{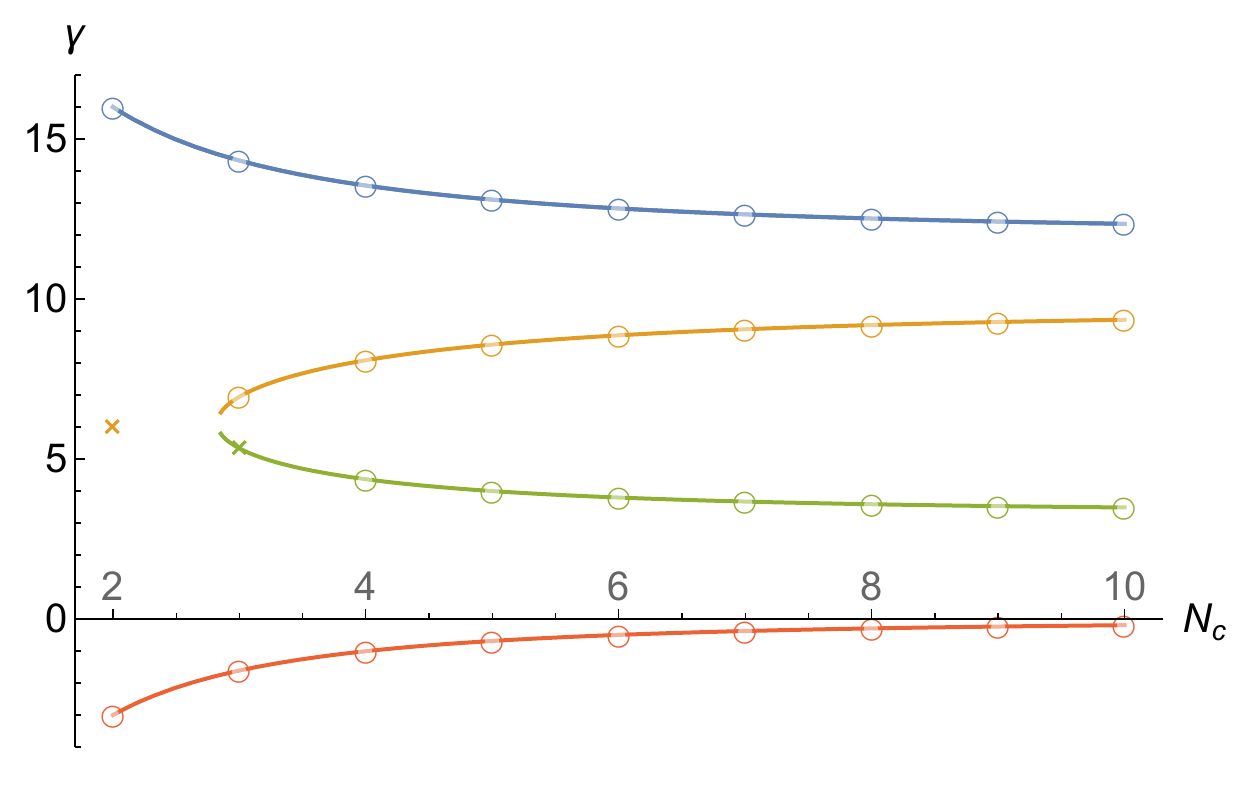}}
\\[5mm]
\subfigure{\includegraphics[scale=0.8]{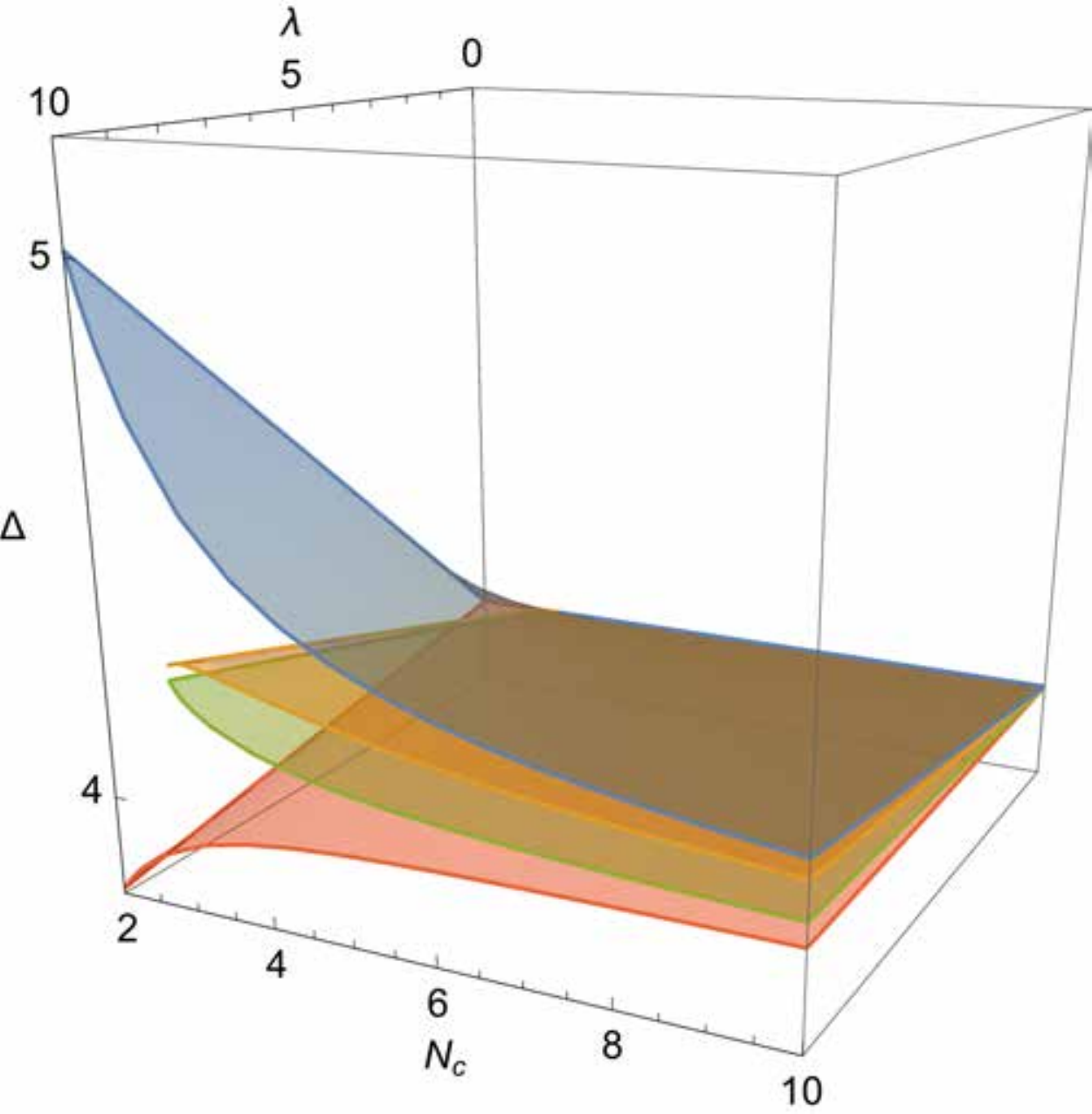}}
\caption{The dimensions of $\alg{so}(6)$ singlets of length $L=4$ at finite $N_c$\,. 
The upper figure shows the one-loop anomalous dimensions, and the eigenvalues corresponding to null eigenvectors are denoted by $\times$.
The lower figure shows the sum of tree and one-loop dimensions as a function of $N_c$ and the 't Hooft coupling $\lambda=N_c \, g_{\rm YM}^2$.}
\label{fig:finiteNc L=4}
\end{center}
\end{figure}

There are four $\alg{so}(N_f)$ singlets at length $L=4$. 
The matrix elements and operator basis for general $N_f$ are given by\footnote{In a computer-friendly format, $M_{IJ}$ is\\ \usebox\fnbfour \\
One can derive submixing matrix elements and various finite $N_c$ plots from this data as presented below.}
\begin{equation}
M_{IJ} = 
\begin{pmatrix}
4 & -2 & -\frac{2 N_f}{N_c} & \frac{2}{N_c} \\[1mm]
-2 & N_f+3 & -\frac{2}{N_c} & \frac{N_f+1}{N_c} \\[1mm]
\frac{2-2N_f}{N_c} & \frac{2 N_f-2}{N_c} & 0 & 2 \\[1mm]
-\frac{12}{N_c} & \frac{12}{N_c} & 0 & 2 N_f 
\end{pmatrix}, \quad
\cO_J =
\begin{pmatrix}
\tr ( \phi _{i_1} \phi _{i_2} \phi _{i_1} \phi_{i_2} ) \\[1mm]
\tr ( \phi _{i_1} \phi _{i_1} \phi _{i_2} \phi_{i_2} ) \\[1mm]
\tr ( \phi _{i_1} \phi _{i_2} ) \, \tr ( \phi _{i_2} \phi _{i_1} ) \\[1mm]
\tr ( \phi_{i_1} \phi _{i_1} ) \, \tr ( \phi _{i_2} \phi_{i_2} ) 
\end{pmatrix} ,
\label{data mmop L=4}
\end{equation}
which is consistent with \cite{BKS03} at $N_f=6$. The spectrum is shown in Figure \ref{fig:finiteNc L=4}. 
The eigenvalue curve is drawn by keeping track of the large $N_c$ eigenvalues down to small $N_c$ as long as they remain real-valued. The complex eigenvalues are not shown, and some of them have null eigenvectors. Whenever the complex eigenvalues show up, the corresponding eigenvectors become null \cite{AB13c}.

There is one large $N_c$ zero mode, so the submixing matrix is one-dimensional:
\begin{equation}
-M_\text{sm} = \( 20 \), \qquad
\cO^\circ = \cC_{ij} \cC_{ij} \,,
\label{L=4 pm-Mat}
\end{equation}
where $\cC_{ij}$ is the symmetric traceless tensor defined in \eqref{def:C-ell}.
The $1/N_c$ corrections to this operator for general $N_f$ are given by
\begin{align}
\gamma = - \frac{(N_f+2)(N_f-1)(N_f-3)}{N_f N_c^2} + \cO(N_c^{-4})
\simeq - 20 N_c^{-2} \qquad (N_f=6).
\label{L=4 neg largeNc} 
\end{align}

\clearpage

\subsubsection{$L=6$}

\newsavebox\myverbS
\begin{lrbox}{\myverbS}\begin{minipage}{1.3\hsize}
\begin{verbatim}
{{6,3,-6,0,0,(3*Nf)/Nc,3/Nc,-6/Nc,0,6/Nc-(3*Nf)/Nc,3/Nc,-6/Nc,0,0,0},
{0,4+Nf,0,-4,2,2/Nc,2/Nc+(2*Nf)/Nc,-6/Nc,0,-4/Nc,3/Nc-Nf/Nc,2/Nc,0,0,0},
{-2,1,6,-2,0,-6/Nc+Nf/Nc,Nc^(-1),4/Nc-(2*Nf)/Nc,2/Nc,-6/Nc,2/Nc+Nf/Nc,4/Nc-Nf/Nc,0,0,0},
{0,-3/2,-2,6+Nf/2,0,(-2*Nf)/Nc,3/(2*Nc),-2/Nc,2/Nc+Nf/(2*Nc),2/Nc,Nc^(-1)+Nf/Nc,-4/Nc,0,0,0},
{0,3/2,0,-3,3+(3*Nf)/2,-6/Nc,15/(2*Nc)+(3*Nf)/(2*Nc),0,-3/Nc,0,0,0,0,0,0},
{2/Nc,2/Nc+Nf/Nc,-4/Nc,(-2*Nf)/Nc,Nf/Nc,3+Nf/2,3/2,-2,0,0,0,0,-2/Nc,Nc^(-1)+Nf/(2*Nc),1/(2*Nc)},
{0,12/Nc,0,-24/Nc,12/Nc,0,3+2*Nf,0,-2,0,0,0,0,-2/Nc,Nc^(-1)+Nf/Nc},
{-4/Nc,-2/Nc,6/Nc-(2*Nf)/Nc,-2/Nc+(2*Nf)/Nc,2/Nc,-2,0,4,1,0,0,0,-(Nf/Nc),Nc^(-1),0},
{0,0,-24/Nc,24/Nc,0,0,-2,0,4+Nf,0,0,0,0,(-2*Nf)/Nc,2/Nc},
{6/Nc-(3*Nf)/Nc,6/Nc,-12/Nc,6/Nc+(3*Nf)/Nc,-6/Nc,0,0,0,0,6,3,-6,0,0,0},
{-4/Nc,-2/Nc-Nf/Nc,-4/Nc,4/Nc,6/Nc+Nf/Nc,0,0,0,0,0,2+Nf,0,0,0,0},
{-6/Nc,(3*Nf)/Nc,12/Nc-(3*Nf)/Nc,-6/Nc,0,0,0,0,0,-6,3,6,0,0,0},
{0,0,0,0,0,-6/Nc+(3*Nf)/Nc,3/Nc,6/Nc-(3*Nf)/Nc,-3/Nc,0,0,0,0,3,0},
{0,0,0,0,0,24/Nc,-2/Nc+(2*Nf)/Nc,-24/Nc,2/Nc-(2*Nf)/Nc,0,0,0,0,Nf,2},
{0,0,0,0,0,0,36/Nc,0,-36/Nc,0,0,0,0,0,3*Nf}}
\end{verbatim}
\end{minipage}\end{lrbox}

There are fifteen $\alg{so}(N_f)$ singlets at length $L=6$. The mixing matrix and operator basis for general $N_f$ are given by

\smallskip \noindent
{\scriptsize $M_{IJ}=$} \ \resizebox{\hsize}{!}{\usebox\myverbS}

\bigskip \noindent
and
{\scriptsize
\begin{align}
\{ \cO_I \} = \Bigl\{ 
&\tr( \phi_{i_1} \phi_{i_2} \phi_{i_3} \phi_{i_1} \phi_{i_2} \phi_{i_3} ), \ 
\tr( \phi_{i_1} \phi_{i_1} \phi_{i_3} \phi_{i_2} \phi_{i_2} \phi_{i_3} ), \ 
\tr( \phi_{i_1} \phi_{i_2} \phi_{i_1} \phi_{i_3} \phi_{i_2} \phi_{i_3} ), \ 
\tr( \phi_{i_1} \phi_{i_1} \phi_{i_2} \phi_{i_3} \phi_{i_2} \phi_{i_3} ), \ 
\notag \\
&\tr( \phi_{i_1} \phi_{i_1} \phi_{i_2} \phi_{i_2} \phi_{i_3} \phi_{i_3} ), \ 
\tr( \phi_{i_1} \phi_{i_2} ) \, \tr( \phi_{i_3} \phi_{i_1} \phi_{i_2} \phi_{i_3} ), \ 
\tr( \phi_{i_3} \phi_{i_3} ) \, \tr( \phi_{i_2} \phi_{i_1} \phi_{i_1} \phi_{i_2} ), \ 
\tr( \phi_{i_1} \phi_{i_3} ) \, \tr( \phi_{i_2} \phi_{i_1} \phi_{i_2} \phi_{i_3} ), \ 
\notag \\
&\tr( \phi_{i_3} \phi_{i_3} ) \, \tr( \phi_{i_1} \phi_{i_2} \phi_{i_1} \phi_{i_2} ), \ 
\tr( \phi_{i_1} \phi_{i_2} \phi_{i_3} ) \, \tr( \phi_{i_3} \phi_{i_1} \phi_{i_2} ), \ 
\tr( \phi_{i_3} \phi_{i_1} \phi_{i_1} ) \, \tr( \phi_{i_2} \phi_{i_2} \phi_{i_3} ), \ 
\tr( \phi_{i_1} \phi_{i_2} \phi_{i_3} ) \, \tr( \phi_{i_3} \phi_{i_2} \phi_{i_1} ), \ 
\notag \\
&\tr( \phi_{i_1} \phi_{i_2} ) \, \tr( \phi_{i_3} \phi_{i_1} ) \, \tr( \phi_{i_2} \phi_{i_3} ), \ 
\tr( \phi_{i_2} \phi_{i_2} ) \, \tr( \phi_{i_1} \phi_{i_3} ) \, \tr( \phi_{i_3} \phi_{i_1} ), \ 
\tr( \phi_{i_1} \phi_{i_1} ) \, \tr( \phi_{i_2} \phi_{i_2} ) \, \tr( \phi_{i_3} \phi_{i_3} )
\Bigr\} .
\end{align}}

\begin{figure}[t]
\begin{center}
\subfigure{\includegraphics[scale=1]{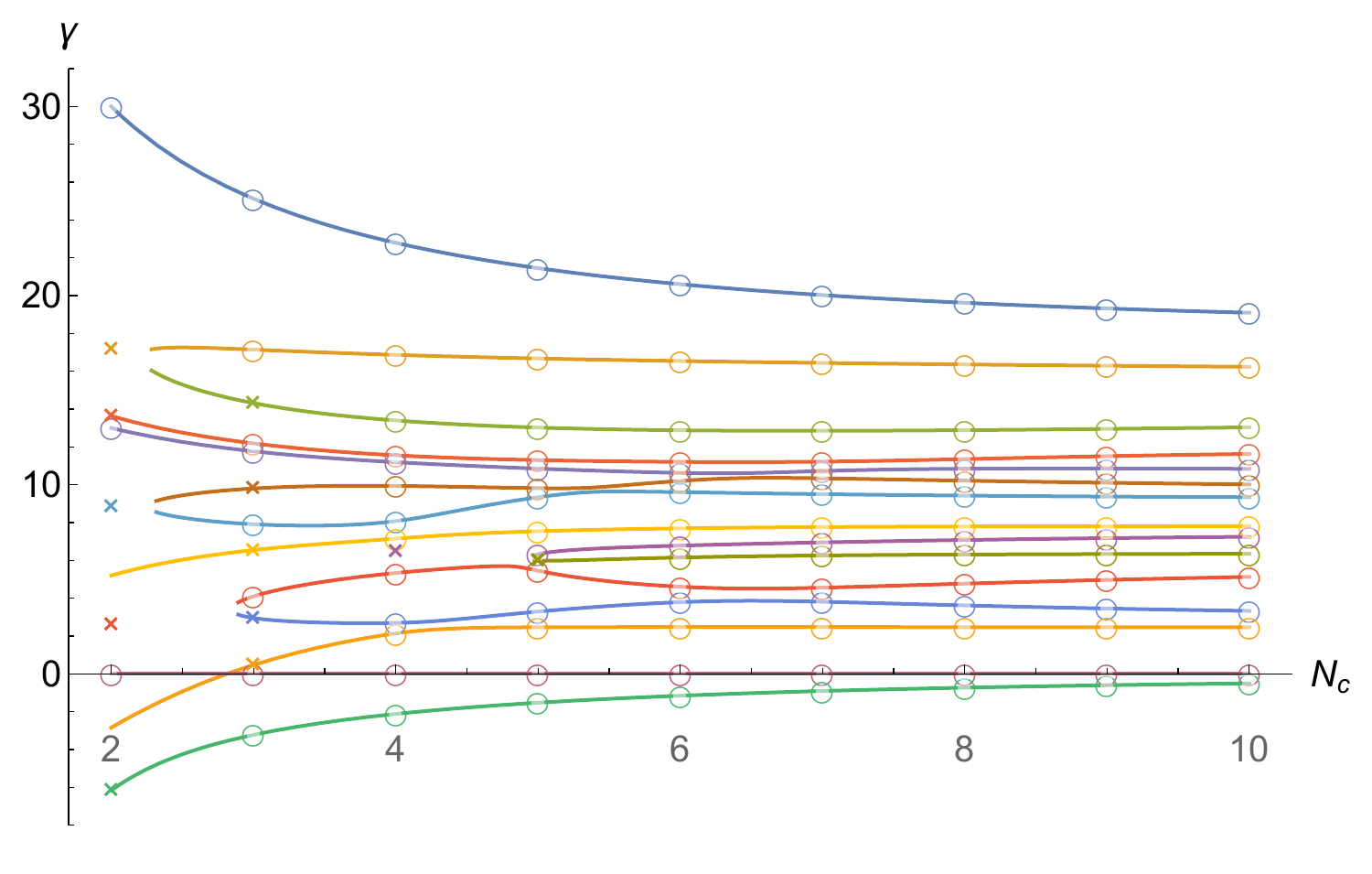}}
\\[5mm]
\subfigure{\includegraphics[scale=0.8]{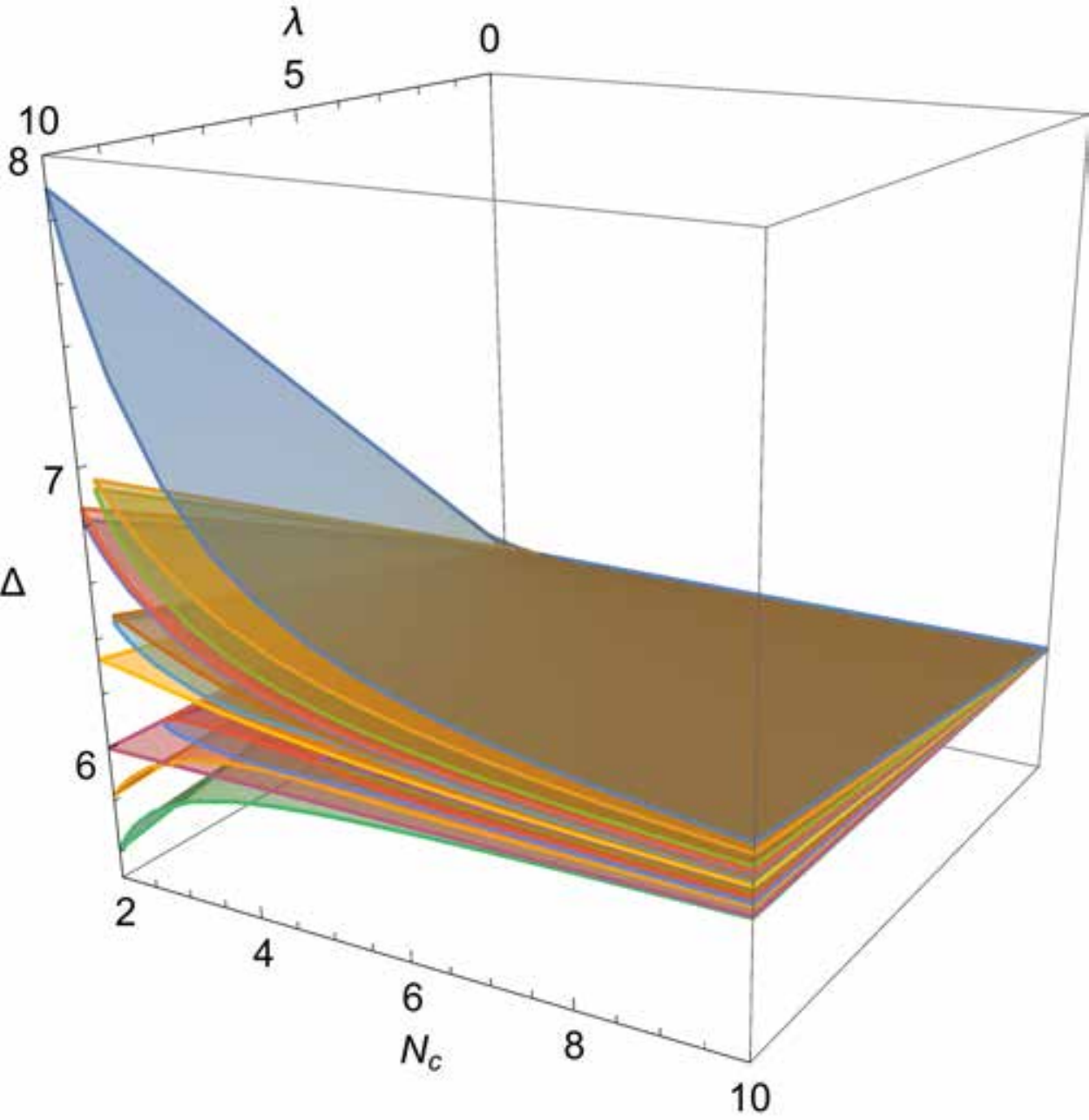}}
\caption{The dimensions of $\alg{so}(6)$ singlets of length $L=6$ at finite $N_c$\,.}
\label{fig:finiteNc L=6}
\end{center}
\end{figure}

\vspace{-3mm} \noindent
The spectrum is shown in Figure \ref{fig:finiteNc L=6}. This mixing matrix is consistent with \cite{BRS03}, which can be shown by computing the characteristic polynomial $\fP (\gamma) = \det (M_{IJ} - \gamma \, \delta_{IJ})$ and substituting $\gamma \to 2 \gamma, N_c \to 1/\nu, N_f \to 6$.

There are two large $N_c$ zero modes. The submixing matrix for general $N_f$ is given by
\begin{equation}
(-M_\text{sm})_{ij} = \begin{pmatrix}
\frac{3 \left(N_f+4\right) \left(3 N_f^4+32 N_f^3-60 N_f^2-132 N_f-8\right)}{4 \left(N_f+2\right) \left(N_f^2+12 N_f+14\right)} & 0 \\
0 & \frac{3 \left(N_f-6\right) \left(N_f-2\right) \left(N_f+4\right)}{4 N_f} 
\end{pmatrix} , \quad
\cO^\circ_j = 
\begin{pmatrix}
\cC_{ijk} \cC_{ijk} \\[3mm]
\cC_{ij} \cC_{jk} \cC_{ki}
\end{pmatrix}.
\label{L=6 pm-Mat}
\end{equation}
The double- and triple-trace zero modes do not submix. 
At $N_f=6$\,, the double-trace mode has $\gamma_2 = - 3675/61$ and the triple-trace mode has precisely zero dimension for any $N_c$ \cite{BRS03}.

\bigskip
We also studied operators other than $\alg{so}(6)$ singlets. No negative modes ($\gamma_2 < 0$) are found in the $\alg{su}(2)$ sector at $L=4$, and the lowest eigenvalue in the $\alg{so}(6)$ sector at $L=4,6$ is same as the lowest eigenvalue of the $\alg{so}(6)$ singlets.

\clearpage

\subsubsection{$L=8$}\label{app:data L8}

\newsavebox\myverbE
\begin{lrbox}{\myverbE}\begin{minipage}{5\hsize}
\begin{verbatim}
{{6+Nf,-4,0,-2,0,0,0,0,0,0,0,0,2,0,0,0,0,1,4/Nc,Nc^(-1),2/Nc+(2*Nf)/Nc,-4/Nc,-4/Nc,-2/Nc,0,0,0,0,0,0,0,Nc^(-1),10/Nc,8/Nc-(2*Nf)/Nc,-8/Nc,-6/Nc,0,-2/Nc,0,5/Nc,Nf/Nc,2/Nc,-4/Nc,0,-4/Nc,0,0,0,0,0,0,0,0,0,0,0,0,0,0,0,0,0,0,0,0,0,0,0,0,0,0},
{-3/2,7+Nf/2,0,0,0,-1,0,0,0,-1,0,0,0,0,0,-2,2,0,-Nc^(-1),2/Nc-(2*Nf)/Nc,2/Nc,Nc^(-1),Nc^(-1),0,2/Nc+Nf/Nc,0,0,-6/Nc,0,0,0,0,-4/Nc,2/Nc+Nf/Nc,6/Nc,0,2/Nc-Nf/Nc,0,-6/Nc,Nc^(-1),-1/(2*Nc),0,4/Nc,Nc^(-1)+Nf/(2*Nc),-2/Nc,0,-2/Nc,-2/Nc,0,0,0,0,0,0,0,0,0,0,0,0,0,0,0,0,0,0,0,0,0,0,0},
{0,0,8,4,0,0,0,0,-8,0,0,0,0,0,0,0,0,0,-8/Nc,4/Nc,0,0,0,8/Nc,4/Nc,-8/Nc,0,0,0,0,0,0,-8/Nc,4/Nc,8/Nc+(4*Nf)/Nc,0,0,0,-8/Nc,0,0,0,0,0,-8/Nc,8/Nc-(4*Nf)/Nc,4/Nc,0,0,0,0,0,0,0,0,0,0,0,0,0,0,0,0,0,0,0,0,0,0,0,0},
{-2,0,0,6+Nf,0,-2,2,0,0,-2,0,0,0,0,0,0,0,1,4/Nc,Nc^(-1),Nc^(-1),0,0,-2/Nc,4/Nc+Nf/Nc,0,0,0,-8/Nc,0,0,-Nc^(-1),-8/Nc,-2/Nc,10/Nc,0,0,4/Nc+(2*Nf)/Nc,-6/Nc,0,Nc^(-1),0,-4/Nc,0,2/Nc,-4/Nc,7/Nc-(2*Nf)/Nc,0,0,0,0,0,0,0,0,0,0,0,0,0,0,0,0,0,0,0,0,0,0,0,0},
{0,0,0,0,8,-1,2,0,0,-1,-4,0,0,0,0,0,0,0,2/Nc,-4/Nc,0,-Nc^(-1),-Nc^(-1),0,2/Nc,2/Nc-(2*Nf)/Nc,2/Nc,0,0,0,0,0,-2/Nc+Nf/Nc,Nc^(-1)+(2*Nf)/Nc,0,4/Nc,2/Nc,0,-8/Nc,2/Nc,0,4/Nc-Nf/Nc,-2/Nc,0,-4/Nc,-2/Nc,0,4/Nc-Nf/Nc,0,0,0,0,0,0,0,0,0,0,0,0,0,0,0,0,0,0,0,0,0,0,0},
{0,-1,0,-1/2,-1,7+Nf/2,-1/2,0,0,1/2,0,-2,1,-1,1/2,0,1/2,0,-Nc^(-1)+Nf/(2*Nc),1/(2*Nc),1/(2*Nc),1/(2*Nc)-Nf/Nc,3/(2*Nc),-Nc^(-1),Nc^(-1),0,2/Nc+Nf/(2*Nc),-Nc^(-1),-3/(2*Nc),-2/Nc,0,1/(2*Nc),6/Nc-Nf/Nc,1/(2*Nc)+Nf/(2*Nc),-6/Nc,-2/Nc,Nc^(-1),2/Nc+Nf/Nc,-2/Nc,Nc^(-1)+Nf/(2*Nc),0,-2/Nc,3/Nc-Nf/Nc,1/(2*Nc),0,0,-Nc^(-1),-Nc^(-1),0,0,0,0,0,0,0,0,0,0,0,0,0,0,0,0,0,0,0,0,0,0,0},
{0,0,0,1,0,-1,6+Nf,0,0,-1,0,0,-1,0,-2,0,1,0,-3/Nc,0,-Nc^(-1),-1/(2*Nc),-1/(2*Nc),0,4/Nc+Nf/Nc,-3/Nc,0,0,3/Nc-Nf/Nc,0,0,Nc^(-1),-8/Nc,9/(2*Nc),4/Nc,2/Nc,-3/Nc,3/Nc+(3*Nf)/(2*Nc),-4/Nc,5/Nc-Nf/Nc,Nc^(-1),0,3/Nc,0,0,0,-5/Nc,-3/Nc,0,0,0,0,0,0,0,0,0,0,0,0,0,0,0,0,0,0,0,0,0,0,0},
{2,0,0,0,0,0,0,8,-4,0,0,2,0,-4,0,0,0,0,0,(2*Nf)/Nc,2/Nc,0,0,0,0,-4/Nc,0,-8/Nc,2/Nc,4/Nc,2/Nc,0,-4/Nc,0,-4/Nc,16/Nc-(4*Nf)/Nc,4/Nc,0,-8/Nc,2/Nc+(2*Nf)/Nc,0,4/Nc,0,0,-8/Nc,0,0,0,0,0,0,0,0,0,0,0,0,0,0,0,0,0,0,0,0,0,0,0,0,0,0},
{1,0,-2,1,0,1,0,-2,8,1,-4,0,0,0,0,0,0,0,-2/Nc,Nc^(-1)+Nf/Nc,Nc^(-1),-3/Nc,-3/Nc,2/Nc,Nc^(-1),0,2/Nc,4/Nc,0,-4/Nc,0,0,4/Nc+(2*Nf)/Nc,0,-6/Nc,-6/Nc,2/Nc,2/Nc,4/Nc-(2*Nf)/Nc,Nc^(-1),0,-2/Nc,2/Nc,0,2/Nc-(2*Nf)/Nc,-2/Nc,Nf/Nc,0,0,0,0,0,0,0,0,0,0,0,0,0,0,0,0,0,0,0,0,0,0,0,0},
{0,-1,0,-1/2,-1,1/2,-1/2,0,0,7+Nf/2,0,-2,1,-1,1/2,0,1/2,0,-Nc^(-1)+Nf/(2*Nc),1/(2*Nc),1/(2*Nc),3/(2*Nc),1/(2*Nc)-Nf/Nc,-Nc^(-1),Nc^(-1),0,2/Nc+Nf/(2*Nc),-Nc^(-1),-3/(2*Nc),-2/Nc,0,1/(2*Nc),6/Nc-Nf/Nc,1/(2*Nc)+Nf/(2*Nc),-6/Nc,-2/Nc,Nc^(-1),2/Nc+Nf/Nc,-2/Nc,Nc^(-1)+Nf/(2*Nc),0,-2/Nc,3/Nc-Nf/Nc,1/(2*Nc),0,0,-Nc^(-1),-Nc^(-1),0,0,0,0,0,0,0,0,0,0,0,0,0,0,0,0,0,0,0,0,0,0,0},
{0,1,0,0,-2,1/2,1,0,-2,1/2,8,-1,0,-2,0,0,0,0,Nc^(-1),-3/Nc,0,-3/(2*Nc)+Nf/(2*Nc),-3/(2*Nc)+Nf/(2*Nc),2/Nc,2/Nc,-Nc^(-1),Nc^(-1),-2/Nc,0,2/Nc-Nf/Nc,Nc^(-1),0,-2/Nc,5/(2*Nc),Nf/Nc,-2/Nc,2/Nc+Nf/(2*Nc),0,(-2*Nf)/Nc,-2/Nc,0,0,3/Nc+Nf/Nc,0,-(Nf/Nc),-2/Nc,2/Nc,-Nc^(-1),0,0,0,0,0,0,0,0,0,0,0,0,0,0,0,0,0,0,0,0,0,0,0},
{1/2,0,0,0,0,-2,0,0,0,-2,-2,8+Nf/2,2,0,-1,0,0,0,Nc^(-1)+Nf/Nc,3/Nc,2/Nc,-2/Nc,-2/Nc,-2/Nc,0,0,-Nc^(-1),-4/Nc,Nf/(2*Nc),0,2/Nc+Nf/(2*Nc),Nc^(-1),-6/Nc,Nc^(-1),6/Nc-(3*Nf)/Nc,-4/Nc,4/Nc,0,2/Nc,2/Nc+Nf/(2*Nc),1/(2*Nc),0,-4/Nc,0,-4/Nc,2/Nc,2/Nc+Nf/Nc,0,0,0,0,0,0,0,0,0,0,0,0,0,0,0,0,0,0,0,0,0,0,0,0},
{1/2,-1,0,0,0,0,-2,0,0,0,0,0,6+(3*Nf)/2,0,0,0,-2,1,Nc^(-1),-2/Nc,4/Nc+Nf/Nc,-3/Nc,-3/Nc,0,-2/Nc,0,0,0,Nc^(-1),0,0,2/Nc+Nf/Nc,4/Nc,Nc^(-1),-8/Nc,0,-2/Nc,6/Nc-Nf/Nc,0,-Nc^(-1),5/(2*Nc)+Nf/(2*Nc),0,-6/Nc,-Nc^(-1),0,0,5/Nc,0,0,0,0,0,0,0,0,0,0,0,0,0,0,0,0,0,0,0,0,0,0,0,0},
{0,1,0,0,0,-1/2,0,-1,0,-1/2,-2,1/2,0,8,-1/2,-1,0,0,0,-Nc^(-1),0,-1/(2*Nc)+Nf/(2*Nc),-1/(2*Nc)+Nf/(2*Nc),-4/Nc,Nc^(-1),0,5/(2*Nc),4/Nc-(2*Nf)/Nc,Nc^(-1),-2/Nc,1/(2*Nc),0,-Nc^(-1),Nc^(-1),-Nc^(-1),2/Nc-(2*Nf)/Nc,3/Nc+Nf/Nc,Nc^(-1),-4/Nc,3/(2*Nc),0,-Nc^(-1),-Nc^(-1)+Nf/Nc,0,0,-Nc^(-1),-3/(2*Nc),2/Nc,0,0,0,0,0,0,0,0,0,0,0,0,0,0,0,0,0,0,0,0,0,0,0},
{1/2,0,0,0,0,1/2,-1,0,0,1/2,0,-1,0,-2,7+Nf/2,0,-1,0,-Nc^(-1),-2/Nc,Nc^(-1),3/Nc-Nf/Nc,3/Nc-Nf/Nc,-4/Nc,2/Nc,-2/Nc,2/Nc+Nf/(2*Nc),0,Nf/(2*Nc),0,-Nc^(-1),0,4/Nc-Nf/Nc,Nf/Nc,-6/Nc,-8/Nc,2/Nc,4/Nc,4/Nc,-5/Nc,1/(2*Nc),0,2/Nc,0,0,0,2/Nc+Nf/(2*Nc),0,0,0,0,0,0,0,0,0,0,0,0,0,0,0,0,0,0,0,0,0,0,0,0},
{0,-3,0,0,0,0,0,0,0,0,0,0,0,-2,1,8,0,0,0,0,0,Nc^(-1),Nc^(-1),(-4*Nf)/Nc,4/Nc,2/Nc,Nc^(-1),-4/Nc,Nc^(-1),-2/Nc,0,0,0,2/Nc,0,-8/Nc,(2*Nf)/Nc,0,4/Nc,0,0,-2/Nc,-2/Nc,-Nc^(-1),-2/Nc,2/Nc,Nc^(-1),2/Nc,2/Nc,0,0,0,0,0,0,0,0,0,0,0,0,0,0,0,0,0,0,0,0,0,0},
{0,1/2,0,0,0,0,1,0,0,0,0,0,-3/2,0,-2,-1,6+Nf,0,3/Nc-(2*Nf)/Nc,-4/Nc,0,-Nc^(-1),-Nc^(-1),-2/Nc,5/Nc+Nf/Nc,0,-2/Nc,0,Nc^(-1),0,0,2/Nc,-8/Nc,3/Nc,4/Nc,0,-4/Nc,4/Nc+Nf/Nc,0,-3/Nc,-1/(2*Nc),0,-2/Nc,7/(2*Nc),0,0,3/Nc,0,-Nc^(-1),0,0,0,0,0,0,0,0,0,0,0,0,0,0,0,0,0,0,0,0,0,0},
{0,0,0,0,0,0,0,0,0,0,0,0,2,0,0,0,-4,4+2*Nf,-8/Nc,0,2/Nc,0,0,0,-4/Nc,0,0,0,0,0,0,8/Nc+(2*Nf)/Nc,0,-12/Nc,0,0,0,12/Nc,0,4/Nc,8/Nc,0,0,-4/Nc,0,0,-8/Nc,0,0,0,0,0,0,0,0,0,0,0,0,0,0,0,0,0,0,0,0,0,0,0,0},
{3/Nc,-6/Nc,0,3/Nc,0,-Nc^(-1),-4/Nc,0,0,-Nc^(-1),0,4/Nc,4/Nc+Nf/Nc,0,-6/Nc,0,5/Nc-(2*Nf)/Nc,-Nc^(-1)+Nf/Nc,3+Nf,1/2,0,-1,-1,-1,0,0,0,0,1,0,0,3/2,0,0,0,0,0,0,0,0,0,0,0,0,0,0,0,0,0,5/Nc+Nf/Nc,5/(2*Nc),-2/Nc,-4/Nc,-2/Nc,-Nc^(-1),0,0,1/(2*Nc),0,2/Nc,-2/Nc,0,0,0,0,0,0,0,0,0,0},
{Nc^(-1)+Nf/Nc,4/Nc-(2*Nf)/Nc,0,-4/Nc,0,1/(2*Nc),Nc^(-1),2/Nc,2/Nc,1/(2*Nc),-6/Nc,6/Nc,Nf/Nc,-6/Nc,-Nc^(-1),0,-Nc^(-1),Nc^(-1),2,4+Nf/2,3/2,-1,-1,-2,0,0,0,0,0,0,0,0,0,0,0,0,0,0,0,0,0,0,0,0,0,0,0,0,0,2/Nc+Nf/Nc,Nf/(2*Nc),0,2/Nc,0,0,-4/Nc,-2/Nc,1/(2*Nc),0,Nc^(-1),2/Nc-Nf/Nc,0,2/Nc,0,-4/Nc,0,0,0,0,0,0},
{12/Nc,-24/Nc,0,0,0,0,-12/Nc,0,0,0,0,0,24/Nc,0,0,0,0,0,0,0,4+2*Nf,0,0,0,-4,0,0,0,0,0,0,2,0,0,0,0,0,0,0,0,0,0,0,0,0,0,0,0,0,2/Nc,0,0,0,-6/Nc,0,0,0,2/Nc+(2*Nf)/Nc,0,0,0,2/Nc,0,-4/Nc,0,3/Nc-Nf/Nc,0,0,0,0,0},
{-3/(2*Nc),3/(2*Nc),0,1/(2*Nc)+Nf/(2*Nc),-3/Nc,4/Nc-Nf/Nc,1/(2*Nc)+Nf/Nc,0,0,5/(2*Nc),2/Nc,-4/Nc,1/(2*Nc),-Nc^(-1),1/(2*Nc)-Nf/Nc,-3/Nc,-1/(2*Nc)+Nf/(2*Nc),Nc^(-1),0,-1,0,5+Nf/2,1,0,1,-1,0,-1,-1/2,0,0,0,0,0,0,0,0,0,0,0,0,0,0,0,0,0,0,0,0,0,1/(2*Nc),-2/Nc,2/Nc-Nf/Nc,2/Nc+Nf/(2*Nc),0,0,-2/Nc,0,0,Nc^(-1)+Nf/(2*Nc),0,0,-4/Nc,0,2/Nc,1/(2*Nc),0,0,0,0,0},
{-3/(2*Nc),3/(2*Nc),0,1/(2*Nc)+Nf/(2*Nc),-3/Nc,5/(2*Nc),1/(2*Nc)+Nf/Nc,0,0,4/Nc-Nf/Nc,2/Nc,-4/Nc,1/(2*Nc),-Nc^(-1),1/(2*Nc)-Nf/Nc,-3/Nc,-1/(2*Nc)+Nf/(2*Nc),Nc^(-1),0,-1,0,1,5+Nf/2,0,1,-1,0,-1,-1/2,0,0,0,0,0,0,0,0,0,0,0,0,0,0,0,0,0,0,0,0,0,1/(2*Nc),-2/Nc,2/Nc-Nf/Nc,2/Nc+Nf/(2*Nc),0,0,-2/Nc,0,0,Nc^(-1)+Nf/(2*Nc),0,0,-4/Nc,0,2/Nc,1/(2*Nc),0,0,0,0,0},
{0,2/Nc+Nf/Nc,2/Nc,Nc^(-1),2/Nc,-3/(2*Nc),Nc^(-1),0,-2/Nc,-3/(2*Nc),0,0,0,-6/Nc,Nc^(-1),2/Nc-(2*Nf)/Nc,Nf/Nc,0,0,-3/2,0,1/2,1/2,5,3/2,0,0,-2,0,0,0,0,0,0,0,0,0,0,0,0,0,0,0,0,0,0,0,0,0,2/Nc,-1/(2*Nc),(-2*Nf)/Nc,0,Nc^(-1),Nc^(-1),-2/Nc,0,0,1/(2*Nc),Nc^(-1),Nf/Nc,0,-4/Nc,0,2/Nc,0,0,0,0,0,0},
{0,6/Nc,0,6/Nc,0,-6/Nc,12/Nc,0,0,-6/Nc,0,0,0,0,-12/Nc,-12/Nc,12/Nc,0,0,0,-3/2,0,0,0,6+(3*Nf)/2,0,-2,0,0,0,0,0,0,0,0,0,0,0,0,0,0,0,0,0,0,0,0,0,0,(-2*Nf)/Nc,0,0,0,-2/Nc,0,0,0,3/(2*Nc),2/Nc+Nf/(2*Nc),0,0,-4/Nc,0,2/Nc,0,Nc^(-1)+Nf/Nc,0,0,0,0,0},
{0,-6/Nc,0,0,6/Nc-(2*Nf)/Nc,Nc^(-1)+Nf/Nc,2/Nc,0,0,Nc^(-1)+Nf/Nc,-8/Nc,0,0,4/Nc,-4/Nc,0,4/Nc,0,0,0,0,-1,-1,0,0,6,1,0,1,-2,0,0,0,0,0,0,0,0,0,0,0,0,0,0,0,0,0,0,0,2/Nc,0,-4/Nc,-2/Nc,2/Nc,0,0,4/Nc-(2*Nf)/Nc,0,0,Nf/Nc,2/Nc,-Nc^(-1),4/Nc,0,-6/Nc,0,0,0,0,0,0},
{0,0,0,0,-12/Nc,12/Nc,0,0,0,12/Nc,0,0,0,-24/Nc,12/Nc,0,0,0,0,0,1,0,0,0,-2,0,6+Nf,0,0,0,-2,0,0,0,0,0,0,0,0,0,0,0,0,0,0,0,0,0,0,-6/Nc+Nf/Nc,0,0,0,4/Nc-(2*Nf)/Nc,0,0,0,Nc^(-1),2/Nc,0,0,4/Nc-Nf/Nc,0,-6/Nc,0,2/Nc+Nf/Nc,0,0,0,0,0},
{Nc^(-1),-Nc^(-1),-2/Nc,Nc^(-1),-2/Nc,-1/(2*Nc)+Nf/(2*Nc),-Nc^(-1),-4/Nc,4/Nc,-1/(2*Nc)+Nf/(2*Nc),0,-2/Nc,Nc^(-1),6/Nc-(2*Nf)/Nc,Nc^(-1)+Nf/Nc,-2/Nc,Nc^(-1),0,0,1,0,0,0,-2,0,0,1,6,0,-2,0,0,0,0,0,0,0,0,0,0,0,0,0,0,0,0,0,0,0,Nc^(-1),0,-4/Nc,-2/Nc+Nf/Nc,Nc^(-1),0,2/Nc-Nf/Nc,2/Nc,0,0,Nc^(-1),Nc^(-1)+Nf/(2*Nc),0,4/Nc-Nf/Nc,0,-6/Nc,1/(2*Nc),0,0,0,0,0},
{0,0,0,2/Nc,0,-6/Nc,8/Nc-(2*Nf)/Nc,0,0,-6/Nc,0,4/Nc,4/Nc+(2*Nf)/Nc,0,4/Nc,0,-12/Nc,2/Nc,2,0,1,-2,-2,0,0,0,0,0,4+Nf,0,0,0,0,0,0,0,0,0,0,0,0,0,0,0,0,0,0,0,0,2/Nc+Nf/Nc,0,0,2/Nc,0,0,0,-6/Nc,Nc^(-1),0,3/Nc,0,0,2/Nc,0,-4/Nc,-Nc^(-1),0,0,0,0,0},
{2/Nc,-4/Nc,0,0,0,-3/Nc,0,4/Nc,-8/Nc,-3/Nc,6/Nc-(2*Nf)/Nc,2/Nc+(2*Nf)/Nc,2/Nc,0,2/Nc,0,0,0,0,2,0,0,0,0,0,-2,0,-4,1,6,1,0,0,0,0,0,0,0,0,0,0,0,0,0,0,0,0,0,0,4/Nc,0,0,(2*Nf)/Nc,0,0,-4/Nc,-2/Nc,0,0,Nc^(-1),2/Nc,0,-6/Nc,-Nc^(-1),6/Nc-(2*Nf)/Nc,0,0,0,0,0,0},
{0,0,0,0,0,0,0,0,0,0,-36/Nc,36/Nc,0,0,0,0,0,0,0,0,3,0,0,0,0,0,-6,0,0,0,6+Nf,0,0,0,0,0,0,0,0,0,0,0,0,0,0,0,0,0,0,(3*Nf)/Nc,0,0,0,-6/Nc,0,0,0,3/Nc,0,0,0,-6/Nc,0,6/Nc-(3*Nf)/Nc,0,3/Nc,0,0,0,0,0},
{0,0,0,0,0,0,0,0,0,0,0,0,18/Nc,0,0,0,-36/Nc,18/Nc,0,0,3/2,0,0,0,-3,0,0,0,0,0,0,3+(5*Nf)/2,0,0,0,0,0,0,0,0,0,0,0,0,0,0,0,0,0,-6/Nc,0,0,0,0,0,0,0,15/(2*Nc)+(3*Nf)/(2*Nc),-3/Nc,0,0,0,0,0,0,0,0,0,0,0,0},
{3/Nc+Nf/Nc,-4/Nc,0,0,0,11/(2*Nc)-Nf/Nc,-4/Nc,-4/Nc,4/Nc,11/(2*Nc)-Nf/Nc,-4/Nc,-6/Nc,Nc^(-1)+(2*Nf)/Nc,-2/Nc,5/Nc-Nf/Nc,0,0,0,0,0,0,0,0,0,0,0,0,0,0,0,0,0,8+Nf/2,1/2,-5,-2,0,2,0,0,0,0,0,0,0,0,0,0,0,0,0,0,0,0,0,0,0,0,0,-Nc^(-1),2/Nc,2/Nc+Nf/(2*Nc),0,-Nc^(-1),-4/Nc,3/(2*Nc),0,0,0,0,0},
{Nc^(-1)-Nf/Nc,-Nc^(-1),0,-Nc^(-1),2/Nc,-1/(2*Nc),3/Nc,0,0,-1/(2*Nc),-4/Nc,-4/Nc,Nc^(-1)+Nf/Nc,-6/Nc,3/Nc,-2/Nc,7/Nc,2/Nc,0,0,0,0,0,0,0,0,0,0,0,0,0,0,0,5+Nf,0,0,-2,0,0,0,0,0,0,0,0,0,0,0,0,0,0,0,0,0,0,0,0,0,0,2/Nc-Nf/Nc,-4/Nc,0,0,0,0,2/Nc+Nf/Nc,0,0,0,0,0},
{2/Nc,5/Nc,2/Nc,3/Nc+Nf/Nc,0,-5/Nc,3/Nc+Nf/Nc,0,-4/Nc,-5/Nc,2/Nc,6/Nc-(3*Nf)/Nc,0,-4/Nc,-4/Nc,-2/Nc,3/Nc+Nf/Nc,-2/Nc,0,0,0,0,0,0,0,0,0,0,0,0,0,0,-5,1/2,8+Nf/2,0,0,2,-2,0,0,0,0,0,0,0,0,0,0,0,0,0,0,0,0,0,0,0,0,-Nc^(-1),2/Nc,-Nc^(-1),-4/Nc,2/Nc+Nf/(2*Nc),0,3/(2*Nc),0,0,0,0,0},
{-Nc^(-1),3/Nc+Nf/Nc,0,-2/Nc,2/Nc,-Nc^(-1)+Nf/(2*Nc),3/Nc+Nf/Nc,4/Nc-Nf/Nc,-4/Nc,-Nc^(-1)+Nf/(2*Nc),-4/Nc,-2/Nc,Nc^(-1),4/Nc-(2*Nf)/Nc,0,-2/Nc,0,0,0,0,0,0,0,0,0,0,0,0,0,0,0,0,-1,1,1,6,1,0,-4,0,0,0,0,0,0,0,0,0,0,0,0,0,0,0,0,0,0,0,0,Nc^(-1),-Nc^(-1)+Nf/(2*Nc),Nc^(-1),-2/Nc-Nf/Nc,Nc^(-1),0,1/(2*Nc),0,0,0,0,0},
{-2/Nc,-Nc^(-1)-Nf/Nc,0,0,-2/Nc,3/(2*Nc),-Nc^(-1),-4/Nc,-4/Nc,3/(2*Nc),-4/Nc,4/Nc,2/Nc,-2/Nc,3/Nc,2/Nc,6/Nc+Nf/Nc,0,0,0,0,0,0,0,0,0,0,0,0,0,0,0,0,-1/2,0,0,4+Nf/2,0,0,0,0,0,0,0,0,0,0,0,0,0,0,0,0,0,0,0,0,0,0,Nf/Nc,-2/Nc-(2*Nf)/Nc,0,0,0,0,3/Nc,0,0,0,0,0},
{-3/Nc,-2/Nc,0,5/Nc,0,-Nc^(-1),6/Nc,0,0,-Nc^(-1),0,-8/Nc,5/Nc-Nf/Nc,0,-6/Nc,0,-Nc^(-1),6/Nc+Nf/Nc,0,0,0,0,0,0,0,0,0,0,0,0,0,0,0,-3/2,0,0,-1,5+(3*Nf)/2,0,0,0,0,0,0,0,0,0,0,0,0,0,0,0,0,0,0,0,0,0,-2/Nc,-2/Nc,0,0,0,0,3/Nc+Nf/Nc,0,0,0,0,0},
{0,0,-2/Nc,Nc^(-1),-2/Nc,1/(2*Nc)+Nf/(2*Nc),2/Nc,-2/Nc,2/Nc-Nf/Nc,1/(2*Nc)+Nf/(2*Nc),2/Nc-(2*Nf)/Nc,-2/Nc+Nf/Nc,2/Nc,-2/Nc,-Nc^(-1)+Nf/Nc,4/Nc,-3/Nc,0,0,0,0,0,0,0,0,0,0,0,0,0,0,0,1,1,-1,-4,1,0,6,0,0,0,0,0,0,0,0,0,0,0,0,0,0,0,0,0,0,0,0,Nc^(-1),-Nc^(-1)+Nf/(2*Nc),Nc^(-1),0,Nc^(-1),-2/Nc-Nf/Nc,1/(2*Nc),0,0,0,0,0},
{6/Nc,-6/Nc,0,-10/Nc,-6/Nc,Nc^(-1),10/Nc-(2*Nf)/Nc,2/Nc,-4/Nc,Nc^(-1),0,4/Nc,4/Nc+Nf/Nc,0,-4/Nc,0,2/Nc,Nf/Nc,0,0,0,0,0,0,0,0,0,0,0,0,0,0,0,0,0,0,0,0,0,6+Nf,1,0,-4,0,0,0,0,0,0,2/Nc+Nf/Nc,0,0,-4/Nc,0,0,0,0,Nc^(-1),0,0,0,0,0,0,0,0,0,0,0,0,0},
{8/Nc,0,0,-16/Nc,0,0,-24/Nc,0,0,0,0,0,28/Nc,0,0,0,-8/Nc,12/Nc,0,0,0,0,0,0,0,0,0,0,0,0,0,0,0,0,0,0,0,0,0,0,6+2*Nf,0,0,-4,0,0,0,0,0,0,-4/Nc,0,0,0,0,0,0,2/Nc+(2*Nf)/Nc,0,0,0,0,0,0,0,0,0,0,0,0,0},
{(4*Nf)/Nc,0,-4/Nc,0,16/Nc-(4*Nf)/Nc,-4/Nc,0,8/Nc,-8/Nc,-4/Nc,0,4/Nc,0,-8/Nc,0,0,0,0,0,0,0,0,0,0,0,0,0,0,0,0,0,0,0,0,0,0,0,0,0,4,0,8,0,0,-8,0,0,0,0,4/Nc,0,0,0,0,0,0,-4/Nc,0,0,0,0,0,0,0,0,0,0,0,0,0,0},
{-2/Nc,3/Nc,0,-Nc^(-1),-4/Nc,Nc^(-1)-Nf/Nc,2/Nc+Nf/Nc,0,-4/Nc,Nc^(-1)-Nf/Nc,0,-4/Nc,Nc^(-1),0,6/Nc,-2/Nc,2/Nc+Nf/Nc,Nc^(-1),0,0,0,0,0,0,0,0,0,0,0,0,0,0,0,0,0,0,0,0,0,-1,0,0,7+Nf/2,1/2,0,0,-1,-2,0,2/Nc,-Nc^(-1),0,-(Nf/Nc),Nc^(-1)+Nf/(2*Nc),0,0,-2/Nc,0,1/(2*Nc),0,0,0,0,0,0,0,0,0,0,0,0},
{0,16/Nc,0,0,0,-16/Nc,8/Nc,0,0,-16/Nc,0,0,0,0,-16/Nc,0,24/Nc,0,0,0,0,0,0,0,0,0,0,0,0,0,0,0,0,0,0,0,0,0,0,0,-2,0,0,7+Nf,0,0,0,0,-2,0,(-2*Nf)/Nc,0,0,0,-2/Nc,0,0,2/Nc,Nc^(-1)+Nf/Nc,0,0,0,0,0,0,0,0,0,0,0,0},
{2/Nc,4/Nc+Nf/Nc,-2/Nc,2/Nc+Nf/Nc,-2/Nc,4/Nc+Nf/Nc,-4/Nc,-2/Nc,2/Nc-(2*Nf)/Nc,4/Nc+Nf/Nc,(-2*Nf)/Nc,-2/Nc,-2/Nc,0,-4/Nc,0,0,0,0,0,0,0,0,0,0,0,0,0,0,0,0,0,0,0,0,0,0,0,0,1,0,-2,2,0,4,-2,1,0,0,2/Nc,0,0,-4/Nc,2/Nc,0,0,0,0,0,0,0,0,0,0,0,0,0,0,0,0,0},
{4/Nc,4/Nc,8/Nc-(4*Nf)/Nc,4/Nc,-4/Nc,0,0,0,-8/Nc,0,-8/Nc,8/Nc+(4*Nf)/Nc,0,0,-8/Nc,8/Nc,-8/Nc,0,0,0,0,0,0,0,0,0,0,0,0,0,0,0,0,0,0,0,0,0,0,0,0,0,0,0,-8,8,4,0,0,4/Nc,0,0,0,0,0,0,-4/Nc,0,0,0,0,0,0,0,0,0,0,0,0,0,0},
{-2/Nc,Nc^(-1),-4/Nc,7/Nc-(2*Nf)/Nc,0,-6/Nc,-8/Nc,0,2/Nc,-6/Nc,-6/Nc,10/Nc,7/Nc+(2*Nf)/Nc,0,4/Nc,-2/Nc,2/Nc,Nc^(-1),0,0,0,0,0,0,0,0,0,0,0,0,0,0,0,0,0,0,0,0,0,0,1,0,-4,0,0,0,6+Nf,0,0,2/Nc+Nf/Nc,0,0,-4/Nc,0,0,0,0,Nc^(-1),0,0,0,0,0,0,0,0,0,0,0,0,0},
{2/Nc,-8/Nc,-2/Nc,0,4/Nc-(2*Nf)/Nc,-4/Nc,4/Nc,-4/Nc,0,-4/Nc,-8/Nc,-2/Nc,0,8/Nc,6/Nc+(2*Nf)/Nc,4/Nc,4/Nc,0,0,0,0,0,0,0,0,0,0,0,0,0,0,0,0,0,0,0,0,0,0,0,0,0,-4,0,0,0,0,8,0,0,0,0,0,4/Nc,-2/Nc,0,(-2*Nf)/Nc,0,0,0,0,0,0,0,0,0,0,0,0,0,0},
{0,0,0,0,-16/Nc,0,0,0,0,0,0,0,0,-32/Nc,16/Nc,32/Nc,0,0,0,0,0,0,0,0,0,0,0,0,0,0,0,0,0,0,0,0,0,0,0,0,0,0,0,-4,0,0,0,0,8,0,0,0,0,0,(-4*Nf)/Nc,0,0,0,4/Nc,0,0,0,0,0,0,0,0,0,0,0,0},
{0,0,0,0,0,0,0,0,0,0,0,0,0,0,0,0,0,0,12/Nc,6/Nc,2/Nc+Nf/Nc,-6/Nc,-6/Nc,-12/Nc,(-2*Nf)/Nc,0,-4/Nc,0,6/Nc,0,2/Nc,Nf/Nc,0,0,0,0,0,0,0,6/Nc,0,0,-12/Nc,0,0,0,6/Nc,0,0,3+(3*Nf)/2,0,0,0,-2,0,0,0,3/2,0,0,0,0,0,0,0,0,Nc^(-1)+Nf/(2*Nc),0,-2/Nc,0,1/(2*Nc)},
{0,0,0,0,0,0,0,0,0,0,0,0,0,0,0,0,0,0,24/Nc,16/Nc,0,-24/Nc,-24/Nc,0,0,0,0,0,8/Nc,0,0,0,0,0,0,0,0,0,0,0,-2/Nc+(2*Nf)/Nc,0,0,2/Nc-(2*Nf)/Nc,0,0,0,0,0,0,3+Nf,0,0,0,-2,0,0,2,0,0,0,0,0,0,0,0,Nc^(-1)+Nf/Nc,-2/Nc,0,0,0},
{0,0,0,0,0,0,0,0,0,0,0,0,0,0,0,0,0,0,(2*Nf)/Nc,4/Nc+(2*Nf)/Nc,0,-2/Nc,-2/Nc,4/Nc-(4*Nf)/Nc,0,4/Nc,0,-12/Nc,0,4/Nc,0,0,0,0,0,0,0,0,0,Nc^(-1),0,2/Nc,0,0,-4/Nc,2/Nc,Nc^(-1),-2/Nc,0,3,0,2,1,0,0,-2,0,0,0,0,0,0,0,0,0,0,Nc^(-1),0,Nc^(-1),-2/Nc,0},
{0,0,0,0,0,0,0,0,0,0,0,0,0,0,0,0,0,0,Nf/Nc,Nc^(-1),Nc^(-1),3/Nc-Nf/Nc,3/Nc-Nf/Nc,-6/Nc,-2/Nc,-6/Nc,0,-2/Nc,3/Nc+Nf/Nc,4/Nc,0,Nc^(-1),0,0,0,0,0,0,0,-Nc^(-1)+Nf/(2*Nc),Nc^(-1),0,2/Nc-Nf/Nc,-Nc^(-1),0,0,-Nc^(-1)+Nf/(2*Nc),0,0,2,1/2,0,3+Nf/2,0,0,0,-2,0,0,0,0,0,0,0,0,0,1/(2*Nc),0,Nc^(-1)+Nf/(2*Nc),-2/Nc,0},
{0,0,0,0,0,0,0,0,0,0,0,0,0,0,0,0,0,0,0,0,-2/Nc,12/Nc,12/Nc,0,-2/Nc+(2*Nf)/Nc,-12/Nc,6/Nc-(2*Nf)/Nc,-12/Nc,0,0,-4/Nc,2/Nc,0,0,0,0,0,0,0,0,0,0,12/Nc,0,0,0,0,-12/Nc,0,-2,0,0,0,4+Nf,0,0,0,0,1,0,0,0,0,0,0,0,Nc^(-1),0,-(Nf/Nc),0,0},
{0,0,0,0,0,0,0,0,0,0,0,0,0,0,0,0,0,0,0,0,0,8/Nc,8/Nc,32/Nc,0,-16/Nc,0,-32/Nc,0,0,0,0,0,0,0,0,0,0,0,0,0,0,0,-2/Nc+(2*Nf)/Nc,0,0,0,0,2/Nc-(2*Nf)/Nc,0,-2,0,0,0,4,0,0,0,2,0,0,0,0,0,0,0,2/Nc,(-2*Nf)/Nc,0,0,0},
{0,0,0,0,0,0,0,0,0,0,0,0,0,0,0,0,0,0,4/Nc,4/Nc,0,2/Nc+(2*Nf)/Nc,2/Nc+(2*Nf)/Nc,-8/Nc,0,0,0,12/Nc-(4*Nf)/Nc,-8/Nc,-8/Nc,0,0,0,0,0,0,0,0,0,2/Nc,0,-4/Nc,-4/Nc,0,8/Nc,-4/Nc,2/Nc,0,0,0,0,-4,2,2,0,4,0,0,0,0,0,0,0,0,0,0,0,0,0,0,0},
{0,0,0,0,0,0,0,0,0,0,0,0,0,0,0,0,0,0,4/Nc,-6/Nc,0,Nf/Nc,Nf/Nc,-4/Nc,2/Nc,4/Nc-(2*Nf)/Nc,-2/Nc,8/Nc,2/Nc,-8/Nc,0,0,0,0,0,0,0,0,0,0,0,0,-2/Nc+Nf/Nc,Nc^(-1),0,0,0,2/Nc-Nf/Nc,-Nc^(-1),0,0,0,-2,2,0,0,4,0,0,0,0,0,0,0,0,0,0,-Nc^(-1),2/Nc,-(Nf/Nc),0},
{0,0,0,0,0,0,0,0,0,0,0,0,0,0,0,0,0,0,0,0,24/Nc,0,0,0,-48/Nc,0,0,0,0,0,0,24/Nc,0,0,0,0,0,0,0,0,12/Nc,0,0,-12/Nc,0,0,0,0,0,0,0,0,0,0,0,0,0,3+3*Nf,-2,0,0,0,0,0,0,0,-2/Nc,0,0,0,Nc^(-1)+Nf/Nc},
{0,0,0,0,0,0,0,0,0,0,0,0,0,0,0,0,0,0,0,0,0,0,0,0,48/Nc,0,-48/Nc,0,0,0,0,0,0,0,0,0,0,0,0,0,0,0,0,12/Nc,0,0,0,0,-12/Nc,0,0,0,0,0,0,0,0,-2,4+2*Nf,0,0,0,0,0,0,0,(-2*Nf)/Nc,0,0,0,2/Nc},
{0,0,0,0,0,0,0,0,0,0,0,0,0,0,0,0,0,0,6/Nc,-7/Nc,-Nc^(-1),3/Nc,3/Nc,-2/Nc,0,2/Nc,0,-6/Nc,5/Nc,-4/Nc,0,Nc^(-1),0,-4/Nc-(2*Nf)/Nc,0,0,-4/Nc,8/Nc+(2*Nf)/Nc,0,0,0,0,0,0,0,0,0,0,0,0,0,0,0,0,0,0,0,0,0,2+Nf,0,0,0,0,0,1,0,0,0,0,0},
{0,0,0,0,0,0,0,0,0,0,0,0,0,0,0,0,0,0,6/Nc+Nf/Nc,-Nc^(-1)-Nf/Nc,0,Nc^(-1),Nc^(-1),2/Nc,0,-2/Nc,0,-2/Nc,-Nc^(-1),-4/Nc,0,0,4/Nc,Nf/Nc,4/Nc,-4/Nc,(-2*Nf)/Nc,Nf/Nc,-4/Nc,0,0,0,0,0,0,0,0,0,0,0,0,0,0,0,0,0,0,0,0,1,1+Nf/2,0,0,0,0,3/2,0,0,0,0,0},
{0,0,0,0,0,0,0,0,0,0,0,0,0,0,0,0,0,0,0,0,(3*Nf)/Nc,0,0,0,-6/Nc,0,12/Nc-(3*Nf)/Nc,0,0,0,-6/Nc,0,36/Nc,0,0,-36/Nc,0,0,0,0,0,0,0,0,0,0,0,0,0,0,0,0,0,0,0,0,0,0,0,0,0,6+Nf,0,-6,0,3,0,0,0,0,0},
{0,0,0,0,0,0,0,0,0,0,0,0,0,0,0,0,0,0,0,Nf/Nc,Nc^(-1),-Nc^(-1),-Nc^(-1),-4/Nc,0,4/Nc,-Nc^(-1),8/Nc-(2*Nf)/Nc,Nf/Nc,-6/Nc,0,0,2/Nc+(2*Nf)/Nc,2/Nc,-4/Nc,-2/Nc-(2*Nf)/Nc,-4/Nc,2/Nc,4/Nc,0,0,0,0,0,0,0,0,0,0,0,0,0,0,0,0,0,0,0,0,1,2,1,6,0,-6,0,0,0,0,0,0},
{0,0,0,0,0,0,0,0,0,0,0,0,0,0,0,0,0,0,0,0,6/Nc,0,0,0,6/Nc+(3*Nf)/Nc,0,-12/Nc,0,0,0,6/Nc-(3*Nf)/Nc,-6/Nc,0,0,36/Nc,0,0,0,-36/Nc,0,0,0,0,0,0,0,0,0,0,0,0,0,0,0,0,0,0,0,0,0,0,-6,0,6+Nf,0,3,0,0,0,0,0},
{0,0,0,0,0,0,0,0,0,0,0,0,0,0,0,0,0,0,-6/Nc,3/Nc,0,2/Nc+Nf/Nc,2/Nc+Nf/Nc,2/Nc,Nc^(-1),-2/Nc,0,-10/Nc,3/Nc,6/Nc-(2*Nf)/Nc,-Nc^(-1),0,-4/Nc,2/Nc,2/Nc+(2*Nf)/Nc,4/Nc,-4/Nc,2/Nc,-2/Nc-(2*Nf)/Nc,0,0,0,0,0,0,0,0,0,0,0,0,0,0,0,0,0,0,0,0,1,2,0,-6,1,6,0,0,0,0,0,0},
{0,0,0,0,0,0,0,0,0,0,0,0,0,0,0,0,0,0,0,0,-2/Nc-Nf/Nc,0,0,0,4/Nc,0,-4/Nc,0,0,0,-4/Nc,6/Nc+Nf/Nc,0,0,0,0,-24/Nc,24/Nc,0,0,0,0,0,0,0,0,0,0,0,0,0,0,0,0,0,0,0,0,0,0,0,0,0,0,0,2+2*Nf,0,0,0,0,0},
{0,0,0,0,0,0,0,0,0,0,0,0,0,0,0,0,0,0,0,0,0,0,0,0,0,0,0,0,0,0,0,0,0,0,0,0,0,0,0,0,0,0,0,0,0,0,0,0,0,48/Nc,12/Nc,0,0,-48/Nc,-12/Nc,0,0,-2/Nc+(2*Nf)/Nc,2/Nc-(2*Nf)/Nc,0,0,0,0,0,0,0,2*Nf,0,0,0,2},
{0,0,0,0,0,0,0,0,0,0,0,0,0,0,0,0,0,0,0,0,0,0,0,0,0,0,0,0,0,0,0,0,0,0,0,0,0,0,0,0,0,0,0,0,0,0,0,0,0,0,-4/Nc+(4*Nf)/Nc,32/Nc,16/Nc,0,4/Nc-(4*Nf)/Nc,-32/Nc,-16/Nc,0,0,0,0,0,0,0,0,0,4,0,0,0,0},
{0,0,0,0,0,0,0,0,0,0,0,0,0,0,0,0,0,0,0,0,0,0,0,0,0,0,0,0,0,0,0,0,0,0,0,0,0,0,0,0,0,0,0,0,0,0,0,0,0,-6/Nc+(3*Nf)/Nc,0,0,36/Nc,6/Nc-(3*Nf)/Nc,0,0,-36/Nc,3/Nc,-3/Nc,0,0,0,0,0,0,0,3,0,Nf,0,0},
{0,0,0,0,0,0,0,0,0,0,0,0,0,0,0,0,0,0,0,0,0,0,0,0,0,0,0,0,0,0,0,0,0,0,0,0,0,0,0,0,0,0,0,0,0,0,0,0,0,4/Nc,4/Nc,-8/Nc,-4/Nc+(4*Nf)/Nc,-4/Nc,-4/Nc,8/Nc,4/Nc-(4*Nf)/Nc,0,0,0,0,0,0,0,0,0,0,0,4,0,0},
{0,0,0,0,0,0,0,0,0,0,0,0,0,0,0,0,0,0,0,0,0,0,0,0,0,0,0,0,0,0,0,0,0,0,0,0,0,0,0,0,0,0,0,0,0,0,0,0,0,0,0,0,0,0,0,0,0,72/Nc,-72/Nc,0,0,0,0,0,0,0,0,0,0,0,4*Nf}}
\end{verbatim}
\end{minipage}\end{lrbox}

There are 71 $\alg{so}(N_f)$ singlets at length $L=8$. 
The mixing matrix and operator basis for general $N_f$ are given by

\medskip \noindent
{\tiny $M_{IJ}=$} \ \resizebox{\hsize}{!}{\usebox\myverbE}

\bigskip \noindent
and
{\tiny
\begin{multline}
\{ \cO_I \} = \Big\{ 
\tr(\phi_{i_1}\phi_{i_1}\phi_{i_4}\phi_{i_3}\phi_{i_2}\phi_{i_2}\phi_{i_3}\phi_{i_4}),\ 
\tr(\phi_{i_1}\phi_{i_1}\phi_{i_4}\phi_{i_2}\phi_{i_3}\phi_{i_2}\phi_{i_3}\phi_{i_4}),\ 
\tr(\phi_{i_1}\phi_{i_2}\phi_{i_3}\phi_{i_4}\phi_{i_1}\phi_{i_2}\phi_{i_3}\phi_{i_4}),\ 
\tr(\phi_{i_1}\phi_{i_1}\phi_{i_3}\phi_{i_4}\phi_{i_2}\phi_{i_2}\phi_{i_3}\phi_{i_4}),\\ 
\tr(\phi_{i_1}\phi_{i_2}\phi_{i_1}\phi_{i_4}\phi_{i_3}\phi_{i_2}\phi_{i_3}\phi_{i_4}),\
\tr(\phi_{i_1}\phi_{i_1}\phi_{i_2}\phi_{i_4}\phi_{i_3}\phi_{i_2}\phi_{i_3}\phi_{i_4}),\
\tr(\phi_{i_1}\phi_{i_1}\phi_{i_2}\phi_{i_4}\phi_{i_2}\phi_{i_3}\phi_{i_3}\phi_{i_4}),\ 
\tr(\phi_{i_1}\phi_{i_3}\phi_{i_2}\phi_{i_1}\phi_{i_4}\phi_{i_2}\phi_{i_3}\phi_{i_4}),\\ 
\tr(\phi_{i_1}\phi_{i_2}\phi_{i_3}\phi_{i_1}\phi_{i_4}\phi_{i_2}\phi_{i_3}\phi_{i_4}),\ 
\tr(\phi_{i_1}\phi_{i_1}\phi_{i_3}\phi_{i_2}\phi_{i_4}\phi_{i_2}\phi_{i_3}\phi_{i_4}),\ 
\tr(\phi_{i_1}\phi_{i_2}\phi_{i_1}\phi_{i_3}\phi_{i_4}\phi_{i_2}\phi_{i_3}\phi_{i_4}),\
\tr(\phi_{i_1}\phi_{i_1}\phi_{i_2}\phi_{i_3}\phi_{i_4}\phi_{i_2}\phi_{i_3}\phi_{i_4}),\\ 
\tr(\phi_{i_1}\phi_{i_1}\phi_{i_2}\phi_{i_2}\phi_{i_4}\phi_{i_3}\phi_{i_3}\phi_{i_4}),\ 
\tr(\phi_{i_1}\phi_{i_2}\phi_{i_1}\phi_{i_3}\phi_{i_2}\phi_{i_4}\phi_{i_3}\phi_{i_4}),\ 
\tr(\phi_{i_1}\phi_{i_1}\phi_{i_2}\phi_{i_3}\phi_{i_2}\phi_{i_4}\phi_{i_3}\phi_{i_4}),\ 
\tr(\phi_{i_1}\phi_{i_2}\phi_{i_1}\phi_{i_2}\phi_{i_3}\phi_{i_4}\phi_{i_3}\phi_{i_4}),\\
\tr(\phi_{i_1}\phi_{i_1}\phi_{i_2}\phi_{i_2}\phi_{i_3}\phi_{i_4}\phi_{i_3}\phi_{i_4}),\ 
\tr(\phi_{i_1}\phi_{i_1}\phi_{i_2}\phi_{i_2}\phi_{i_3}\phi_{i_3}\phi_{i_4}\phi_{i_4}),\ 
\tr(\phi_{i_1}\phi_{i_2}) \, \tr(\phi_{i_4}\phi_{i_1}\phi_{i_2}\phi_{i_3}\phi_{i_3}\phi_{i_4}),\ 
\tr(\phi_{i_2}\phi_{i_1}) \, \tr(\phi_{i_4}\phi_{i_3}\phi_{i_1}\phi_{i_2}\phi_{i_3}\phi_{i_4}),\\ 
\tr(\phi_{i_4}\phi_{i_4}) \, \tr(\phi_{i_3}\phi_{i_2}\phi_{i_1}\phi_{i_1}\phi_{i_2}\phi_{i_3}),\ 
\tr(\phi_{i_1}\phi_{i_2}) \, \tr(\phi_{i_4}\phi_{i_1}\phi_{i_3}\phi_{i_2}\phi_{i_3}\phi_{i_4}),\ 
\tr(\phi_{i_3}\phi_{i_1}) \, \tr(\phi_{i_4}\phi_{i_2}\phi_{i_1}\phi_{i_2}\phi_{i_3}\phi_{i_4}),\ 
\tr(\phi_{i_2}\phi_{i_1}) \, \tr(\phi_{i_3}\phi_{i_4}\phi_{i_1}\phi_{i_2}\phi_{i_3}\phi_{i_4}),\\ 
\tr(\phi_{i_4}\phi_{i_4}) \, \tr(\phi_{i_3}\phi_{i_1}\phi_{i_2}\phi_{i_1}\phi_{i_2}\phi_{i_3}),\ 
\tr(\phi_{i_1}\phi_{i_3}) \, \tr(\phi_{i_2}\phi_{i_1}\phi_{i_2}\phi_{i_4}\phi_{i_3}\phi_{i_4}),\ 
\tr(\phi_{i_4}\phi_{i_4}) \, \tr(\phi_{i_1}\phi_{i_3}\phi_{i_2}\phi_{i_1}\phi_{i_2}\phi_{i_3}),\ 
\tr(\phi_{i_3}\phi_{i_1}) \, \tr(\phi_{i_2}\phi_{i_4}\phi_{i_1}\phi_{i_2}\phi_{i_3}\phi_{i_4}),\\ 
\tr(\phi_{i_1}\phi_{i_3}) \, \tr(\phi_{i_4}\phi_{i_1}\phi_{i_2}\phi_{i_2}\phi_{i_3}\phi_{i_4}),\ 
\tr(\phi_{i_1}\phi_{i_3}) \, \tr(\phi_{i_2}\phi_{i_1}\phi_{i_4}\phi_{i_2}\phi_{i_3}\phi_{i_4}),\ 
\tr(\phi_{i_4}\phi_{i_4}) \, \tr(\phi_{i_1}\phi_{i_2}\phi_{i_3}\phi_{i_1}\phi_{i_2}\phi_{i_3}),\ 
\tr(\phi_{i_4}\phi_{i_4}) \, \tr(\phi_{i_3}\phi_{i_1}\phi_{i_1}\phi_{i_2}\phi_{i_2}\phi_{i_3}),\\ 
\tr(\phi_{i_2}\phi_{i_4}\phi_{i_3}) \, \tr(\phi_{i_1}\phi_{i_1}\phi_{i_2}\phi_{i_3}\phi_{i_4}),\ 
\tr(\phi_{i_4}\phi_{i_4}\phi_{i_2}) \, \tr(\phi_{i_1}\phi_{i_1}\phi_{i_3}\phi_{i_2}\phi_{i_3}),\ 
\tr(\phi_{i_1}\phi_{i_2}\phi_{i_3}) \, \tr(\phi_{i_4}\phi_{i_1}\phi_{i_2}\phi_{i_3}\phi_{i_4}),\ 
\tr(\phi_{i_1}\phi_{i_4}\phi_{i_3}) \, \tr(\phi_{i_2}\phi_{i_1}\phi_{i_2}\phi_{i_3}\phi_{i_4}),\\ 
\tr(\phi_{i_4}\phi_{i_4}\phi_{i_2}) \, \tr(\phi_{i_1}\phi_{i_3}\phi_{i_1}\phi_{i_2}\phi_{i_3}),\ 
\tr(\phi_{i_4}\phi_{i_4}\phi_{i_2}) \, \tr(\phi_{i_3}\phi_{i_1}\phi_{i_1}\phi_{i_2}\phi_{i_3}),\ 
\tr(\phi_{i_1}\phi_{i_2}\phi_{i_4}) \, \tr(\phi_{i_3}\phi_{i_1}\phi_{i_2}\phi_{i_3}\phi_{i_4}),\ 
\tr(\phi_{i_2}\phi_{i_2}\phi_{i_3}\phi_{i_4}) \, \tr(\phi_{i_4}\phi_{i_3}\phi_{i_1}\phi_{i_1}),\\ 
\tr(\phi_{i_2}\phi_{i_1}\phi_{i_1}\phi_{i_2}) \, \tr(\phi_{i_4}\phi_{i_3}\phi_{i_3}\phi_{i_4}),\ 
\tr(\phi_{i_2}\phi_{i_1}\phi_{i_3}\phi_{i_4}) \, \tr(\phi_{i_4}\phi_{i_3}\phi_{i_1}\phi_{i_2}),\ 
\tr(\phi_{i_2}\phi_{i_2}\phi_{i_3}\phi_{i_4}) \, \tr(\phi_{i_4}\phi_{i_1}\phi_{i_3}\phi_{i_1}),\ 
\tr(\phi_{i_1}\phi_{i_2}\phi_{i_1}\phi_{i_2}) \, \tr(\phi_{i_4}\phi_{i_3}\phi_{i_3}\phi_{i_4}),\\ 
\tr(\phi_{i_1}\phi_{i_3}\phi_{i_2}\phi_{i_4}) \, \tr(\phi_{i_4}\phi_{i_3}\phi_{i_1}\phi_{i_2}),\ 
\tr(\phi_{i_1}\phi_{i_2}\phi_{i_3}\phi_{i_4}) \, \tr(\phi_{i_4}\phi_{i_1}\phi_{i_2}\phi_{i_3}),\ 
\tr(\phi_{i_3}\phi_{i_3}\phi_{i_2}\phi_{i_4}) \, \tr(\phi_{i_4}\phi_{i_1}\phi_{i_1}\phi_{i_2}),\ 
\tr(\phi_{i_3}\phi_{i_1}\phi_{i_3}\phi_{i_4}) \, \tr(\phi_{i_4}\phi_{i_2}\phi_{i_1}\phi_{i_2}),\\ 
\tr(\phi_{i_1}\phi_{i_2}\phi_{i_1}\phi_{i_2}) \, \tr(\phi_{i_3}\phi_{i_4}\phi_{i_3}\phi_{i_4}),\ 
\tr(\phi_{i_1}\phi_{i_2}) \, \tr(\phi_{i_4}\phi_{i_4}) \, \tr(\phi_{i_3}\phi_{i_1}\phi_{i_2}\phi_{i_3}),\ 
\tr(\phi_{i_3}\phi_{i_4}) \, \tr(\phi_{i_4}\phi_{i_3}) \, \tr(\phi_{i_2}\phi_{i_1}\phi_{i_1}\phi_{i_2}),\\ 
\tr(\phi_{i_1}\phi_{i_2}) \, \tr(\phi_{i_4}\phi_{i_3}) \, \tr(\phi_{i_4}\phi_{i_1}\phi_{i_2}\phi_{i_3}),\
\tr(\phi_{i_1}\phi_{i_4}) \, \tr(\phi_{i_4}\phi_{i_2}) \, \tr(\phi_{i_3}\phi_{i_1}\phi_{i_2}\phi_{i_3}),\ 
\tr(\phi_{i_1}\phi_{i_3}) \, \tr(\phi_{i_4}\phi_{i_4}) \, \tr(\phi_{i_2}\phi_{i_1}\phi_{i_2}\phi_{i_3}),\\ 
\tr(\phi_{i_3}\phi_{i_4}) \, \tr(\phi_{i_4}\phi_{i_3}) \, \tr(\phi_{i_1}\phi_{i_2}\phi_{i_1}\phi_{i_2}),\ 
\tr(\phi_{i_1}\phi_{i_3}) \, \tr(\phi_{i_4}\phi_{i_2}) \, \tr(\phi_{i_4}\phi_{i_1}\phi_{i_2}\phi_{i_3}),\ 
\tr(\phi_{i_1}\phi_{i_4}) \, \tr(\phi_{i_4}\phi_{i_3}) \, \tr(\phi_{i_2}\phi_{i_1}\phi_{i_2}\phi_{i_3}),\\ 
\tr(\phi_{i_3}\phi_{i_3}) \, \tr(\phi_{i_4}\phi_{i_4}) \, \tr(\phi_{i_2}\phi_{i_1}\phi_{i_1}\phi_{i_2}),\ 
\tr(\phi_{i_3}\phi_{i_3}) \, \tr(\phi_{i_4}\phi_{i_4}) \, \tr(\phi_{i_1}\phi_{i_2}\phi_{i_1}\phi_{i_2}),\ 
\tr(\phi_{i_3}\phi_{i_2}) \, \tr(\phi_{i_1}\phi_{i_1}\phi_{i_2}) \, \tr(\phi_{i_4}\phi_{i_3}\phi_{i_4}),\\ 
\tr(\phi_{i_1}\phi_{i_2}) \, \tr(\phi_{i_3}\phi_{i_3}\phi_{i_4}) \, \tr(\phi_{i_4}\phi_{i_1}\phi_{i_2}),\ 
\tr(\phi_{i_4}\phi_{i_4}) \, \tr(\phi_{i_1}\phi_{i_2}\phi_{i_3}) \, \tr(\phi_{i_3}\phi_{i_2}\phi_{i_1}),\ 
\tr(\phi_{i_3}\phi_{i_1}) \, \tr(\phi_{i_2}\phi_{i_3}\phi_{i_4}) \, \tr(\phi_{i_2}\phi_{i_4}\phi_{i_1}),\\ 
\tr(\phi_{i_4}\phi_{i_4}) \, \tr(\phi_{i_1}\phi_{i_2}\phi_{i_3}) \, \tr(\phi_{i_3}\phi_{i_1}\phi_{i_2}),\ 
\tr(\phi_{i_3}\phi_{i_2}) \, \tr(\phi_{i_1}\phi_{i_3}\phi_{i_4}) \, \tr(\phi_{i_4}\phi_{i_1}\phi_{i_2}),\ 
\tr(\phi_{i_4}\phi_{i_4}) \, \tr(\phi_{i_2}\phi_{i_2}\phi_{i_3}) \, \tr(\phi_{i_3}\phi_{i_1}\phi_{i_1}),\\ 
\tr(\phi_{i_1}\phi_{i_4}) \, \tr(\phi_{i_2}\phi_{i_2}) \, \tr(\phi_{i_3}\phi_{i_3}) \, \tr(\phi_{i_4}\phi_{i_1}),\ 
\tr(\phi_{i_1}\phi_{i_4}) \, \tr(\phi_{i_2}\phi_{i_3}) \, \tr(\phi_{i_3}\phi_{i_2}) \, \tr(\phi_{i_4}\phi_{i_1}),\ 
\tr(\phi_{i_1}\phi_{i_2}) \, \tr(\phi_{i_2}\phi_{i_4}) \, \tr(\phi_{i_3}\phi_{i_3}) \, \tr(\phi_{i_4}\phi_{i_1}),\\ 
\tr(\phi_{i_1}\phi_{i_4}) \, \tr(\phi_{i_2}\phi_{i_1}) \, \tr(\phi_{i_2}\phi_{i_3}) \, \tr(\phi_{i_4}\phi_{i_3}),\ 
\tr(\phi_{i_1}\phi_{i_1}) \, \tr(\phi_{i_2}\phi_{i_2}) \, \tr(\phi_{i_3}\phi_{i_3}) \, \tr(\phi_{i_4}\phi_{i_4}) \Big\}.
\end{multline}}

\begin{figure}[t]
\begin{center}
\subfigure{\includegraphics[scale=0.74]{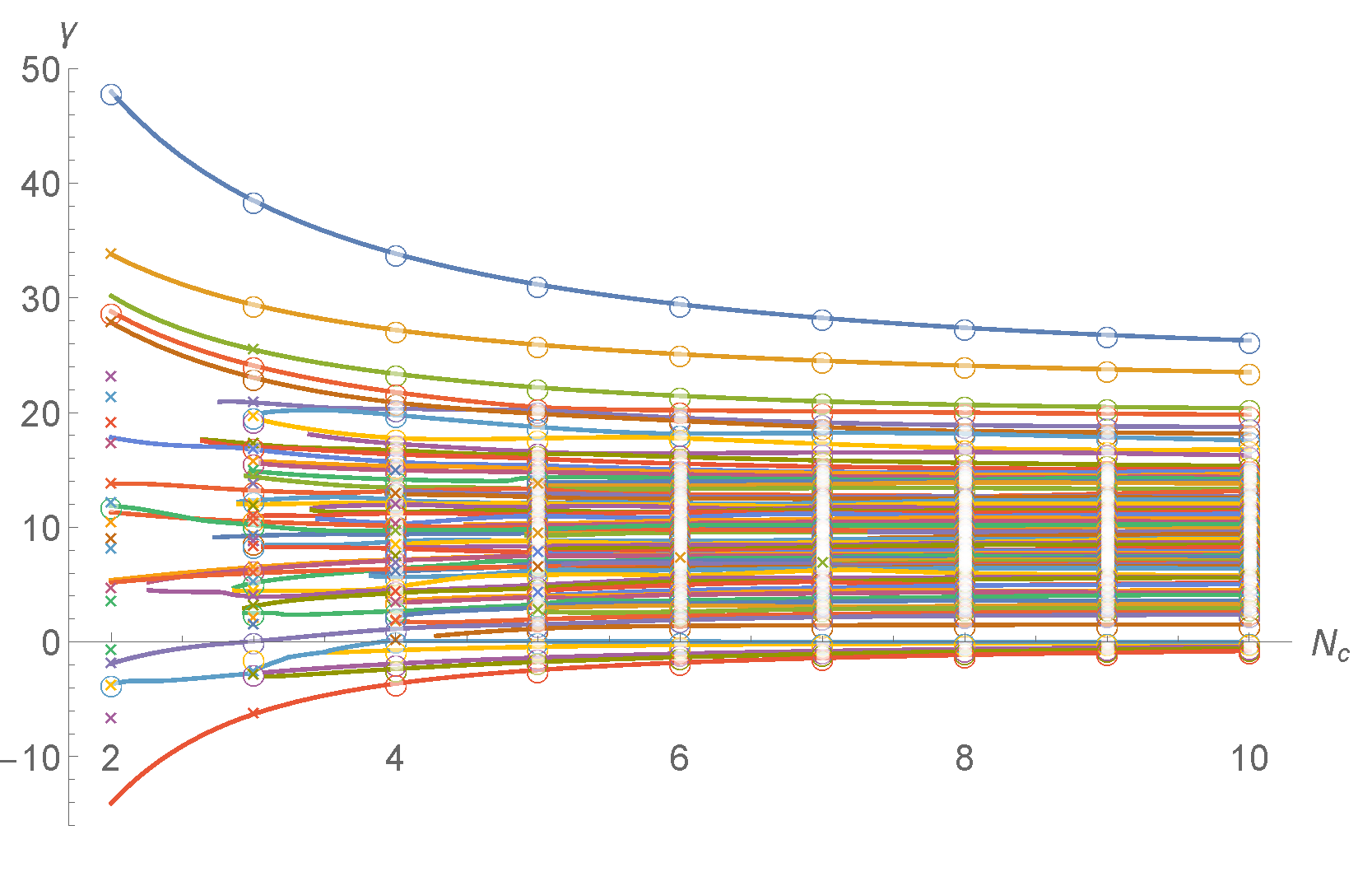}}
\\[-2mm]
\subfigure{\includegraphics[scale=0.74]{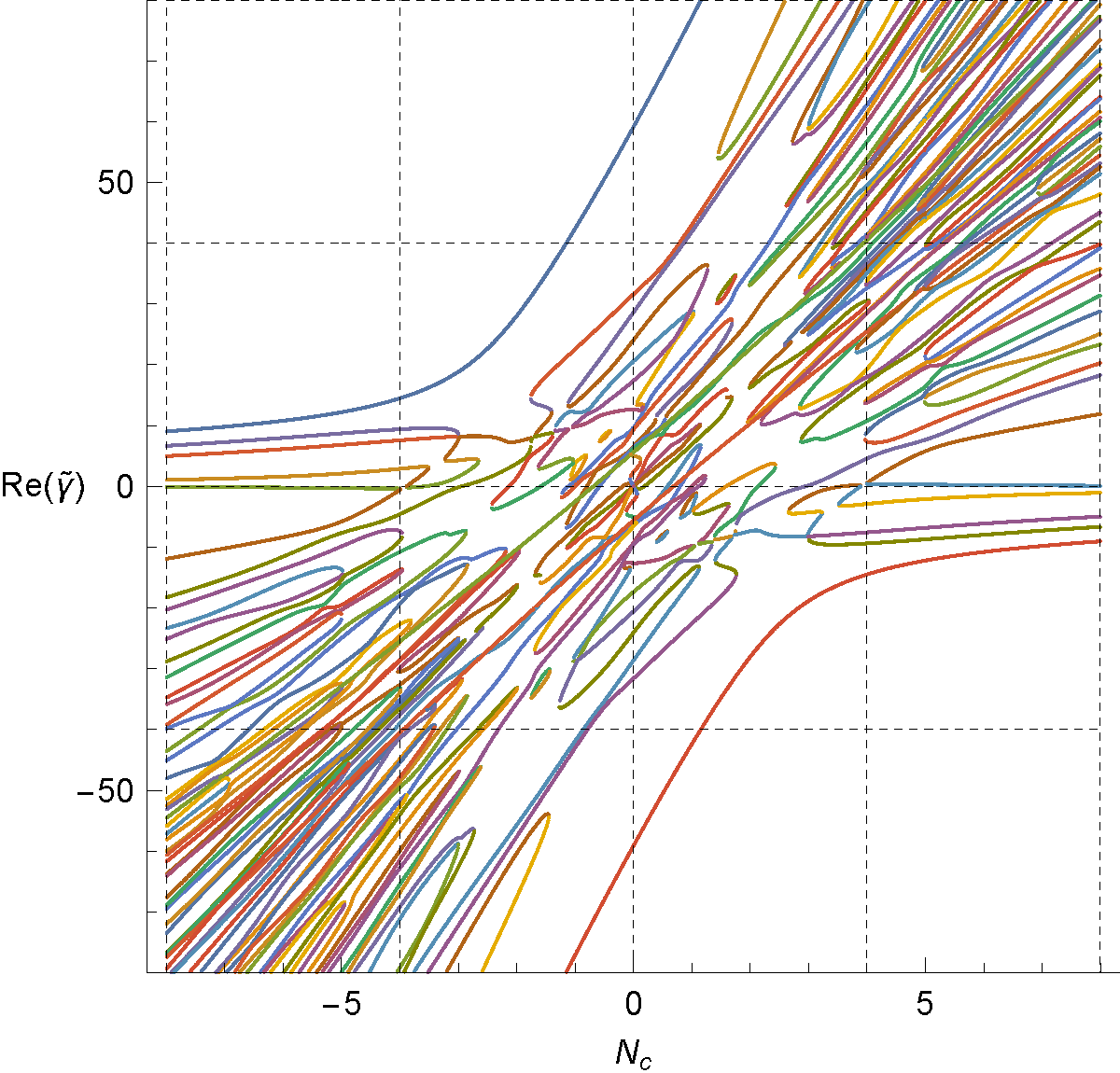}}
\caption{The one-loop anomalous dimensions $\gamma$ and the rescaled dimensions $\tilde \gamma = N_c \, \gamma$ of $\alg{so}(6)$ singlets of length $L=8$ at finite $N_c$\,. 
In the upper figure, eigenvalue curves do not cross for $N_c \ge L=8$ and continuously connected to the large $N_c$ spectrum. Some of the eigenvalue curves have a loose end at small $N_c$\,, due to creations and annihilations of complex eigenvalues as shown in the lower figure.}
\label{fig:finiteNc L=8}
\end{center}
\end{figure}

\noindent
The spectrum is shown in Figure \ref{fig:finiteNc L=8}.

\clearpage

Also, there is a pair of exceptional operators at $L=8$, which take the form
\begin{align}
\cO^{(\pm)} &= \tr( \phi_{i_1} \phi_{i_2} ) \, \tr( \phi_{i_1} \[ \phi_{i_3} \phi_{i_2} \phi_{i_3} \,, \phi_{i_4} \phi_{i_4} \] )
+ a^{(\pm)} \, \tr( \phi_{i_1} \phi_{i_1} \phi_{i_2} \phi_{i_3} \phi_{i_4} \[ \phi_{i_3} \,, \phi_{i_2} \phi_{i_4} \] ),
\notag \\[1mm]
a^{(\pm)} &= \frac{1}{18} \(5 N_c \pm \sqrt{25 N_c^2+504}\) ,
\end{align}
and their one-lop dimensions are exactly given by
\begin{equation}
\gamma^{(\pm)} = \frac{1}{4} \(33 \pm \frac{\sqrt{25 N_c^2+504}}{N_c} \).
\end{equation}

\bigskip
There are five large $N_c$ zero modes. The submixing matrix for general $N_f$ is given by 
{\scriptsize
\begin{align}
(-M_\text{sm})_{ij} &=
\begin{pmatrix}
\frac{(N_f+1) (N_f+6) \, \varkappa_{11}}{3 (N_f+2) (N_f+4) \, \varkappa_{10} } 
& 0 & 0 & 0 & 0 \\[2mm]
0 & \frac{2 \left(N_f+1\right) \left(N_f+6\right) \varkappa_{21}}{3 \left(N_f+2\right) \left(N_f+4\right) \varkappa_{20}}
& \frac{2 N_f \left(N_f+1\right) \left(N_f+6\right) \varkappa_{22}}{3 \left(N_f+2\right) \left(N_f+4\right) \varkappa_{20}} 
& 0 & 0 \\[2mm]
0 & \frac{\left(N_f-2\right) \left(N_f+1\right) \left(N_f+6\right) \varkappa_{22}}{2 \left(N_f+2\right){}^2 \varkappa_{20}} 
& \frac{\left(N_f+1\right) \left(N_f+6\right) \varkappa_{23}}{2 N_f \left(N_f+2\right){}^2 \varkappa_{20}}
& 0 & 0 \\[2mm]   
0 & 0 & 0 &  \left(N_f-6\right) \left(N_f+3\right) & \frac{6 \left(N_f^2-N_f+6\right)}{N_f} \\[2mm]
0 & 0 & 0 & 4 \left(N_f-6\right) & \frac{2 \left(N_f^3-2 N_f^2-9 N_f+42\right)}{N_f} 
\end{pmatrix},
\label{L=8 pm-Mat} \\[3mm]
\kappa_{11} &= 
20823 N_f^{11}+1675098 N_f^{10}+55911035 N_f^9+1008163871 N_f^8
+10662939982 N_f^7 +66380603808 N_f^6 +219710331899 N_f^5
\notag \\
&\qquad +178394761318 N_f^4
-1267103763092 N_f^3-4642599661272 N_f^2
-6369708861792 N_f-3243010514688 ,
\notag \\
\varkappa_{10} &= 
1971 N_f^9+158898 N_f^8+5341569 N_f^7+97989373 N_f^6+1077096339 N_f^5+7325782233 N_f^4
+30674558032 N_f^3
\notag \\
&\qquad +75803848404 N_f^2+99816658464 N_f+53440944192 ,
\notag \\
\kappa_{21} &= 124 N_f^5+2271 N_f^4+8447 N_f^3-15528 N_f^2-88236 N_f-81648,
\notag \\
\kappa_{22} &= 63 N_f^4+1153   N_f^3+4522 N_f^2-2208 N_f-12960,
\notag \\
\kappa_{23} &= 163 N_f^6+2459 N_f^5+876 N_f^4-63460 N_f^3-94656 N_f^2+136800   N_f+217728,
\notag \\
\kappa_{20} &= 37 N_f^3+727 N_f^2+3714 N_f+4536
\label{def:varkappa L=8}
\end{align}}

\vspace{-3mm} \noindent
These matrix elements are defined on the basis\footnote{The off-diagonal elements of $M_\text{sm}$ depend on the normalization of $\cC$. Here we use the convention: $\cC_{i_1 \dots i_\ell} \equiv \frac{1}{\ell!} \sum_{\sigma \in \cS_\ell} \tr (\Phi_{i_{\sigma(1)}} \dots \Phi_{i_{\sigma(\ell)}}) - (\text{flavor contractions})$.}
\begin{equation}
\cO^\circ_j =
\(
\cC_{ijkl} \, \cC_{ijkl} ,\ 
\cC_{ijkl} \, \cC_{ij} \, \cC_{kl} ,\ 
\cC_{ijk} \, \cC_{ijl} \, \cC_{kl} ,\ 
\[ \cC_{ij} \, \cC_{jk} \, \cC_{kl} \, \cC_{li} \]' ,\ 
\cC_{ij} \, \cC_{ji} \, \cC_{kl} \, \cC_{lk}
\)^T,
\end{equation}
where $\[ \cC_{ij} \, \cC_{jk} \, \cC_{kl} \, \cC_{li} \]'$ is the following combination of two quadruple traces,
\begin{align}
&\[ \cC_{ij} \, \cC_{jk} \, \cC_{kl} \, \cC_{li} \]' = 
\tr( \phi_{i_1} \phi_{i_2} ) \, \tr( \phi_{i_2} \phi_{i_3} ) \, \tr( \phi_{i_3} \phi_{i_4} ) \, \tr( \phi_{i_4} \phi_{i_1} )
\\[1mm]
&\quad 
+ \frac{3}{N_f} \, \tr( \phi_{i_1} \phi_{i_4} ) \, \tr( \phi_{i_4} \phi_{i_1} ) \, \tr( \phi_{i_2} \phi_{i_3} ) \, \tr( \phi_{i_3} \phi_{i_2} )
- \frac{4}{N_f} \, \tr( \phi_{i_4} \phi_{i_4} ) \, \tr( \phi_{i_1} \phi_{i_2} ) \, \tr( \phi_{i_3} \phi_{i_1} ) \, \tr( \phi_{i_2} \phi_{i_3} ) .
\notag 
\end{align}

By diagonalizing the submixing matrix at $N_f=6$, one finds
\begin{equation}
\{ \gamma_2^{(2)} , \gamma_2^{(3)-} , \gamma_2^{(4)-} , \gamma_2^{(3)+} , \gamma_2^{(4)+} \}
\approx \pare{-101.276, -66.6009, -44, -6.89905, 0 },
\end{equation}
where $\gamma_2^{(m)}$ are $\gamma_2$ for $m$-trace operators.
The operator corresponding to $\gamma_2^{(4)+}$ is not protected, and receives corrections at $\cO(N_c^{-4})$ with a positive sign.

\bigskip
\paragraph{Degeneracy of non-zero modes.}

There are six degenerate positive eigenvalues at $L=8$, namely $\{ 6,7,8,9.5,12,14 \}$. 
The eigenvalues receive finite $N_c$ corrections at the order $1/N_c$\,, and the large $N_c$ expansion of these modes can submix with the operators having different number of traces. 
They make a contrast to the properties of the large $N_c$ zero modes \eqref{L=8 pm-Mat}.
For example, the eigenstates corresponding to the eigenvalue 6 take the form
\begin{equation}
\cO_1 \sim O_{5,3} + O_{3,3,2} \,, \quad
\cO_2 \sim O_{4,2,2} \,, \quad
\cO_3 \sim O_{2,2,2,2} \,, \quad
\cO_4 \sim O_{5,3} + O_{3,3,2} \,,
\end{equation}
where $O_{L_1,L_2, \dots}$ is a multi-trace operator with length $\{ L_1,L_2, \dots \}$.

\subsubsection{$L=10$}

\begin{figure}[t]
\begin{center}
\includegraphics[scale=1]{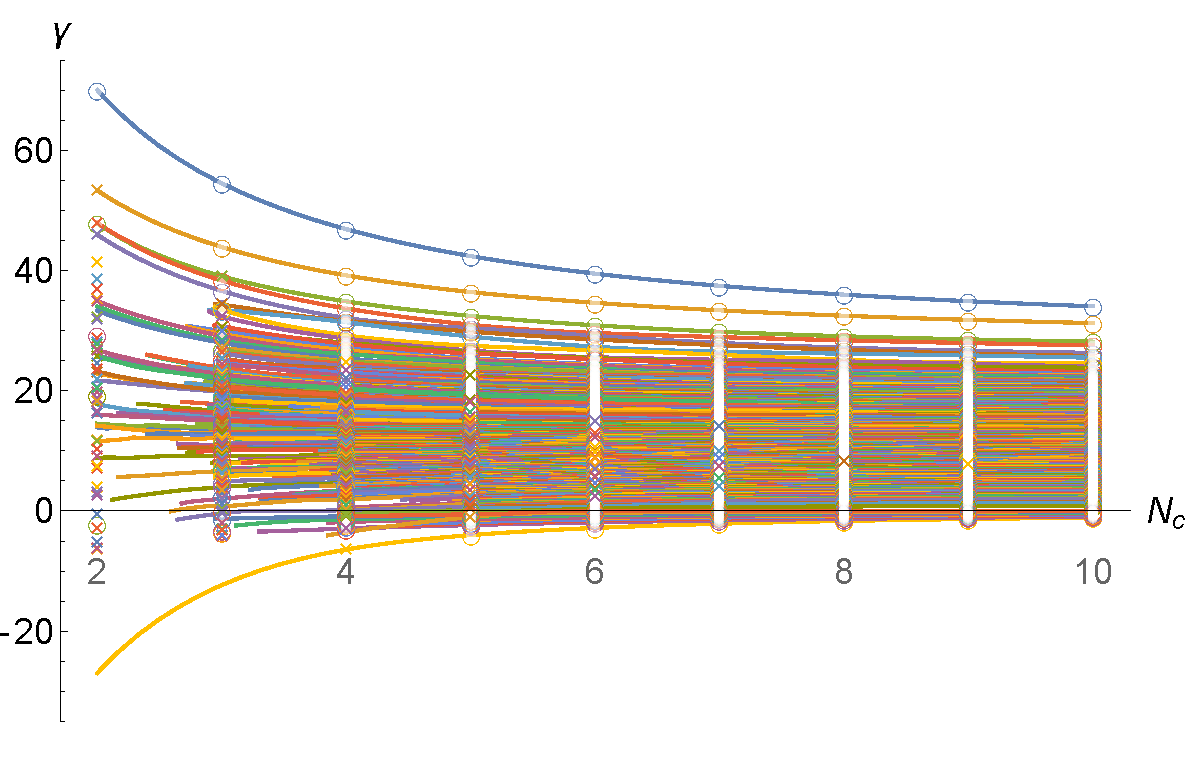}
\caption{Spectrum of the one-loop dimensions among the $\alg{so}(6)$ singlets of length $L=10$ at finite $N_c$\,.
Each eigenvalue curve is continuously connected to the large $N_c$ spectrum.}
\label{fig:finiteNc L=10}
\end{center}
\end{figure}

\begin{figure}[h]
\begin{center}
\subfigure{\includegraphics[scale=0.9]{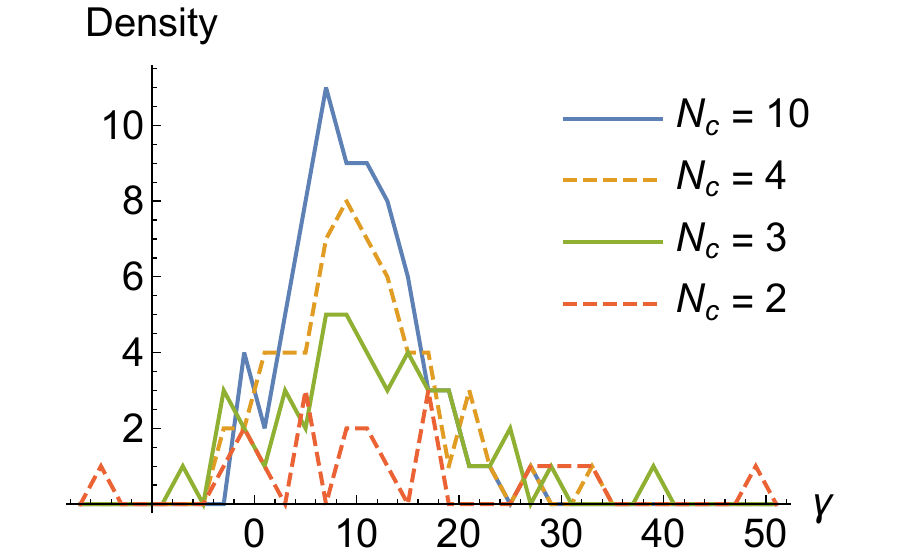}}
\subfigure{\includegraphics[scale=0.9]{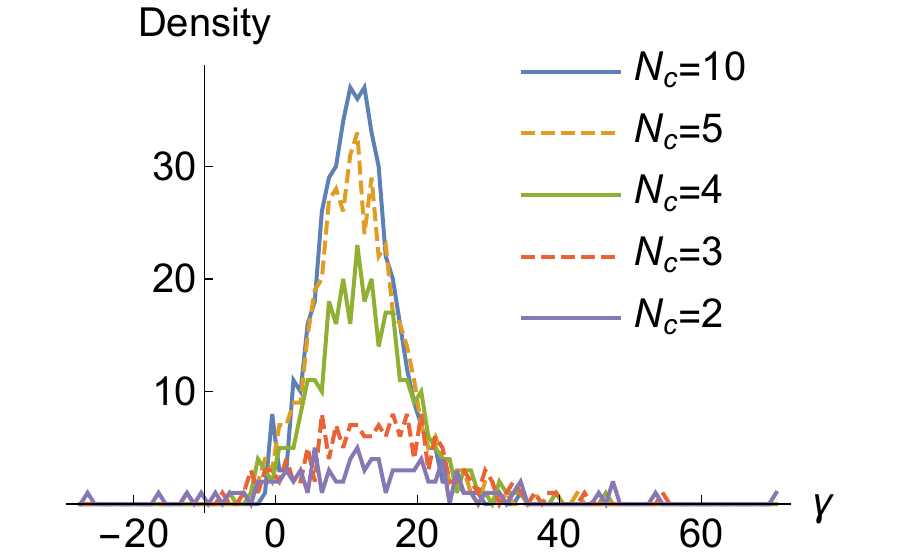}}
\caption{The eigenvalue density of the mixing matrix length $L=8$ (left) and $L=10$ (right) at finite $N_c$\,. The vertical axis shows the number of real-valued eigenvalues in each bin $[\gamma-\Delta\gamma/2, \gamma+\Delta \gamma/2)$. The bin width $\Delta \gamma$ is adjusted for each $L$.}
\label{fig:ev density}
\end{center}
\end{figure}

There are 469 $\alg{so}(N_f)$ singlets at length $L=10$. 
The mixing matrix and operator basis for general $N_f$ are shown in the attached files {\tt AncillaryNegative.nb} and {\tt AncillaryL10Data.txt}.
The spectrum is shown in Figure \ref{fig:finiteNc L=10}.\footnote{We have not checked potential level-crossing with exceptional eigenvalues due to the complexity.}
To highlight the structure of the spectrum lying in the middle, we plot the eigenvalue density in Figure \ref{fig:ev density}.

There are 11 large $N_c$ zero modes, as shown in \eqref{products sym trsless 5}.
There are no finite\,-$N_c$ zero modes. At $N_f=6$, the submixing matrix has two zero modes corresponding to combinations of quadruple- and quintuple-trace operators.
The one-loop dimensions of two operators are $\cO(N_c^{-4})$ and positive.
The submixing matrix at $N_f=6$ is given by 

\medskip\noindent
{\scriptsize
\begin{equation}
(-M_\text{sm})_{ij} = 
\left(
\begin{array}{ccccccccccc}
132.315 & 0 & 0 & 0 & 0 & 0 & 0 & 0 & 0 & 0 & 0 \\
0 & 87.4095 & 36.7996 & 40.8718 & 0 & 0 & 0 & 0 & 0 & 0 & 0 \\
0 & 34.5075 & 50.803 & 0 & 0 & 0 & 0 & 0 & 0 & 0 & 0 \\
0 & 25.2692 & 0 & 77.1909 & 0 & 0 & 0 & 0 & 0 & 0 & 0 \\
0 & 0 & 0 & 0 & 36.6545 & -0.0000105607 & -12.6725 & -21.2528 & 0 & 0 & 0 \\
0 & 0 & 0 & 0 & -2.64168 & 1.97708 & 2.62507 & 2.22822 & -9.00914 & 0 & 0 \\
0 & 0 & 0 & 0 & -12.0246 & -6.72677 & 59.5308 & -34.3732 & 30.6522 & 0 & 0 \\
0 & 0 & 0 & 0 & -14.1498 & -1.17112 & -23.5379 & 40.8351 & 5.33674 & 0 & 0 \\
0 & 0 & 0 & 0 & -10.4861 & -13.2197 & 7.32801 & 6.00623 & 60.2393 & 0 & 0 \\
0 & 0 & 0 & 0 & 0 & 0 & 0 & 0 & 0 & 26 & 0 \\
0 & 0 & 0 & 0 & 0 & 0 & 0 & 0 & 0 & 16.5923 & 0 \\
\end{array}
\right),
\label{L=10 pm-Mat}
\end{equation}}
which is block-diagonal.
These matrix elements are defined on the basis
\begin{equation}
\cO^\circ_j = \{ \cC^2, \ 3 \times \cC^3, \ 5 \times \cC^4, \ 2 \times \cC^5 \},
\label{L=10 LNZM basis}
\end{equation}
where $p \times \cC^{m}$ means $p$ linear combinations of the large $N_c$ zero modes with $m$-traces, and explicitly given in \eqref{products sym trsless 5}. The eigenvalues of the submixing matrix is
\begin{equation}
\gamma_2 = - \{ 132.315,\ 125.673,\ 83.7144,\ 67.7793,\ 66.8384,\ 42.3268,\ 26,\ 22.892,\ 5.41638,\ 0,\ 0 \}.
\end{equation}
Only numerical values are shown in \eqref{L=10 pm-Mat} because their precise forms are quite complicated. For example, the eigenvalue for the double-trace mode is
\begin{equation}
\gamma_2 = -\frac{36834342860563635266164905898354615349151994033997}{278383379192667725097781339828443592271084673096} 
\approx -132.315.
\label{lambda2 double-trace L=10}
\end{equation}

\subsubsection{$L=12$}

There are 4477 $\alg{so}(N_f)$ singlets at length $L=12$.
We have not made a detailed analysis of the spectrum due to a huge amount of computation involved.
There are 34 large $N_c$ zero modes,
\begin{equation}
\cO^\circ_j = \{ \cC^2, \ 5 \times \cC^3, \ 14 \times \cC^4, \ 10 \times \cC^5, \ 4 \times \cC^6 \},
\label{L=12 LNZM basis}
\end{equation}
in the notation of \eqref{L=10 LNZM basis}.

\subsection{Spectrum at general $N_f$}

We consider the spectrum of submixing matrix for $\alg{so}(N_f)$ singlets at general $N_f$\,, which will highlight the special properties of $\cN=4$ SYM from the non-planar spectrum of other gauge theories.
The value $N_f=6$ corresponds to $\cN=4$ SYM.

\begin{figure}[t]
\begin{center}
\includegraphics[scale=0.92]{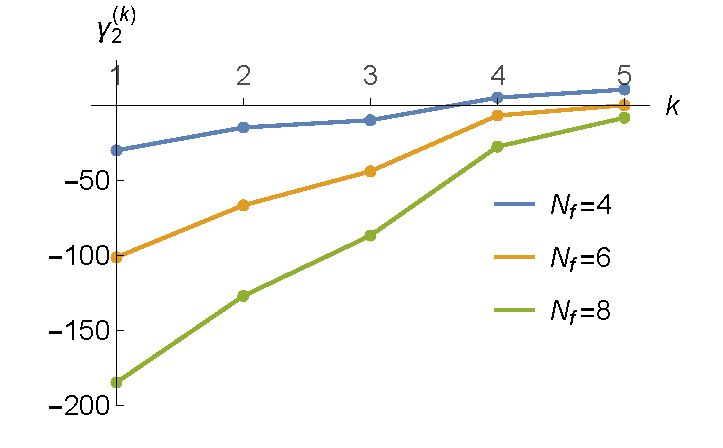}
\hspace{-2mm}
\includegraphics[scale=0.92]{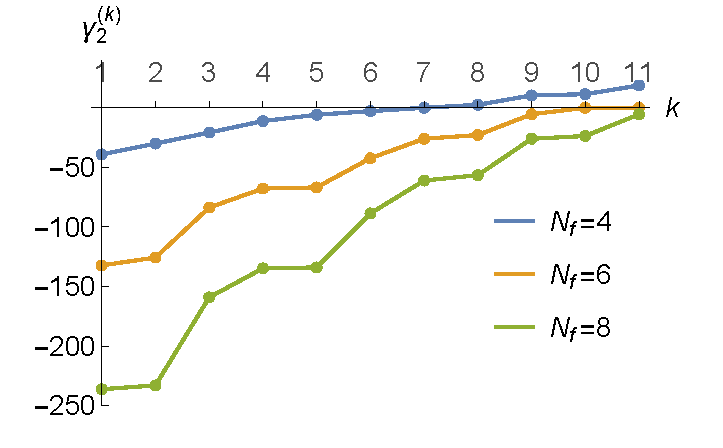}
\caption{The one-loop dimensions of the $\alg{so}(N_f)$ singlet large $N_c$ zero modes with $L=8,10$ at $N_c=10$. Blue, orange, green curves correspond to $N_f=4,6,8$, respectively.}
\label{fig:low-lying}
\end{center}
\end{figure}

Let us outline two properties of the spectrum of the submixing Hamiltonian at general $N_f$\,.
First, $\gamma_2$ decrease as $N_f$ increases,
\begin{equation}
\frac{d \gamma_2}{d N_f} < 0, \qquad (\gamma_2 \le 0, \ N_f \sim 6).
\end{equation}
It shows that the leading $1/N_c$ corrections to the large $N_c$ zero modes can be positive when $N_f < 6$ as shown in Figure \ref{fig:low-lying}.

\begin{figure}[t]
\begin{center}
\includegraphics[scale=1]{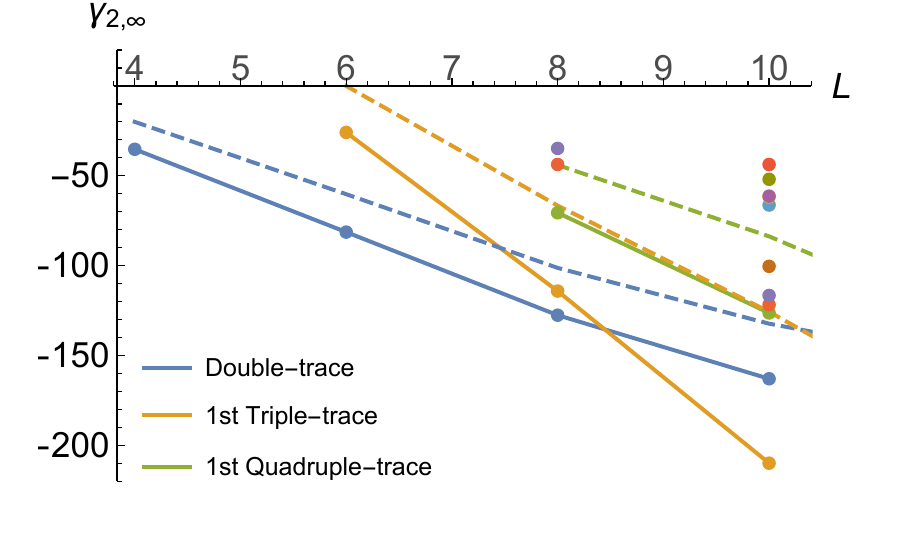}
\caption{Plot of $\gamma_{2,\infty} \equiv \gamma_2 \, (6/N_f)^2$ for large $N_c$ zero modes at $N_f=100$. The dashed lines represent the $N_f=6$ data.}
\label{fig:lam2 Nf}
\end{center}
\end{figure}

Second, all eigenvalues of the submixing matrix scale as $(N_f/N_c)^2$ in the large $N_f$ limit,\footnote{The limit $N_f \gg 1$ does not commute with $N_c \gg 1$, so we consider $N_c \gg N_f \gg 1$.}
\begin{equation}
\gamma = \frac{\gamma_2}{N_c^2} + \cO(N_c^{-3}) 
\sim c (L) \, \frac{N_f^2}{N_c^2} , \qquad (\gamma_2 < 0),
\label{LNZM scaling}
\end{equation}
with some coefficient $c(L)$. This scaling behavior can be seen from the submixing matrix at $L=4,6,8$ given in \eqref{L=4 neg largeNc}, \eqref{L=6 pm-Mat}, \eqref{L=8 pm-Mat}.
Moreover, the large $N_f$ approximation is numerically not bad. 
Figure \ref{fig:lam2 Nf} shows that the rescaled eigenvalues $\gamma_{2,\infty} = \gamma_2 \(6/N_f\)^2$ do not deviate a lot from the $N_f=6$ data.

\section{Correlation functions at large $N_f$}
\label{app:large_Nf_correlator}

We explain the details of the computation in Section \ref{sec:correlator}.

\subsection{A proof of $\langle \psi_0 \fD_1 \psi_0 \rangle<0$ at large $N_f$}
\label{proof_negative_firstterm_lambda2}

In this subsection, we will show that 
the first term of $\gamma_2$, 
$\langle \psi_0 \fD_1 \psi_0 \rangle$, is always non-positive  
by making use of the limit 
$N_f \gg L$.\footnote{
In general $\psi_0$ is given by a linear combination of 
the large $N_c$ zero modes as $\psi_0=\sum_i a_i \psi_0^{(i)}$. 
Because off-diagonal expectation values 
$\langle \psi_0^{(i)}\fD_1 \psi_0^{(j)}\rangle$ ($i\neq j$) cannot have 
the leading power with respect to both 
$1/N_c$ and $1/N_f$, we may consider only 
diagonal expectation values $\langle \psi_0^{(i)}\fD_1 \psi_0^{(i)}\rangle$.
} 

Since $\fD_0 \psi_0 = 0$, we can consider $\langle \psi_0 \fD_\text{one-loop} \psi_0 \rangle$.
For convenience, we rename each term of the one-loop dilatation operator by 
\begin{eqnarray}
\fD_\text{one-loop} =
\frac{1}{N_c} :\! \left(
D^{(1)}+D^{(2)}
\right) \! :
\end{eqnarray}
where 
\begin{eqnarray}
D^{(1)}=-\frac{1}{2} \tr [\Phi_{m},\Phi_{n}][\check{\Phi}^{m},\check{\Phi}^{n}], 
\quad 
D^{(2)} = -\frac{1}{4} \tr [\Phi_{m},\check{\Phi}^{n}] [\Phi_{m},\check{\Phi}^{n}],
\end{eqnarray}
and 
\begin{eqnarray}
D^{(2)} = D^{(21)}+D^{(22)}
=
-\frac{1}{2} \tr (\Phi_{m}\check{\Phi}^{n}\Phi_{m}\check{\Phi}^{n})
+\frac{1}{2} \tr (\Phi_{m}\Phi_{m}\check{\Phi}^{n}\check{\Phi}^{n}).
\end{eqnarray}

Suppose that 
$D^{(2)}$ acts on 
\begin{eqnarray}
\varphi\equiv \tr(\Phi_{(b_1}\Phi_{b_2}\cdots \Phi_{b_s}\Phi_{a_{1}}\cdots \Phi_{a_{p})}) \,
\tr(\Phi_{(b_1}\Phi_{b_2}\cdots \Phi_{b_s}\Phi_{c_{1}}\cdots \Phi_{c_{q})}), 
\end{eqnarray} 
where one derivative in $D^{(2)}$ acts on one of the $\Phi$'s 
in the first trace 
and the other derivative acts on one of the $\Phi$'s in the second trace. 
We denote the remaining traces of $\psi_0$ by $\tilde{\varphi}$, i.e. 
$\psi_{0}=\varphi \tilde{\varphi}$. 

We can find that the leading behavior of $N_f$ in 
the correlator $\langle \psi_0 \fD_1 \psi_0 \rangle$ is $N_f^{L/2+1}$. 
The leading power arises if the following two conditions are satisfied:
(a) 
the two derivatives in 
$\fD_\text{one-loop}$ act on two $\Phi_{b_i}$'s, giving 
$\delta_{b_i b_i}=N_f$ and replacing them with 
$\Phi_{m}$'s, (b)
when a $\Phi_{b_j}$(or $\Phi_{m}$)
is Wick-contracted with a $\Phi_{a_k}$,  
the other $\Phi_{b_j}$(or $\Phi_{m}$) is also Wick-contracted with the other $\Phi_{a_k}$\,,
giving $N_f^{L/2}$.  
Because 
the factor $N_f^{L/2+1}$ cannot arise from the action of $D^{(1)}$, 
only the action of $D^{(2)}$ may be considered in what follows. 
At large $N_f$, the traceless part of $\psi_0$ does not give 
the leading contribution.

When 
the conditions (a) and (b) are satisfied, 
the correlator can be reduced as 
\begin{eqnarray}
\langle 
\varphi \tilde{\varphi} (\fD\varphi) \tilde{\varphi}
\rangle 
=
\langle 
 \varphi (\fD\varphi) 
\rangle 
\langle 
\tilde{\varphi}  \tilde{\varphi}
\rangle \rightarrow \langle 
\varphi (\fD\varphi) 
\rangle .
\end{eqnarray} 
At the last step, we performed Wick-contractions in 
$\langle \tilde{\varphi} \tilde{\varphi} \rangle$ 
and simplified the flavor indices using the $\delta$'s.
In the reduced correlator, all flavor indices are contracted.
Then, the dilatation operator makes a single trace from 
$\varphi$, removing two $\Phi_{b_i}$'s and putting two $\Phi_{m}$'s. 
More precisely, $D^{(21)}$ put two $\Phi_{m}$'s away from each other while $D^{(22)}$ put them next to each other, as shown in (\ref{D-action 2a}) and (\ref{D-action 2c}).
When two $\Phi_{m}$'s are adjacent in the reduced correlator, we have a smaller 
symmetry factor than the non-adjacent case.
This difference in symmetry factor is the reason the correlator $\langle \psi_0 \fD_1 \psi_0 \rangle$ is non-positive. 

The following is an example. 
Consider a reduced correlator
\begin{multline}
\frac{1}{s}
\langle 
:\tr(\Phi_{a_1}\Phi_{a_2}\cdots \Phi_{a_s}) \,
\tr(\Phi_{a_1}\Phi_{a_2}\cdots \Phi_{a_s} ):
D^{(2)}:
\tr(\Phi_{b_1}\Phi_{b_2}\cdots \Phi_{b_s}) \,
\tr(\Phi_{b_1}\Phi_{b_2}\cdots \Phi_{b_s} ):
\rangle . 
\label{correlator_negative_example_p0q0}
\end{multline}
This is a case of $p=0$ and $q=0$.
It has the following two contributions:\footnote{
The symmetry factor $2s$ in 
(\ref{correlator_negative_example_p0q0_D21}) can be explained as follows. 
When $a_1$ in the first trace is Wick-contracted with 
the first $m$, $a_1$ in the second trace must be Wick-contracted with 
the second $m$. 
To keep the planarity, we should Wick-contract the first $a_2$ with the first $b_2$, 
and the second $a_2$ with the second $b_2$, and so on.  
We say this that  
$(a_1,a_2\cdots,a_s)$ are Wick-contracted with $(m,b_2,\cdots,b_s)$. 
Likewise $(a_1,a_2\cdots,a_s)$ can be Wick-contracted with 
$(b_{k+1},\cdots,b_s,m,b_2,\cdots,b_k)$ for any $k$ while keeping the planarity. 
In this way, we get the factor $s$. 
The factor $2$ comes from exchanging two single traces. }
\begin{eqnarray}
&&
\frac{1}{s}
\langle 
:\tr(\Phi_{a_1}\cdots \Phi_{a_s}) \,
\tr(\Phi_{a_1}\cdots \Phi_{a_s} ):
D^{(21)}:
\tr(\Phi_{b_1}\cdots \Phi_{b_s}) \,
\tr(\Phi_{b_1}\cdots \Phi_{b_s} ):
\rangle
\nonumber \\ 
&=&
-N_f
\langle 
:\tr(\Phi_{a_1}\cdots \Phi_{a_s})
\tr(\Phi_{a_1}\cdots \Phi_{a_s} ):
:\tr(\Phi_{m}\Phi_{b_2}\cdots \Phi_{b_s}
\Phi_{m}\Phi_{b_2}\cdots \Phi_{b_{s}} ):
\rangle
\nonumber \\ 
&=&
-2sN_c^{2s-1} N_f^{s+1} + \cO(N_c^{2s-2}) , 
\label{correlator_negative_example_p0q0_D21}
\end{eqnarray} 
and\footnote{The symmetry factor 
(\ref{correlator_negative_example_p0q0_D22}) comes from 
the Wick-contractions between 
$(a_1,a_2\cdots,a_s)$ and $(m, b_2,\cdots,b_s)$ only. 
It can be more manifest if we write the correlator as \\
$\langle 
:\tr(\Phi_{a_1}\Phi_{a_2}\cdots \Phi_{a_s}) \,
\tr(\Phi_{a_2}\cdots \Phi_{a_s}\Phi_{a_1}): \,
:\tr(\Phi_{m}\Phi_{b_2}\cdots \Phi_{b_s}
\Phi_{b_2}\cdots \Phi_{b_s}\Phi_{m}): 
\rangle$.}
\begin{eqnarray}
&&\frac{1}{s}
\langle 
:\tr(\Phi_{a_1}\Phi_{a_2}\cdots \Phi_{a_s}) \,
\tr(\Phi_{a_1}\Phi_{a_2}\cdots \Phi_{a_s}):
D^{(22)}:
\tr(\Phi_{b_1}\Phi_{b_2}\cdots \Phi_{b_s}) \,
\tr(\Phi_{b_1}\Phi_{b_2}\cdots \Phi_{b_s}):
\rangle
\nonumber \\ 
&=& N_f
\langle 
:\tr(\Phi_{a_1}\Phi_{a_2}\cdots \Phi_{a_s}) \,
\tr(\Phi_{a_1}\Phi_{a_2}\cdots \Phi_{a_s}): \,
:\tr(\Phi_{m}\Phi_{m} \Phi_{b_2}\cdots \Phi_{b_s} \, 
\Phi_{b_2}\cdots \Phi_{b_s} ): 
\rangle
\nonumber \\ 
&=&
2N_c^{2s-1} N_f^{s+1} + \cO(N_c^{2s-2}), 
\label{correlator_negative_example_p0q0_D22}
\end{eqnarray} 
where we have kept only the leading term with respect to $N_f$ and $N_c$. 
Because $s>1$, the correlator (\ref{correlator_negative_example_p0q0}) 
is negative.


\subsection{$N_f$ and $L$ dependence of $\gamma_2$}
\label{lambda2_double_trace_order_estimate}

In this subsection, we will consider the $1/N_f$ expansion to 
study the $L$-dependence of 
$\gamma_2$
for the operators presented in Section \ref{define_correlator_group_theory}.
General properties of $\gamma_2$ will be first discussed, and then 
concrete examples are shown. 

Recall the expression of $\gamma_2$:
\begin{eqnarray}
\gamma_2=\frac{
N_c \langle \psi_0 \fD_1 \psi_0 \rangle 
}{\langle \psi_0 \psi_0 \rangle } 
-
\frac{
\langle \psi_1 \fD_0 \psi_1 \rangle 
}{\langle \psi_0 \psi_0 \rangle } .
\end{eqnarray} 
As we discussed in the last subsection, 
the first term behaves as $O(N_f)$, and as we will see below   
the second term has a more dominant behavior $O(N_f^2)$. 

The relation between $\psi_0$ and $\psi_1$ is given by 
the equation $\fD_0 \psi_1+\fD_1\psi_0 \sim 0$ \footnote{
$\eta/N_c$ in (\ref{constraint_for_psi_1_operator}) is not important at the leading order of the 
$1/N_c$ expansion in $\gamma_2$, so we can use $\fD_0 \psi_1+\fD_1\psi_0 \sim 0$. 
} obtained in (\ref{constraint_for_psi_1_operator}), which can rewrite 
the second term as 
\begin{eqnarray}
-
\frac{
\langle \psi_1 \fD_0 \psi_1 \rangle 
}{\langle \psi_0 \psi_0 \rangle } 
\sim 
\frac{
\langle \psi_1 \fD_1 \psi_0 \rangle 
}{\langle \psi_0 \psi_0 \rangle }. 
\end{eqnarray} 
Focusing only on the leading of the $1/N_f$ expansion, 
we find that 
$\fD_1\psi_0$ can be expanded as 
\begin{eqnarray}
\fD_1\psi_0 \sim 
N_f  F_i
+
N_f  G_i +O(1), 
\label{D1_on_psi0_appendix}
\end{eqnarray}
where $O(1)$ is multi-traces without the factor $N_f$. 
$F_i$ is a multi-trace that does not contain two adjacent matrices with the flavor indices contracted, 
and 
$G_i$ is a multi-trace that does contain such matrices. 
They appeared in the image of $D^{(21)}$ and $D^{(22)}$ respectively; see the second lines of \eqref{correlator_negative_example_p0q0_D21} and \eqref{correlator_negative_example_p0q0_D22} as well as (\ref{D_2_on_double_trace}) for the double-trace operator.
The $O(1)$ comes from the action of $D^{(1)}$.

The planar dilatation operator has the form 
$\fD_0 =I-P+C/2$, where $I$, $P$ and $C$ are the identity, the transposition and the contraction  
acting on nearest neighbor matrices. 
Only the contraction can bring a factor of $N_f$\,.
In order to determine the form of $\psi_1$,  
we need to solve $\fD_0 \psi_1\sim -N_f(F_i+G_i)$. 
Since $C \cdot F_i \sim G_i$ and $C \cdot G_i \sim N_f G_i$\,, the solution is roughly written as
\begin{eqnarray}
\psi_1 \sim 
-\alpha N_f  F_i -\beta G_i + \cdots, 
\end{eqnarray}
where $\alpha$ and $\beta$ are $N_f$-independent coefficients.  
Substituting these equations into the form of $\gamma_2$, 
we have 
\begin{eqnarray}
\gamma_2 =
-N_f^2  \alpha
\frac{
\langle F_i F_i \rangle 
}{\langle \psi_0 \psi_0 \rangle } +O(N_f) ,
\end{eqnarray} 
where 
the ratio of the two-point functions is independent of $N_f$. 
The $N_f^2$ behavior of $\gamma_2$ is consistent with (\ref{LNZM scaling}). 
It is emphasized that $G_i$ in (\ref{D1_on_psi0_appendix}) does not matter 
for the evaluation of $\gamma_2$ at the leading order.

\bigskip
Our next goal is to estimate the $L$-dependence of the correlators. The double-trace operator and $L/2$-trace operators are studied below.

Consider 
the singlet double trace operator (\ref{define_double_trace_operator}). 
We will use the property of the symmetric representation 
$\langle \sigma(\vec{a})|P_{[L/2]}|\vec{b}\rangle
=\langle \vec{a}|\sigma P_{[L/2]}|\vec{b}\rangle=
\langle \vec{a}|P_{[L/2]}|\vec{b}\rangle$, and 
\begin{eqnarray} 
\delta_{c_{L/2},d_{L/2}}
\langle \vec{c}|P_{[L/2]}|\vec{d}\rangle
&\sim& \frac{1}{L/2}\left(N_f+L/2-1\right) \langle \vec{c}_{L/2-1}|P_{[L/2-1]}|\vec{d}_{L/2-1}\rangle
\nonumber \\
&\sim& \frac{1}{L/2}N_f\langle \vec{c}_{L/2-1}|P_{[L/2-1]}|\vec{d}_{L/2-1}\rangle,
\end{eqnarray} 
where $\vec{c}_{L/2-1}=(c_1,\cdots,c_{L/2-1})$, and the factor $\frac{1}{L/2}$ comes  
from the normalization of the projector. 

Acting with $D^{(2)}$ on $\psi_0$, we have 
\begin{eqnarray}
D^{(21)}\psi_0
&=&
-\frac{1}{2} \tr(\Phi_{m}\check{\Phi}_n\Phi_{m}\check{\Phi}_n)\psi_0  
\nonumber \\
&=&-\left(\frac{L}{2}\right)^2\delta_{c_{L/2},d_{L/2}}\langle \vec{c}|P_{[L/2]}|\vec{d}\rangle
\tr(
\Phi_{m}
\Phi_{c_1} \cdots  \Phi_{c_{L/2-1}}\Phi_{m}
\Phi_{d_1} \cdots  \Phi_{d_{L/2-1}})
\nonumber \\
&=&-\frac{L}{2}N_f \langle \vec{c}_{L/2-1}|P_{[L/2-1]}|\vec{d}_{L/2-1}\rangle
\tr(
\Phi_{m}
\Phi_{c_1} \cdots  \Phi_{c_{L/2-1}}\Phi_{m}
\Phi_{d_1} \cdots  \Phi_{d_{L/2-1}})
\nonumber \\
&=&-\frac{L}{2}N_f 
\tr(
\Phi_{m}
\Phi_{(c_1} \cdots  \Phi_{c_{L/2-1})}\Phi_{m}
\Phi_{(c_1} \cdots  \Phi_{c_{L/2-1})}),
\nonumber \\
&\equiv &-\frac{L}{2}N_f 
f_1,
\nonumber \\
D^{(22)}\psi_0
&=&
\frac{1}{2} \tr(\Phi_{m}\Phi_{m}\check{\Phi}_n\check{\Phi}_n)\psi_0  
\nonumber \\
&=&\frac{L}{2}N_f 
\tr(
\Phi_{m}\Phi_{m}
\Phi_{(c_1} \cdots  \Phi_{c_{L/2-1})}
\Phi_{(c_1} \cdots  \Phi_{c_{L/2-1})})
\nonumber \\
&\equiv &\frac{L}{2}N_f 
f_2.
\label{D_2_on_double_trace}
\end{eqnarray} 
The multi-traces $f_1$ and $f_2$ belong to $F_i$ and $G_i$ respectively. 
Note that the flavor indices are in the irreducible representation $[L/2-1]\otimes [1]$, 
where $[p]$ represents the rank-$p$ symmetric traceless representation.

We have to solve the equation $\fD_0 \psi_1+\fD_1\psi_0  \sim 0$. 
Suppose that the solution is given by
\begin{eqnarray}
\psi_1 \sim
\frac{L}{2}N_f \alpha f_1 +\cdots  ,
\label{psi_1_form_double}
\end{eqnarray}
where $\alpha$ is an $N_f$-independent positive constant that may depend on $L$. 
Other terms are denoted by $\cdots$, expecting that they do not change the $L$-dependence at the leading order of the $1/N_f$ expansion.
We will leave it as a future problem to determine $\psi_1$. We then have 
\begin{eqnarray}
\gamma_2\sim
-\frac{N_f^2 L^2 }{4} \alpha
\frac{
\langle f_1 f_1 \rangle 
}{\langle \psi_0 \psi_0 \rangle } +O(N_f) .
\end{eqnarray} 
The normalization was computed in 
(\ref{normalisation_double_trace}), which 
is evaluated at large $N_f$ as
\footnote{
The dimension of the rank-$L/2$ traceless symmetric representation of 
$\alg{so}(N_f)$ is 
\begin{eqnarray}
{\rm Dim}_{L/2}
=\frac{1}{(N_f-1)!}\frac{(L/2+N_f-3)!}{(L/2)!}(N_f-1)(L+N_f-2)\sim \frac{N_f^{L/2}}{(L/2)!},
\end{eqnarray}
where we have approximated using Stirling's formula for $N_f\gg L$. 
The result (\ref{normalisation_psi0_large_N_f}) can also be obtained  
by considering only Wick-contractions that satisfy the condition (b)
in the last subsection. }
\begin{eqnarray}
\langle \psi_0 \psi_0\rangle 
= \frac{L^2}{2} \, {\rm Dim}_{L/2} \, N_c^{L}
\sim \frac{L^2}{2} \frac{N_f^{L/2}}{(L/2)!} \, N_c^{L}, 
\label{normalisation_psi0_large_N_f}
\end{eqnarray} 
and 
we also have 
\begin{eqnarray}
&&
\langle f_1 f_1\rangle
\sim 
L \frac{N_f^{L/2}}{(L/2-1)!}N_c^{L} . 
\label{two_pt_f1f1}
\end{eqnarray} 
to obtain 
\begin{eqnarray}
\gamma_2\sim 
-\alpha\frac{N_f^2 L^2 }{4} +O(N_f) .
\label{appendix_doubletrace_gamma2}
\end{eqnarray} 
Therefore the double-trace operator scale as $\gamma_2 \sim \alpha L^2$.

\quad

The next study is about $L/2$-trace operators such as 
$\cC_{ij}\cC_{jk}\cC_{kl}\cdots \cC_{mi}$ and 
$\cC_{ij}\cC_{ji}\cC_{kl}\cdots \cC_{mi}$. 
These multi-traces have the same color structure, but the  
flavor indices are contracted differently. 
As we mentioned 
in Section \ref{define_correlator_group_theory}, 
such multi-traces can be  
classified by the partitions of $L/2$, and we write them by $\psi_0^{(T)}$, 
where $T$ expresses a partition of $L/2$.

In general, a submixing eigenstate $\psi_0$ is a linear combination of $\psi_0^{(T)}$'s. This situation simplifies at large $N_f$\,, where the two-point functions are orthogonal $\langle  \psi_0^{(T)} \psi_0^{(T^{\prime})} \rangle \propto \delta_{TT'}$ and the submixing Hamiltonian acts diagonally on $\psi_0^{(T)}$. Then $\gamma_2$ is expressed by
\begin{eqnarray}
\gamma_2^{(T)}=
- 
\frac{
\langle \psi_1^{(T)} \fD_0 \psi_1^{(T)} \rangle 
}{ \langle  \psi_0^{(T)} \psi_0^{(T)} \rangle } +O(N_f) .
\label{gamma2_L/2traces}
\end{eqnarray}

For $\psi_0^{(L/2)}= \cC_{ij}\cC_{jk}\cC_{kl}\cdots \cC_{mi}=\cC^{[L/2]}$,\footnote{
Introduced the notation $\cC^{[p]}\equiv \cC_{a_{1}a_{2}}\cC_{a_{2}a_{3}}\cC_{a_{3}a_{4}}\cdots \cC_{a_{p}a_{1}}$.
}  
we obtain  
\begin{eqnarray}
D^{(21)}\psi_0^{(L/2)} =-\frac{L}{2}N_f 
\tr(\Phi_{m}\Phi_{i}\Phi_{m}\Phi_{k}) \,
\cC_{kl} \cC_{ls} \cdots \cC_{ti}\equiv  -\frac{L}{2}N_f f_1^{(L/2)}.
\end{eqnarray}
The two-point functions can be computed 
\begin{eqnarray}
&&
\langle \psi_0^{(L/2)} \psi_0^{(L/2)}\rangle \sim L N_f ^{L/2}N_c^{L}, 
\nonumber \\
&&
\langle f_1^{(L/2)} f_1^{(L/2)}\rangle\sim 
2(1+\delta_{L,4})N_f ^{L/2}N_c^{L}
\end{eqnarray}
to obtain 
\begin{eqnarray}
\gamma_2^{(L/2)}= 
- \frac{N_f^2}{2} (1+\delta_{L,4}) \alpha_{(L/2)} L +O(N_f), 
\label{appendix_multitrace_gamma2}
\end{eqnarray} 
where $\alpha_{(L/2)}$ is determined by the equation 
\begin{equation}
\fD_0 \psi_1+\fD_1 \psi_0 \approx 0 \qquad \Longrightarrow \qquad
\psi_1^{(L/2)} = \frac{N_f}{2} \, \alpha_{(L/2)} L f_1^{(L/2)}+\cdots.
\end{equation}
The two expressions 
(\ref{appendix_doubletrace_gamma2}) and (\ref{appendix_multitrace_gamma2})
agree at $L=4$ as expected. 
The equation \eqref{appendix_multitrace_gamma2} shows that the $L/2$-trace scale like $\gamma_2 \sim \alpha_{(L/2)} L$ as a function of $L$.

Consider 
more general $L/2$-trace operators  
$\psi_0^{(L_1/2,\cdots,L_p/2)}= \cC^{[L_1/2]}\cC^{[L_2/2]}\cdots \cC^{[L_p/2]}$, where $L=\sum_{s}L_s$, 
and we assume $L_i \neq L_j \ (i\neq j)$ for simplicity. 
We then have 
\begin{eqnarray}
D^{(21)}\psi_0^{(L_1/2,\cdots,L_p/2)} &=&
-
\sum_s \frac{L_s}{2}N_f 
\cC^{[L_1/2]}\cC^{[L_2/2]} \cdots \cC^{[L_{s-1}/2]}
\nonumber \\
&& \quad \times 
\left( \tr(\Phi_{m}\Phi_{a_1}\Phi_{m}\Phi_{a_3})
\cC_{a_3a_4}\cdots \cC_{a_{L_s/2}a_1}\right)
\cC^{[L_{s+1}/2]}\cdots \cC^{[L_{p}/2]}
\nonumber \\
& \equiv & -\sum_s \frac{L_s}{2}N_f f_1^{(L_1/2,\cdots,L_p/2;s)},
\end{eqnarray}
and 
\begin{eqnarray}
&&
\langle \psi_0^{(L_1/2,\cdots,L_p/2)} \psi_0^{(L_1/2,\cdots,L_p/2)}\rangle \sim 
L_T N_f ^{L/2}N_c^{L}, 
\nonumber \\
&&
\langle f_1^{(L_1/2,\cdots,L_p/2;s)}f_1^{(L_1/2,\cdots,L_p/2;s)}
\rangle\sim 
2(1+\delta_{L_s,4})L_1 \cdots L_{s-1}L_{s+1}\cdots L_p N_f ^{L/2}N_c^{L}
\nonumber \\
&&
\hspace{5.4cm}
=
2(1+\delta_{L_s,4})\frac{L_T}{L_s}N_f ^{L/2}N_c^{L},
\end{eqnarray}
where $L_T \equiv L_1L_2\cdots L_p$. 
Substituting these two-point functions into the 
expression of $\gamma_2$, we will obtain 
\begin{eqnarray}
\gamma_2^{(T)}&\sim&
- \frac{N_f^2}{4}
\frac{
\sum_s L_s^2 \alpha_{(T;s)}\langle  f_1^{(T;s)}f_1^{(T;s)}  \rangle
}{\langle  \psi_0^{(T)} \psi_0^{(T)} \rangle } +O(N_f)
\nonumber \\
&= &
- \frac{N_f^2}{2}
\sum_s L_s (1+\delta_{L_s,4}) \alpha_{(T;s)} +O(N_f), 
\label{gamma_2_L/2_appendix}
\end{eqnarray} 
where $\alpha_{(T;s)}$ is another positive constant in 
$\psi_1^{(T)} =(N_f/2)\sum_s\alpha_{(T;s)}L_s f_1^{(T;s)}+\cdots$,  
which is determined by the equation 
$\fD_0 \psi_1+\fD_1 \psi_0=0$.
We leave it for a future problem to determine the precise $L$ dependence.


\section{Relation to Mandelstam Variables}\label{app:Mandelstam}

We pursue the coincidence between the number of $\alg{so}(6)$ singlet large $N_c$ zero-modes in Table \ref{tab:cZL} in Section \ref{sec:LN zero modes} and the number of the completely symmetric polynomial of Mandelstam variables under certain conditions in Table 2 of \cite{Boels13}.
This fact allows us to construct an explicit basis for the latter.

We defined $\cZ_{L}$ as the number of $\alg{so}(6)$ singlet large $N_c$ zero modes with length $L$.
These zero-modes can be written explicitly in terms of the traceless symmetric single-trace operators $\cC_{i_1i_2 \dots i_\ell}$ as
\begin{equation}
\begin{gathered}
\{ \, \},
\\
\{ \cC_{ij} \, \cC_{ij} \},
\\
\{ \cC_{ijk} \, \cC_{ijk} \,, \ 
\cC_{ij} \, \cC_{jk} \, \cC_{ki} \},
\\
\{ \cC_{ijkl} \, \cC_{ijkl} \,, \ 
\cC_{ijkl} \, \cC_{ij} \, \cC_{kl} \,, \ 
\cC_{ijk} \, \cC_{ijl} \, \cC_{kl} \,, \ 
\cC_{ij} \, \cC_{ji} \, \cC_{kl} \, \cC_{lk} \,, \ 
\cC_{ij} \, \cC_{jk} \, \cC_{kl} \, \cC_{li} \},
\end{gathered}
\label{products sym trsless 2-4}
\end{equation}
corresponding to $\{ \cZ_2, \cZ_4, \cZ_6, \cZ_8 \} = \{ 0,1,2,5 \}$.

Another series of positive integers is computed in \cite{Boels13}, which is the number of the completely symmetric polynomial of Mandelstam variables of degree $K$, subject to massless momentum conservation, for a sufficiently large number of particles $n$. If we denote this series by $\cZ'_K$\,, they are given by $\{ \cZ'_1, \cZ'_2, \cZ'_3, \cZ'_4 \} = \{ 0,1,2,5 \}$.
The Mandelstam variables are defined by
\begin{equation}
s_{ij} = \( p_i^\mu + p_j^\mu \)^2 = 2 \, p_i \cdot p_j \,, \qquad 
(i,j=1,2,\dots, n),
\end{equation}
and they satisfy
\begin{equation}
s_{ij} = s_{ji} \,, \qquad
\sum_{j=1}^n s_{ij} = 0, \qquad
\( \sum_{i=1}^n p_i^\mu \)^2 = \sum_{i<j}^n s_{ij} = 0.
\label{Mandelstam constraints}
\end{equation}
A basis of the completely symmetric polynomials of $\{ s_{ij} \}$ can be given as
\begin{equation}
\begin{gathered}
\pare{ \sum_{\sigma} s_{\sigma(1) \sigma(2)}^2 },
\\
\pare{ \sum_{\sigma} s_{\sigma(1) \sigma(2)}^3 \,, \ 
\sum_{\sigma} s_{\sigma(1) \sigma(2)} s_{\sigma(2) \sigma(3)} s_{\sigma(3) \sigma(1)} },
\\
\Bigl\{ \sum_{\sigma} s_{\sigma(1) \sigma(2)}^4 \,, \ 
\sum_{\sigma} s_{\sigma(1) \sigma(2)}^2 s_{\sigma(1) \sigma(3)}^2 \,, \quad
\sum_{\sigma} s_{\sigma(1) \sigma(2)}^2 s_{\sigma(1) \sigma(3)} s_{\sigma(2) \sigma(3)} \,,
\\
\hspace{50mm} 
\sum_{\sigma} s_{\sigma(1) \sigma(2)}^2 s_{\sigma(3) \sigma(4)}^2 \,, \quad
\sum_{\sigma} s_{\sigma(1) \sigma(2)} s_{\sigma(2) \sigma(3)} s_{\sigma(3) \sigma(4)} s_{\sigma(4) \sigma(1)} \Bigr\},
\end{gathered}
\label{Mandelstam poly 2-4}
\end{equation}
where we sum $\sigma$ over the permutation group of $n$-th order $\cS_n$\,. The constraints \eqref{Mandelstam constraints} must be imposed after the summation.

The relation between \eqref{products sym trsless 2-4} and \eqref{Mandelstam poly 2-4} can be explained graphically. Note that we neglect dimensionality constraints; namely the finite $N_f$ constraints in \eqref{products sym trsless 2-4} and the Gram determinant constraints in \eqref{Mandelstam poly 2-4}.\footnote{The latter comes from the linear relations among $\{ p_i^\mu \}$ when $n$ is greater than the spacetime dimensions.}

We begin by the $\alg{so}(6)$ singlet large $N_c$ zero modes. 
Since $\cC_{i_1i_2 \dots i_\ell}$ is symmetric traceless, a flavor index of $\cC^{(\ell_i)}$ should be paired with the flavor index of another $\cC^{(\ell_j)}$. Since $\cC^{(\ell)}$ is symmetric, the position of the flavor index inside $\cC^{(\ell)}$ is irrelevant.
When we contract $m$ indices of $\cC^{(\ell_i)}$ and $\cC^{(\ell_j)}$, we draw $m$ lines in between as
\ifusetikz
\begin{equation}
\cC_{i_1i_2} \cC_{i_1 i_2} =
\begin{tikzpicture}[baseline=-\the\dimexpr\fontdimen22\textfont2\relax]
\node (r1) at (0,1.2) [rectangle, minimum width=8mm] {$C^{(2)}$};
\node at (0,1.2) [draw, rectangle, minimum width=16mm] {$C^{(2)}$};
\node (r2) at (0,-1.2) [rectangle, minimum width=8mm] {$C^{(2)}$};
\node at (0,-1.2) [draw, rectangle, minimum width=16mm] {$C^{(2)}$};
\node at (-0.9,0.6) {$i_1$};
\node at (-0.9,-0.55) {$i_1$};
\node at (0.9,0.6) {$i_2$};
\node at (0.9,-0.55) {$i_2$};
\draw (r1.south west) to (r2.north west);
\draw (r1.south east) to (r2.north east);
\end{tikzpicture}
\ = \ 
\begin{tikzpicture}[baseline=-\the\dimexpr\fontdimen22\textfont2\relax]
\node (r1) at (0,1) [rectangle, minimum width=8mm] {};
\node at (0,1) [draw, fill=black!50, rectangle, minimum width=16mm] {};
\node (r2) at (0,-1) [rectangle, minimum width=8mm] {};
\node at (0,-1) [draw, fill=black!50, rectangle, minimum width=16mm] {};
\draw (r1.south west) to (r2.north west);
\draw (r1.south east) to (r2.north east);
\end{tikzpicture} \ .
\end{equation}
\else
\begin{equation}
\cC_{i_1i_2} \cC_{i_1 i_2} = \adjustbox{valign=c}{\includegraphics{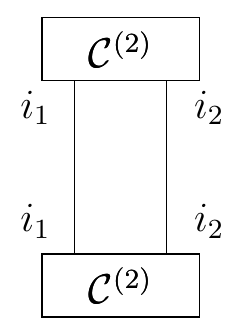}} 
\ = \ \adjustbox{valign=c}{\includegraphics{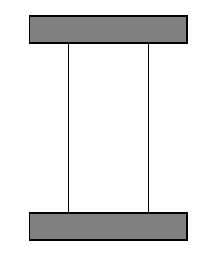}} \ .
\end{equation}
\fi
Similarly, the zero modes with length six are expressed as
\ifusetikz
\begin{equation}
\cC_{i_1i_2i_3} \cC_{i_1 i_2 i_3} =
\begin{tikzpicture}[baseline=-\the\dimexpr\fontdimen22\textfont2\relax, scale=0.6,
 every node/.style={scale=0.6}]
\node (r1) at (0,1) [rectangle, minimum width=10mm] {};
\node at (0,1) [draw, fill=black!50, rectangle, minimum width=18mm] {};
\node (r2) at (0,-1) [rectangle, minimum width=10mm] {};
\node at (0,-1) [draw, fill=black!50, rectangle, minimum width=18mm] {};
\draw (r1.south west) to (r2.north west);
\draw (r1) to (r2);
\draw (r1.south east) to (r2.north east);
\end{tikzpicture} \ , \qquad
\cC_{i_1 i_2} \cC_{i_2 i_3} \cC_{i_3 i_1} =
\begin{tikzpicture}[baseline=-\the\dimexpr\fontdimen22\textfont2\relax, scale=0.6,
 every node/.style={scale=0.6}]
\node (r1) at (0,1) [rectangle, minimum width=5mm] {};
\node at (0,1) [draw, fill=black!50, rectangle, minimum width=12mm] {};
\node (r2) at (-1.2,-0.8) [minimum width=5mm, rotate=120] {};
\node at (-1.2,-0.8) [draw, fill=black!50, rectangle, minimum width=12mm, rotate=120] {};
\node (r3) at (1.2,-0.8) [rectangle, minimum width=5mm, rotate=240] {};
\node at (1.2,-0.8) [draw, fill=black!50, rectangle, minimum width=12mm, rotate=240] {};
\draw[bend left=15] (r1.south west) to (r2.south east);
\draw[bend left=15] (r2.south west) to (r3.south east);
\draw[bend left=15] (r3.south west) to (r1.south east);
\end{tikzpicture} \ ,
\end{equation}
\else
\begin{equation}
\cC_{i_1i_2i_3} \cC_{i_1 i_2 i_3} = \adjustbox{valign=c}{\includegraphics{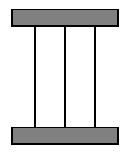}} \ , \qquad
\cC_{i_1 i_2} \cC_{i_2 i_3} \cC_{i_3 i_1} = \adjustbox{valign=c}{\includegraphics{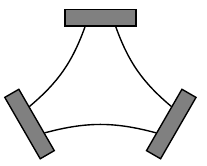}} \ ,
\end{equation}
\fi
and with length eight as,
\ifusetikz
\begin{equation}
\begin{tikzpicture}[baseline=-\the\dimexpr\fontdimen22\textfont2\relax, scale=0.6,
 every node/.style={scale=0.6}]
\node (r1) at (0,1) [rectangle, minimum width=7mm] {};
\node (r2) at (0,1) [rectangle, minimum width=18mm] {};
\node at (0,1) [draw, fill=black!50, rectangle, minimum width=24mm] {};
\node (r3) at (0,-1) [rectangle, minimum width=7mm] {};
\node (r4) at (0,-1) [rectangle, minimum width=18mm] {};
\node at (0,-1) [draw, fill=black!50, rectangle, minimum width=24mm] {};
\draw (r1.south west) to (r3.north west);
\draw (r2.south west) to (r4.north west);
\draw (r1.south east) to (r3.north east);
\draw (r2.south east) to (r4.north east);
\end{tikzpicture} \ , \quad
\begin{tikzpicture}[baseline=-\the\dimexpr\fontdimen22\textfont2\relax, scale=0.6,
 every node/.style={scale=0.6}]
\node (r1) at (0,1) [rectangle, minimum width=7mm] {};
\node (r2) at (0,1) [rectangle, minimum width=18mm] {};
\node at (0,1) [draw, fill=black!50, rectangle, minimum width=24mm] {};
\node (r3) at (-1.2,-0.8) [minimum width=5mm, rotate=120] {};
\node at (-1.2,-0.8) [draw, fill=black!50, rectangle, minimum width=12mm, rotate=120] {};
\node (r4) at (1.2,-0.8) [rectangle, minimum width=5mm, rotate=240] {};
\node at (1.2,-0.8) [draw, fill=black!50, rectangle, minimum width=12mm, rotate=240] {};
\draw[bend left=15] (r1.south west) to (r3.south west);
\draw[bend left=15] (r2.south west) to (r3.south east);
\draw[bend right=15] (r1.south east) to (r4.south east);
\draw[bend right=15] (r2.south east) to (r4.south west);
\end{tikzpicture} \ , \quad
\begin{tikzpicture}[baseline=-\the\dimexpr\fontdimen22\textfont2\relax, scale=0.6,
 every node/.style={scale=0.6}]
\node (r1) at (0,1) [rectangle, minimum width=5mm] {};
\node at (0,1) [draw, fill=black!50, rectangle, minimum width=12mm] {};
\node (r2) at (-1.2,-0.8) [minimum width=10mm, rotate=120] {};
\node at (-1.2,-0.8) [draw, fill=black!50, rectangle, minimum width=18mm, rotate=120] {};
\node (r3) at (1.2,-0.8) [rectangle, minimum width=10mm, rotate=240] {};
\node at (1.2,-0.8) [draw, fill=black!50, rectangle, minimum width=18mm, rotate=240] {};
\draw[bend left=15] (r1.south west) to (r2.south east);
\draw[bend right=15] (r1.south east) to (r3.south west);
\draw[bend left=15] (r2.south west) to (r3.south east);
\draw[bend left=15] (r2) to (r3);
\end{tikzpicture} \ , \quad
\begin{tikzpicture}[baseline=-\the\dimexpr\fontdimen22\textfont2\relax, scale=0.6,
 every node/.style={scale=0.6}]
\node (r1) at (0,1.2) [rectangle, minimum width=5mm] {};
\node at (0,1.2) [draw, fill=black!50, rectangle, minimum width=12mm] {};
\node (r2) at (-1.2,0) [minimum width=5mm, rotate=90] {};
\node at (-1.2,0) [draw, fill=black!50, rectangle, minimum width=12mm, rotate=90] {};
\node (r3) at (0,-1.2) [rectangle, minimum width=5mm] {};
\node at (0,-1.2) [draw, fill=black!50, rectangle, minimum width=12mm] {};
\node (r4) at (1.2,0) [minimum width=5mm, rotate=270] {};
\node at (1.2,0) [draw, fill=black!50, rectangle, minimum width=12mm, rotate=270] {};
\draw[bend left=15] (r1.south west) to (r2.south east);
\draw[bend left=15] (r2.south west) to (r3.north west);
\draw[bend left=15] (r3.north east) to (r4.south east);
\draw[bend left=15] (r4.south west) to (r1.south east);
\end{tikzpicture} \ , \quad 
\begin{tikzpicture}[baseline=-\the\dimexpr\fontdimen22\textfont2\relax, scale=0.6,
 every node/.style={scale=0.6}]
\node (r1) at (0,1.2) [rectangle, minimum width=5mm] {};
\node at (0,1.2) [draw, fill=black!50, rectangle, minimum width=12mm] {};
\node (r2) at (-1.2,0) [minimum width=5mm, rotate=90] {};
\node at (-1.2,0) [draw, fill=black!50, rectangle, minimum width=12mm, rotate=90] {};
\node (r3) at (0,-1.2) [rectangle, minimum width=5mm] {};
\node at (0,-1.2) [draw, fill=black!50, rectangle, minimum width=12mm] {};
\node (r4) at (1.2,0) [minimum width=5mm, rotate=270] {};
\node at (1.2,0) [draw, fill=black!50, rectangle, minimum width=12mm, rotate=270] {};
\draw[bend left=15] (r1.south west) to (r2.south east);
\draw[bend right=15] (r2.south west) to (r1.south east);
\draw[bend left=15] (r3.north west) to (r4.south west);
\draw[bend right=15] (r4.south east) to (r3.north east);
\end{tikzpicture} \ .
\label{LNZM L=8 graphs}
\end{equation}
\else
\begin{equation}
\adjustbox{valign=c}{\includegraphics{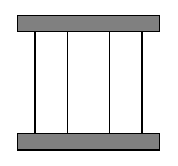}} \ , \quad
\adjustbox{valign=c}{\includegraphics{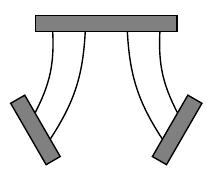}} \ , \quad
\adjustbox{valign=c}{\includegraphics{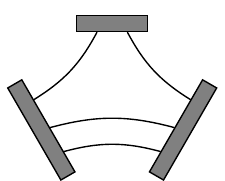}} \ , \quad
\adjustbox{valign=c}{\includegraphics{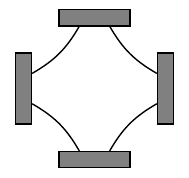}} \ , \quad
\adjustbox{valign=c}{\includegraphics{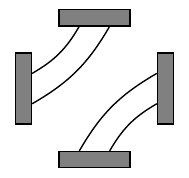}} \ .
\label{LNZM L=8 graphs}
\end{equation}
\fi

\bigskip
Let us turn to the completely symmetric polynomials of Mandelstam variables
Given a graph representing the large $N_c$ zero modes, we label each ``single-trace'' by $i=1,2,\dots, n$. 
Then for each line connecting the $i$-th and $j$-th trace, we associate Mandelstam variable $s_{ij}$\,, as
\ifusetikz
\begin{equation}
\begin{tikzpicture}[baseline=-\the\dimexpr\fontdimen22\textfont2\relax]
\node (r1) at (0,1) [rectangle, minimum width=8mm] {};
\node at (0,1) [draw, fill=black!50, rectangle, minimum width=16mm] {};
\node (r2) at (0,-1) [rectangle, minimum width=8mm] {};
\node at (0,-1) [draw, fill=black!50, rectangle, minimum width=16mm] {};
\draw (r1.south west) to (r2.north west);
\draw (r1.south east) to (r2.north east);
\end{tikzpicture}
\ = \ 
\begin{tikzpicture}[baseline=-\the\dimexpr\fontdimen22\textfont2\relax]
\node (r1) at (0,1.2) [rectangle, minimum width=8mm] {$p_1^\mu$};
\node at (0,1.2) [draw, rectangle, minimum width=16mm] {$p_1^\mu$};
\node (r2) at (0,-1.2) [rectangle, minimum width=8mm] {$p_2^\mu$};
\node at (0,-1.2) [draw, rectangle, minimum width=16mm] {$p_2^\mu$};
\node at (-0.75,0) {$s_{12}$};
\node at (0.75,0) {$s_{12}$};
\draw (r1.south west) to (r2.north west);
\draw (r1.south east) to (r2.north east);
\end{tikzpicture}
\ = s_{12}^2 + \text{permutations},
\end{equation}
\else
\begin{equation}
\adjustbox{valign=c}{\includegraphics{diag2-2}} \ = \ 
\adjustbox{valign=c}{\includegraphics{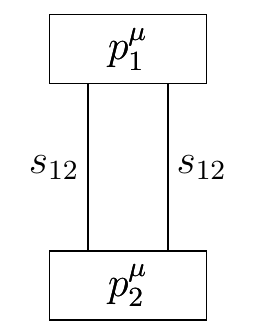}} 
\ = s_{12}^2 + \text{permutations},
\end{equation}
\fi
which gives the first line of \eqref{Mandelstam poly 2-4}. Similarly, a basis of the completely symmetric polynomials of degree three is expressed as
\ifusetikz
\begin{equation}
\begin{tikzpicture}[baseline=-\the\dimexpr\fontdimen22\textfont2\relax, scale=0.6,
 every node/.style={scale=0.6}]
\node (r1) at (0,1) [rectangle, minimum width=10mm] {};
\node at (0,1) [draw, fill=black!50, rectangle, minimum width=18mm] {};
\node (r2) at (0,-1) [rectangle, minimum width=10mm] {};
\node at (0,-1) [draw, fill=black!50, rectangle, minimum width=18mm] {};
\draw (r1.south west) to (r2.north west);
\draw (r1) to (r2);
\draw (r1.south east) to (r2.north east);
\end{tikzpicture}
\ = s_{12}^3 + \text{permutations}, \qquad
\begin{tikzpicture}[baseline=-\the\dimexpr\fontdimen22\textfont2\relax, scale=0.6,
 every node/.style={scale=0.6}]
\node (r1) at (0,1) [rectangle, minimum width=5mm] {};
\node at (0,1) [draw, fill=black!50, rectangle, minimum width=12mm] {};
\node (r2) at (-1.2,-0.8) [minimum width=5mm, rotate=120] {};
\node at (-1.2,-0.8) [draw, fill=black!50, rectangle, minimum width=12mm, rotate=120] {};
\node (r3) at (1.2,-0.8) [rectangle, minimum width=5mm, rotate=240] {};
\node at (1.2,-0.8) [draw, fill=black!50, rectangle, minimum width=12mm, rotate=240] {};
\draw[bend left=15] (r1.south west) to (r2.south east);
\draw[bend left=15] (r2.south west) to (r3.south east);
\draw[bend left=15] (r3.south west) to (r1.south east);
\end{tikzpicture}
\ = s_{12} s_{23} s_{31} + \text{permutations}.
\end{equation}
\else
\begin{equation}
\adjustbox{valign=c}{\includegraphics{diag3-3}} \ = \ s_{12}^3 + \text{permutations}, \qquad
\adjustbox{valign=c}{\includegraphics{diag2-2-2}} \ = \ s_{12} \, s_{23} \, s_{31} + \text{permutations}.
\end{equation}
\fi
One can check that the length eight graphs \eqref{LNZM L=8 graphs} reproduce the polynomials of degree four in \eqref{Mandelstam poly 2-4}.

This pattern continues. At $L=10$ we have $\cZ'_5 =11$. The correspondence can be seen from
\begin{align}
&\cC_{ijklm} \, \cC_{ijklm} \,, \quad 
\cC_{ijklm} \, \cC_{ijk} \, \cC_{lm} \,, \quad 
\cC_{ijkl} \, \cC_{ijkm} \, \cC_{lm} \,, \quad 
\cC_{ijkl} \, \cC_{ijm} \, \cC_{klm} \,,
\notag \\
&\cC_{ijkl} \, \cC_{ij} \, \cC_{km} \, \cC_{lm} \,, \quad
\cC_{ijk} \, \cC_{ijl} \, \cC_{km} \, \cC_{lm} \,,\quad 
\cC_{ijk} \, \cC_{ilm} \, \cC_{jl} \, \cC_{km} \,,\quad 
\cC_{ijk} \, \cC_{ijk} \, \cC_{lm} \, \cC_{lm} \,,
\notag\\
&\cC_{ijk} \, \cC_{lmk} \, \cC_{ij} \, \cC_{lm} \,,\quad
\cC_{ij} \, \cC_{ji} \, \cC_{kl} \, \cC_{lm} \, \cC_{mk} \,,\quad 
\cC_{ij} \, \cC_{jk} \, \cC_{kl} \, \cC_{lm} \, \cC_{mi} \,,
\label{products sym trsless 5}
\end{align}
and
\begin{align}
&s_{12}^5 + \dots, \quad
s_{12}^3 s_{13}^2 + \dots, \quad
s_{12}^3 s_{13} s_{23} + \dots, \quad
s_{12}^2 s_{13}^2 s_{23} + \dots,
\notag \\
&s_{12}^2 s_{13} s_{14} s_{34} + \dots, \quad
s_{12}^2 s_{13} s_{24} s_{34} + \dots, \quad
s_{12} s_{13} s_{14} s_{23} s_{24} + \dots, \quad
s_{12}^3 s_{34}^2 + \dots,
\notag \\
&s_{12}^2 s_{13} s_{34}^2 + \dots,\quad
s_{12}^2 s_{34} s_{35} s_{45} + \dots, \quad
s_{12} s_{23} s_{34} s_{45} s_{51} + \dots,
\label{Mandelstam poly 5}
\end{align}
where $+ \dots$ means the sum over the permutations $\cS_n$\,.
Note that the all terms add up with the same sign.

\bigskip
It is not obvious to prove the linear independence when the massless momentum conservations \eqref{Mandelstam constraints} are imposed.
For example, there are three candidates of the symmetric polynomials at degree two,
\begin{equation}
P_{2,1} = s_{12}^2 + \dots , \qquad
P_{2,2} = s_{12} s_{13} + \dots , \qquad
P_{2,3} = s_{12} s_{34} + \dots .
\label{degree two polys}
\end{equation}
It turns out that $P_{2,2} = - P_{2,1}$ and $P_{2,3} = 2 \, P_{2,1}$ when the constraints \eqref{Mandelstam constraints} are imposed.
This can be shown, for example, by applying the condition $\sum_j s_{ij} = 0$ repeatedly, until there are no indices which appear only once. 
We checked that the polynomials written in our ``graphical basis'' are linearly independent up to $K \le 5$. It also follows that our basis is complete thanks to $\cZ_{2K} = \cZ'_K$\,.

We also emphasize that it is non-trivial to construct an explicit basis of the completely symmetric polynomials of $\{ s_{ij} \}$.
A candidate is given by taking multi-traces of the matrix of Mandelstam variables,
\begin{equation}
P (\vec k, \vec \ell, \dots ) = \tr (S_{k_{1}} S_{k_{2}} \dots ) \, \tr (S_{\ell_{1}} S_{\ell_{2}} \dots ), \qquad
S_{k} = 
\begin{pmatrix}
0 & s_{12}^k & \dots & s_{1n}^k \\
s_{12}^k & 0 & \dots & s_{2n}^k \\
\vdots & \vdots & \ddots & \vdots \\
s_{1n}^k & s_{2n}^k & \dots & 0
\end{pmatrix}, \quad
(k \ge 0).
\end{equation}
These polynomials are manifestly invariant under permutations, but highly redundant even before imposing the massless momentum conservation. It is not clear how to take a linearly independent set.

When we turn on masses, the momentum conservation becomes
\begin{equation}
\sum_{j=1}^n s_{ij} = M_i^2 \,, \qquad
\( \sum_{i=1}^n p_i^\mu \)^2 = \sum_{i<j}^n s_{ij} = 2 \sum_{i=1}^n M_i \,,
\end{equation}
and the number of independent completely symmetric polynomials increases. For example, three degree-two polynomials in \eqref{degree two polys} are no longer proportional. 
The corresponding generalization in $\cN=4$ SYM will be to consider $U(N_c)$ gauge group instead of $SU(N_c)$.

It is interesting to see how the above relation can be proven for general $K$, and whether the correspondence can be extended to massive cases.

\section{Polynomial notation for multi-traces}\label{app:polynote}

A new concise notation for multi-trace operators is invented to reduce the computational workload significantly in {\tt Mathematica}. We call it polynomial notation, which can be used to describe any gauge invariant operators of $\cN=4$ SYM.

We associate a polynomial to each multi-trace operator. The basic idea is to describe the position inside a trace at which a given SYM field appears, rather than to describe which SYM field appears at a given position inside a trace.

First, consider single-trace operators. If a field $F$ stands at the $p$-th position inside the trace, we write $F^p$, then sum over all fields as
\begin{equation}
\tr \big( X Y Z Z Y^\dagger Z^\dagger \big) \ \leftrightarrow \ P (X + Y^2 + Z^3 + Z^4 + \olY^5 + \olZ^6 ),
\end{equation}
where $P$ means that this polynomial is defined modulo cyclic permutation of a trace,
\begin{equation}
P \( f + g^2 + \dots + y^{L-1} + z^L \) = P \( f^2 + g^3 + \dots + y^L + z \).
\label{P for st}
\end{equation}
For later purposes, we rewrite this equation as 
\begin{equation}
P=T \cdot P,
\end{equation}
where $T$ is the shift operator defined by
\begin{equation}
T \cdot P \( \sum_p F^p \) \equiv P \( \sum_p F^{[p+1]_L} \), \qquad
[a]_{L} = \begin{cases}
a+L &\quad (a \le 0)
\\
a &\quad (1 \le a \le L)
\\
a-L &\quad (L < a)
\end{cases}.
\label{def:shift operator}
\end{equation}

Next, we generalize this notation to the operators with flavor indices contracted, like $X_i X_i^\dagger$ for the $\alg{u}(3)$ singlets and $\Phi_I \Phi_I$ for the $\alg{so}(6)$ singlets, where $X_i = \Phi_{2i-1} + i \, \Phi_{2i}$\,.
If $X_i$ is at the $p$-th position and $X_i^\dagger$ is at the $q$-th position, we write
\begin{equation}
\tr \big( \dots X_i \dots X_i^\dagger \dots \big) \ \leftrightarrow \ 
P ( \phi^p \, \olphi^q + \dots ).
\label{polynote contract}
\end{equation}
The $\alg{so}(6)$ singlets can be expressed by imposing the relation $\phi^a \olphi^b \sim \phi^b \olphi^a$.\footnote{We keep $\olphi$ to avoid confusion between $\phi^a \phi^b$ and $\phi^{a+b}$.}
This polynomial notation is significantly simpler than the usual one because no flavor subscripts are used anymore. To see this, consider the following $\alg{u}(3)$ singlet operator of length four,
\begin{equation}
\tr ( X_{i_1} X_{i_2} X_{i_2}^\dagger X_{i_1}^\dagger )
= \tr( X_{i_2} X_{i_1} X_{i_1}^\dagger X_{i_2}^\dagger ) 
\ \leftrightarrow \ P (\phi \olphi^4 + \phi^2 \olphi^3 ).
\end{equation}
On the left-hand side, we wrote two identical operators generated by the permutation of flavor indices $(i_1 i_2 \dots i_{L/2}) \to (i_{\sigma(1)} i_{\sigma(2)} \dots i_{\sigma(L/2)})$ with $\sigma \in \cS_{L/2}$\,.
As $L$ increases, the order of the permutation group grows factorially. 
It becomes quite cumbersome to keep removing redundant elements generated by $\cS_{L/2}$\,. In this sense, the polynomial notation is factorially simpler to describe single-trace operators than the conventional notation.\footnote{The polynomial notation becomes less efficient if we consider multi-trace operators, particularly those containing single-traces of the same length. The computation using the polynomial notation is faster overall.}

Second, consider multi-trace operators. 
If a field $F$ stands at the $p$-th position inside the $m$-th trace, we write $F_m^p$, then sum over all fields as
\begin{equation}
\tr (Y X_i) \, \tr (\olY X_i^\dagger) \ \leftrightarrow \ P (Y_1 + \olY_2 + \phi_1^2 \, \olphi_2^2),
\end{equation}
where $\phi_m^p \olphi_n^q$ means that we contract $X_i$ at the $p$-th position in the $m$-th trace and $X_i^\dagger$ at the $q$-th position in the $n$-th trace. 
We impose the cyclicity condition $P$ as
\begin{equation}
P = T_m \cdot P \qquad \text{for each} \ m,
\label{P for mt}
\end{equation}
where $T_m$ is the shift operator \eqref{def:shift operator} for the $m$-th trace.

The polynomial notation is related to a graphical representation of multi-trace operators as follows:
\ifusetikz
\begin{gather}
\tr (\Phi_{i_1} \Phi_{i_2} \Phi_{i_3} \Phi_{i_1} \Phi_{i_2} \Phi_{i_3} ) = 
\begin{tikzpicture}[baseline=-\the\dimexpr\fontdimen22\textfont2\relax, scale=0.6, every node/.style={scale=0.6}]
\node (n1) at (0:\nrad) [wnode] {}; \node (n2) at (60:\nrad) [wnode] {};
\node (n3) at (120:\nrad) [wnode] {}; \node (n4) at (180:\nrad) [wnode] {};
\node (n5) at (240:\nrad) [wnode] {}; \node (n6) at (300:\nrad) [wnode] {};
\draw[br3] (n2) to (n5); \draw[br3] (n4) to (n1); \draw[br3] (n6) to (n3); 
\trfigVIc
\end{tikzpicture} \ 
= P \( \phi \olphi^4 + \phi^2 \olphi^5 + \phi^3 \olphi^6 \),
\label{diagram example1} \\[1mm]
\tr (\Phi_{i_1} \Phi_{i_2}) \, \tr (\Phi_{i_2} \Phi_{i_3}^\dagger \Phi_{i_3} \Phi_{i_1}) = 
\begin{tikzpicture}[baseline=-\the\dimexpr\fontdimen22\textfont2\relax, scale=0.6,
 every node/.style={scale=0.6}]
\node (n1) at (0:\nrad) [wnode] {}; \node (n2) at (60:\nrad) [wnode] {};
\node (n3) at (120:\nrad) [wnode] {}; \node (n4) at (180:\nrad) [wnode] {};
\node (n5) at (240:\nrad) [wnode] {}; \node (n6) at (300:\nrad) [wnode] {};
\draw[dashed] (n2) -- (n3);
\draw[dashed] (n1) -- (n4);
\draw[br6] (n3) to (n4); \draw[br6] (n5) to (n6); \draw[br6] (n1) to (n2);
\begin{pgfonlayer}{background}
\draw[dashed] (60:\nrad) arc (60:120:\nrad);
\draw[dashed] (180:\nrad) arc (180:240:\nrad);
\draw[dashed] (240:\nrad) arc (240:300:\nrad);
\draw[dashed] (300:\nrad) arc (300:360:\nrad);
\end{pgfonlayer}
\end{tikzpicture} 
= \ 
\begin{tikzpicture}[baseline=-\the\dimexpr\fontdimen22\textfont2\relax, scale=0.6,
 every node/.style={scale=0.6}]
\node (n1) at (0:\nrad) [wnode] {}; \node (n2) at (60:\nrad) [wnode] {};
\node (n3) at (120:\nrad) [wnode] {}; \node (n4) at (180:\nrad) [wnode] {};
\node (n5) at (240:\nrad) [wnode] {}; \node (n6) at (300:\nrad) [wnode] {};
\draw[dashed] (n2) -- (n3);
\draw[dashed] (n1) -- (n4);
\draw[bl3] (n2) to (n4); \draw[br6] (n5) to (n6); \draw[bl3] (n1) to (n3);
\begin{pgfonlayer}{background}
\draw[dashed] (60:\nrad) arc (60:120:\nrad);
\draw[dashed] (180:\nrad) arc (180:240:\nrad);
\draw[dashed] (240:\nrad) arc (240:300:\nrad);
\draw[dashed] (300:\nrad) arc (300:360:\nrad);
\end{pgfonlayer}
\end{tikzpicture} \ 
\label{diagram example2} \ 
= P \(\phi_1 \olphi_2^4 + \phi_1^2 \olphi_2 + \phi_2^2 \olphi_2^3 \),
\end{gather}
\else
\begin{gather}
\tr (\Phi_{i_1} \Phi_{i_2} \Phi_{i_3} \Phi_{i_1} \Phi_{i_2} \Phi_{i_3} ) 
= \adjustbox{valign=c}{\includegraphics{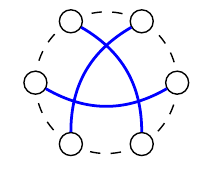}} \ 
= P \( \phi \olphi^4 + \phi^2 \olphi^5 + \phi^3 \olphi^6 \),
\label{diagram example1} \\[1mm]
\tr (\Phi_{i_1} \Phi_{i_2}) \, \tr (\Phi_{i_2} \Phi_{i_3} \Phi_{i_3} \Phi_{i_1}) 
= \adjustbox{valign=c}{\includegraphics{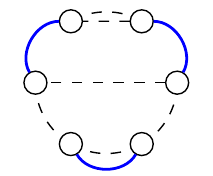}} 
= \adjustbox{valign=c}{\includegraphics{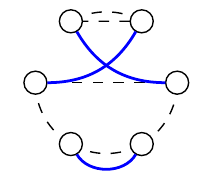}} \ 
= P \(\phi_1 \olphi_2^4 + \phi_1^2 \olphi_2 + \phi_2^2 \olphi_2^3 \),
\label{diagram example2}
\end{gather}
\fi
where the black dashed lines represent the $SU(N_c)$ color traces, and the blue solid lines represent $\alg{so}(6)$ flavor contractions. The middle equality in the last line comes from the cyclic symmetry of the trace.
In short, the polynomial notation describes the configuration of the blue lines.
Note also that a similar graphical method was useful in finding the relation to Mandelstam variables in Appendix \ref{app:Mandelstam}.

We generated all $\alg{so}(6)$ singlet multi-trace operators up to $L=12$ using {\tt Mathematica}, and implemented the action of the non-planar dilatation operator in the polynomial notation. The details are given in the attached file.

\let\oldbibliography\thebibliography
\renewcommand{\thebibliography}[1]{\oldbibliography{#1}
\setlength{\itemsep}{-1pt}}
\renewcommand{\baselinestretch}{1.0}

\end{document}